\documentclass[twocolumn]{aastex631}
\usepackage{graphicx}
\usepackage{tabularx}
\usepackage{subfigure}

\begin{document}

\title{Studying Outflows with Synthetic Absorption Line Spectra from High Resolution Simulations}

\author[0009-0005-1612-9946]{Jake Magee}
\affiliation{Department of Physics and Astronomy, University of Pittsburgh, 3941 O’Hara St, Pittsburgh, PA 15260}
\email{jpm263@pitt.edu}

\author[0000-0001-9735-7484]{Evan Schneider}
\affiliation{Department of Physics and Astronomy, University of Pittsburgh, 3941 O’Hara St, Pittsburgh, PA 15260}

\author[0000-0001-6248-1864]{Kate H. R. Rubin}
\affiliation{Department of Astronomy, San Diego State University, San Diego, CA 92182 USA}

\author[0000-0002-2491-8700]{S. Alwin Mao}
\affiliation{Department of Physics and Astronomy, University of Pittsburgh, 3941 O’Hara St, Pittsburgh, PA 15260}

\author[0000-0002-3817-8133]{Cameron Hummels}]
\affiliation{TAPIR, California Institute of Technology, Pasadena, CA 91125, USA}

\begin{abstract}

Down-the-barrel absorption line spectra are a powerful probe of the properties of multiphase galactic outflows. In this work, we demonstrate a method to generate realistic synthetic spectra using high resolution ($\Delta x=5-20 \ \mathrm{pc}$), 20-kiloparsec-scale outflow simulations from the next generation of the CGOLS project. Aiming to mimic modern observational surveys of nearby star-forming galaxies, we generate a suite of UV absorption lines, from which we measure common observables such as equivalent widths and velocity statistics, and compare them to observed values. The velocity statistics from the simulated data agree well with high-resolution, far-UV HST treasury data, but the equivalent widths (EWs) display more variability. Our fiducial EWs are systematically lower than the empirical predictions by $\sim50-97\%$, particularly in the lines with higher ionization potentials, though these values increase with inclination angle, star formation rate, and time. We find a strong correlation between the measured outflow velocity and ionization potential, reminiscent of the known correlation between velocity and temperature from the simulations. We also study the effect of simulation resolution on our statistics, finding the EWs to be much more sensitive to the simulation resolution than the velocities. Overall, we find that we can generate realistic synthetic spectra from high-resolution hydrodynamic simulations that are qualitatively similar to observations. In future work, we will further improve the physical realism of our synthetic spectra by implementing a more accurate ionizing background spectrum, with the ultimate goal of improving constraints on derived outflow properties from observations.

\end{abstract}

\keywords{}

\section{Introduction} \label{sec:intro}

Galactic outflows are central to our current understanding of the feedback processes that drive galaxy evolution, and are seen ubiquitously across cosmic time and galaxy morphologies \citep{rupke2018review}. Observational studies aimed at investigating the relationship between outflows and their hosts have been conducted since the 1960s \citep{lynds1963evidence}, and researchers have clearly demonstrated that the structure and evolution of interstellar and circumgalactic media (ISM, CGM) are greatly impacted by outflows \citep{saintonge2022cold, naab2017theoretical, somerville2015physical}. Though it is clear that various evolutionary mechanisms, such as gas accretion, ejection, and star formation regulation, must rely on feedback processes driven by outflows, their exact role is not well constrained \citep{heckman2017galactic, veilleux2020cool}. 

The ubiquity and observed importance of outflows necessitates the development of theoretical models in order to better interpret observations and allow us to put better constraints on the inferred properties (e.g. mass, energy, and momentum outflow rates). Outflows are inherently multiphase, and may be driven by a hot ($\rm T\simeq10^7\ K$) phase \cite[see][for a foundational model of a hot phase outflow]{chevalier1985wind}, but the properties of this hot phase are difficult to directly constrain observationally, as there are few X-ray luminous objects driving these winds that are close enough to produce sufficient signal to be detected, and only one for which the hot phase velocity has been measured \citep{strickland2009supernova, audard2026fast}. 

While the cooler phases are easier to observe, it remains unclear if these observations alone are sufficient to describe the entire multiphase outflow. Using observations of the cool phase combined with theoretical models, we might hope to extrapolate outflow properties from measurements of features in galaxy spectra, for example, UV absorption lines measured from ``down-the-barrel" observations. The ionic species traced by these absorption lines (most commonly silicon, carbon, and oxygen) span a large range of excitation potential, and thus probe a large range of gas temperatures ($\rm T \simeq [10^4, \ 10^6] \ K$). Much observational effort has been dedicated to measuring properties of these absorption lines, such as velocity statistics, full width half maxima (FWHMs), and equivalent widths (EWs), for large samples of galaxies \cite[e.g.][]{rubin2014evidence, heckman2015systematic, zhu2015near, chisholm2016shining, chisholm2016robust, chisholm2017galaxies, carr2021effects, xu2022classy}. 

In order to build a more complete picture of the underlying physics, we would ideally be able to accurately approximate the mass, energy, and momentum outflow rates based on derivations of optical depth, covering fractions and column densities from observational data \citep{chisholm2015scaling}. Though absorption line properties (like EWs and central velocities) are relatively straight-forward to measure accurately, extrapolating these more complex quantities from the data necessitates making assumptions about the underlying structure of the outflow (e.g. full versus partial covering models;  \citealt{chisholm2016shining}). Additionally, the quantities derived via these assumptions most directly, the optical depth and covering fraction, are typically degenerate with one another (unless measured from a multiplet transition). Another major challenge is the general lack of observational information about the spatial extent of the outflow \citep{burchett2021circumgalactic}, which is necessary for computing outflow rates, but cannot be inferred directly from absorption line studies. Without an associated emission line tracer of the extent and morphology of the outflow, one must make an assumption without any real method for testing its validity, further exacerbating the uncertainties and limitations of these empirical models. Disentangling these uncertainties therefore necessitates a method for reliably checking the validity of the inferred properties, which is difficult to accomplish through observation alone.

Simulations are a promising avenue for studying outflows and investigating the limitations of current empirical models, however they must have high enough resolution at large enough scales to accurately model the physics driving the evolution of outflows. Cosmological simulations typically have to assume outflow properties like mass outflow rates and velocities, making them less suitable for investigation into feedback-driven processes from first principles. Kiloparsec-scale tall-box simulation suites with higher resolution \citep[e.g.][]{walch2015silcc, kim2018numerical} can directly resolve supernova-driven feedback in the ISM and are thus able to capture the manner in which stellar feedback drives gas out of galaxies, but their simulation domain sizes are not large enough to make direct comparisons to observational data that averages over entire galaxies. 

The Cholla Galactic Outflow Simulations (CGOLS) project \citep{schneider2018introducing} comprises a happy medium between these options. CGOLS is a suite of uniformly high-resolution ($\Delta x\simeq5-20 \ \rm pc$) hydrodynamic simulations of isolated galaxies designed specifically to investigate the physics of multiphase galactic outflows. This uniformly high resolution is implemented in large simulation boxes ($\rm 20-40 \ kpc$ in all directions) which capture the full extent of the galaxies they study, and the simulations include a realistic model for supernova-driven feedback \citep{schneider2020physical, schneider2024cgols}. The next generation of CGOLS expands on the study of disk-wide supernova-driven galactic outflows by exploring the effects of star-formation rate (SFR), as well as expanding the physical bounds of the simulation. These simulations, dubbed CGOLS SFR (Schneider et al., \textit{in prep}), implement the same galactic initial conditions as previous CGOLS models, but host the galaxy in a simulation domain 32-times the volume of the previous generation. Even within this vastly-larger simulation domain size, CGOLS SFR still uses a uniformly-high resolution of $\Delta x\simeq20 \ \rm pc$, thus capturing the critical hydrodynamic interactions between phases in the outflows while following their evolution over many kiloparsecs. 

CGOLS provides a perfect testbed to investigate the link between simulated and observed properties of outflows, but in order to do so, we must first generate realistic mock observables from the simulated data. \textit{Trident} \citep{hummels2017trident} was created to do just this. Built atop the \textit{YT} \citep{turk2010yt} framework for analyzing hydrodynamical simulations, \textit{Trident} enables the generation of absorption line spectra from simulations. Initially created to model quasar absorption line spectra used to study the CGM, like those studied in the COS-Halos survey \citep{werk2013cos, tumlinson2013cos, werk2016cos}, \textit{Trident} generates pencil-beam LightRay objects and populates Voigt profiles along their paths according to the gas properties in the simulation. \textit{Trident} has enabled the development of physically-motivated models informed by simulations that can better constrain the uncertainties in observations \citep{lehner2020project, hafen2024halo21}. Other theoretical studies have utilized \textit{Trident}, or similar analytical frameworks (e.g. semianalytic line transfer and partial covering models), to generate mock down-the-barrel observations utilizing data from cosmological zoom-in simulations \citep[e.g.][]{carr2025evaluating}, as well as smaller-scale, higher-resolution data from cloud-wind simulations \citep[e.g.][]{de2021synthetic, casavecchia2024imprint}. While these studies have made important connections between mock observations and real data, they either lack sufficient resolution or physical scale to adequately address the uncertainties in empirically derived models of multiphase outflows.  Ideally, we would like to leverage this tool to make similar comparisons for down-the-barrel observations, with the hope of improving the accuracy of the models used to interpret observations \citep[e.g.][]{carr2025evaluating}. Though \textit{Trident} has demonstrated its strength as a tool for enabling such analyses for background quasar surveys, we must expand its current capabilities for our specific use-case.

The pencil-beam spectra generated by \textit{Trident} cannot immediately be used to model down-the-barrel observations, as the underlying assumption of \textit{Trident} is that gas structures are much larger in physical extent than the background light source. By contrast, the gas structures we observe in outflows are cloud-like and much smaller than their background light source, which in the case of mock down-the-barrel observations is the integrated light of the whole galaxy. However, we can generate and average together a suite of pencil-beam spectra that emulate the distributed light sources we would observe in down-the-barrel observations. Leveraging the existing spectrum generation framework in \textit{Trident}, we have implemented a novel method for emulating galaxy-wide observations, leading to more realistic synthetic spectra that can be meaningfully compared with observational data.

Here, we present our initial analysis of the CGOLS SFR datasets, and our current ability to produce realistic mock down-the-barrel spectra from high-resolution simulations. In \S\ref{sec:simulation}, we describe the CGOLS SFR simulation set-up, and in \S\ref{sec:methods} we introduce our methods for spectrum generation and analysis, including our novel methods for expanding the current capabilities of \textit{Trident} to model galaxy-scale down-the-barrel observations. In \S\ref{sec:results} we present our mock observables and an analysis of the effects of spectral and simulation resolution on these observables, as well as an investigation into the trends seen in these statistics as a function of the ionization potentials of the atomic species producing the lines. Additionally, we study how the spectral features evolve with time, inclination angle, and SFR, as well as how the fiducial spectrum differs from that of a simulation run in a smaller box at higher resolution. We discuss these mock observables in the context of real observations in \S\ref{sec:discussion}, as well as the current shortcomings of our spectrum generation and analysis pipeline. We summarize our results and outline future goals in \S\ref{sec:conclusions}.

\section{The Simulations: CGOLS SFR} \label{sec:simulation}

The CGOLS SFR suite of simulations host an M82-like ($\rm10^{10} \ M_\odot$ starburst) galaxy, with initial conditions very similar to that of CGOLS V \citep{schneider2024cgols}. A companion paper to this work (Schneider et al., \textit{in prep}) further details the setup of this suite of simulations, so we focus here on the features of CGOLS SFR that are most pertinent to this work. First is the size of the simulation box, which has dimensions of $L_x=L_y=L_z=40 \ \mathrm{kpc}$. This volume is 32 times larger than that of the previous generations of CGOLS simulations, allowing the outflows to evolve further in three dimensions before reaching the edge of the simulation box. Despite the substantial increase in simulation box volume, CGOLS SFR still maintains a constant physical resolution of $\Delta x \approx20 \ \mathrm{pc}$. The fiducial data is also rebinned in post-processing at 3 lower resolutions ($\Delta x \approx40, \ 80, \& \ 160 \ \mathrm{pc}$), enabling us to study the effects of simulation resolution on mock observables.\footnote{We emphasize that this rebinning allows us to test the effects of simulation resolution on mock spectra while ensuring that the underlying physical properties of the data remain fixed. This is different than measuring the effects of running the simulation with a different native resolution, which could change the underlying structure of the outflow. We address this point in Section \ref{subsec: sim comp}.}

Like CGOLS V, the CGOLS SFR simulations implement a distributed cluster model for the sources of supernova feedback. Clusters have a radius $R_{\rm cl}=30 \ \mathrm{pc}$, follow a cluster mass function of $PDF\propto \rm M_{cl}^{-2}$ (with masses between $10^4\rm M_\odot$ and $10^{5.5}M_\odot$), and are distributed throughout the disk according to $N_{\rm cl}\propto \mathrm{R}e^{-\rm R}$, where R is the radial coordinate in the galaxy (with a maximum radius of $4.5$kpc). The ISM consists of a smooth, single phase disk at $10^4\rm \ K$\footnote{The cooling curve cuts off at this temperature, although gas can still achieve lower temperatures through adiabatic expansion.} that is initially in rotational and vertical hydrostatic equilibrium \cite[see][for more details]{schneider2024cgols}. Though the implementation of the clusters is identical to that of CGOLS V, the total star formation rate varies between simulations. 

\begin{figure*}[!t]
  \centering
  \subfigure[]{\includegraphics[scale=0.41]{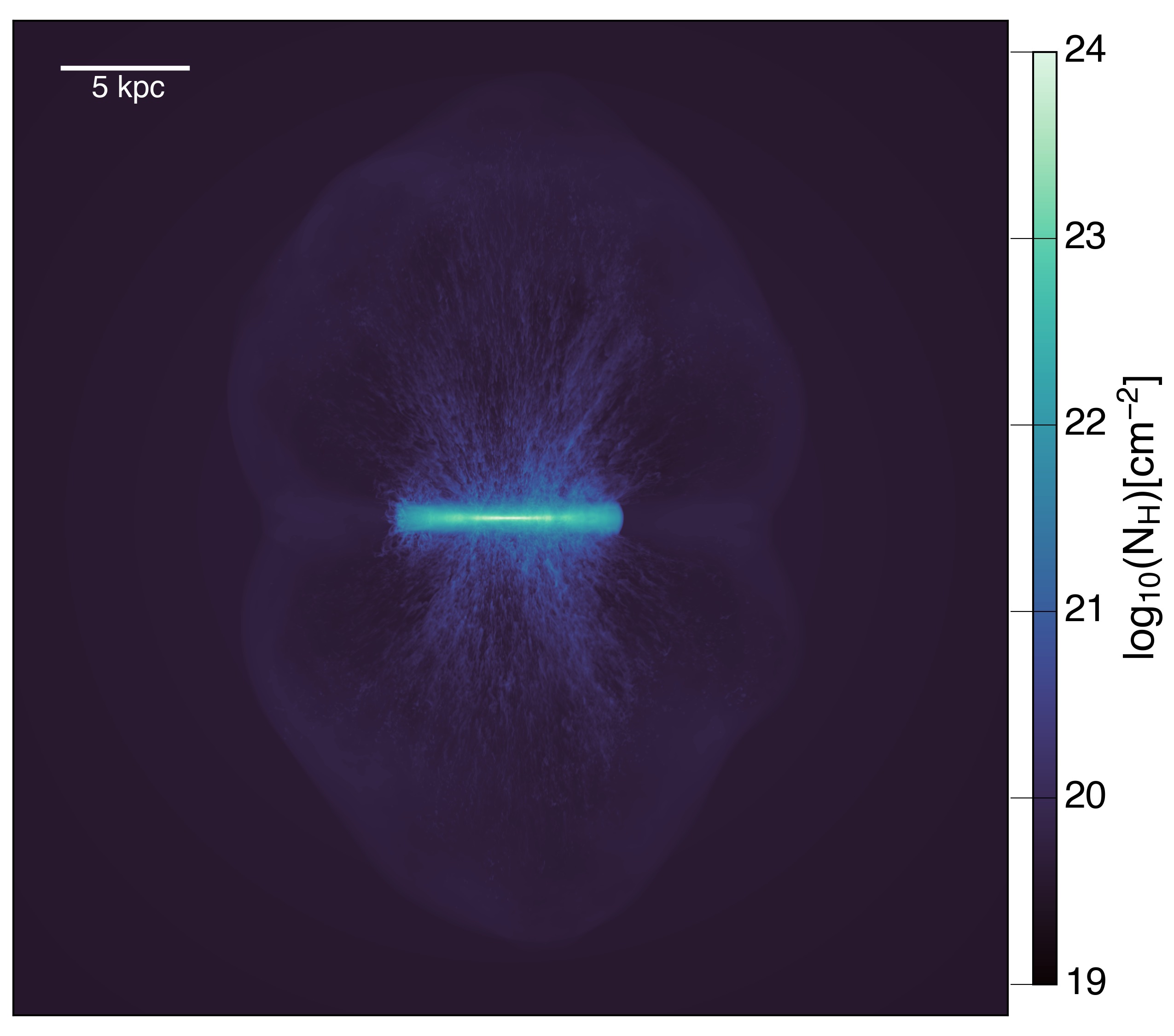}}
  \subfigure[]{\includegraphics[scale=0.41]{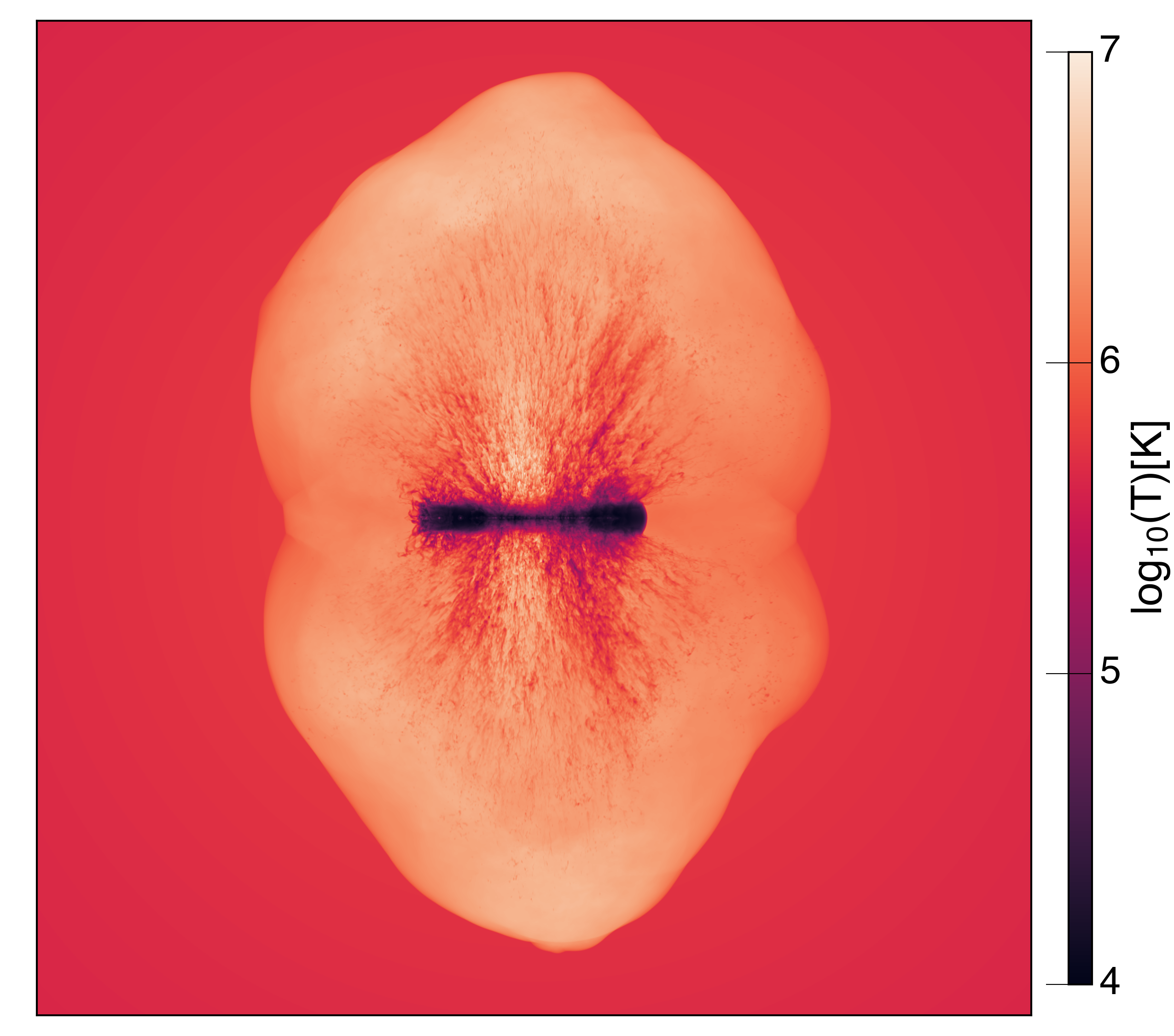}}
  \caption{(a) x-z projection of gas column density. (b) Density-weighted x-z projection of gas temperature. Projections are from a snapshot of the fiducial simulation ($\Delta x = 20 \ \rm pc$, SFR = $20 \ \rm M_\odot \ yr^{-1}$), 30 Myr after the onset of feedback.}
  \label{fig:proj plots}
\end{figure*}

In Figure \ref{fig:proj plots}, we show x-z plane projections of the gas density and density-weighted gas temperature for the $\Delta x =20 \ \rm pc$ resolution simulation, $30 \ \rm Myr$ after the onset of feedback from the star-forming clusters with a constant SFR of $20 \ \rm M_{\odot} \ \mathrm{yr^{-1}}$. By this time, the outflow has a well-developed multiphase structure, and we use this as the fiducial simulation resolution, SFR, and time-snapshot for the analysis in this work. Though this snapshot will be the primary focus of the analysis in this work, we will also investigate the effects of all of these parameters (resolution, time, and SFR) on the spectral absorption features. The left panel of Figure \ref{fig:proj plots} shows the column density distribution for the fiducial snapshot, with the highest density gas concentrated in the ISM, as well as lower-density filamentary structures evolving in a bi-conical outflow. A bubble of low-density, diffuse gas has nearly reached the edge of the simulation domain, with the aforementioned filamentary structure residing within this volume-filling phase, and higher projected densities closer to the disk. Looking at the temperature projection in panel (b), we see that the diffuse gas filling the bulk of the volume of the outflow is primarily hot at temperatures of $\rm T\sim10^{6-7}\ K$, while the denser filamentary structures are at substantially lower temperatures, ranging between $\rm T\sim10^{4-5.5}\ K$.

\section{Methods} \label{sec:methods}

In this section we outline our procedure for generating mock down-the-barrel spectra from key snapshots of the CGOLS SFR simulation suite. In \S\ref{subsec:spec-gen} we describe our method for generating synthetic spectra, including the weighting scheme and some comments on the ionizing background. We outline our method for fitting the spectra in \S\ref{subsec: spec fit}

\subsection{Spectrum Generation} \label{subsec:spec-gen}

``Down-the-barrel" absorption spectra arise from systems in which the integrated background light from the galaxy propagates through the surrounding ISM and CGM, and are ideally measured from ``face-on" inclination galaxies, but galaxies in observational studies often have a range of inclination angles \cite[e.g.][]{rubin2014evidence}. As this light interacts with the multiphase ISM / CGM, continuum flux is absorbed by intervening gas along the line-of-sight, populating the observed spectrum with absorption lines. These absorption lines are sensitive to the density, temperature, and relative velocity of the absorbing gas that produce them, thus making these spectral features a valuable tool for probing the multiphase structure of the ISM and CGM of the galaxies being observed. In particular, these observations are well-suited to measure the properties of star-formation-driven outflows, which span orders of magnitude in their density, temperature, and velocity structure. The velocity information in the spectra is the primary feature that makes ``down-the-barrel" observations so valuable, as outflows typically have velocities several times larger than typical ISM velocities, so the two components can be separated based on the velocity structure of the absorption features.

In order to generate integrated mock down-the-barrel spectra, we must first generate many individual line-of-sight LightRay objects at the resolution of the simulation that propagate through the ISM and CGM material of the simulated galaxy. We define the aperture for our mock observation as the region in the x-y plane of the simulation box that contributes to the background stellar continuum; in this case, a circular region with a radius of $5 \rm  \ kpc$ (five-times the scale radius of the star cluster distribution) centered at $\rm x=y=0 \ kpc$. This ensures we capture most of the background light that contributes to the ionizing radiation, while also minimizing the number of sightlines in the galaxy outskirts, in order to reduce the computational expense associated with generating the spectra. This approach is predicated on the assumption that sightlines in the outskirts of the galaxy will have little-to-no contribution to the average spectrum.

\textit{Trident} was originally developed with angularly-small background sources in mind, such as quasars at great distance. Its 1-dimensional `LightRay` objects emulate this behavior to probe the gas along the line of sight connecting a distant object and the observer. As it travels, the LightRay object stores the information from each cell it passes through, and a spectrum is then generated based upon that information for each ray. The gas-density, gas-temperature, gas-velocity, and elemental abundance (if available) in each cell are processed as \textit{YT} field parameters, and \textit{Trident} then uses this information to query a table of ionic abundances generated from \textit{Cloudy} (assuming a solar abundance pattern for a single metallicity field, and a metagalactic UV ionizing background). After querying this table for the atomic species specified by the user, \textit{Trident} then deposits a Voigt profile for each cell at the corresponding location in velocity (or wavelength) space (accounting for both doppler redshift in the gas and cosmological redshift from the Hubble flow), resulting in a final spectrum which features absorption lines from a variety of ions at a variety of temperatures, densities, and velocities along the path of the LightRay.

For each cell in our aperture, we generate a \textit{Trident} LightRay object that propagates from the mid-plane of the simulation box to the top edge of the box in the z-direction. For each LightRay object, we generate a separate spectrum accounting for absorption from oxygen, carbon, and silicon lines. When generating these spectra, we use the spectral resolution of HST's Cosmic Origins Spectrograph (COS), which has been used to measure high-sensitivity medium-resolution UV spectroscopy of star-forming galaxies in multiple HST surveys. This results in a suite of thousands to hundreds of thousands of 1-dimensional absorption line spectra to cover the full aperture. In order to generate a composite spectrum for the aperture as a whole, we must then average together all of the individual line-of-sight spectra in a physically motivated way to produce a realistic, COS-like down-the-barrel spectrum.

\subsubsection{Weighting Function} \label{subsubsec:weighting}

Once armed with the full suite of line-of-sight spectra for each xy-location in our aperture, we can then construct the composite spectrum for the entire aperture. As outlined previously, the way in which we construct this composite spectrum must be physically motivated, since not every line-of-sight will contribute the same amount of continuum flux. Because the extended background source of a star-forming galaxy galaxy is not uniform in its brightness, we apply a weight to the individual spectra prior to averaging the full spectral suite that reflects this intrinsic light profile. In our fiducial model, this weight is assigned according to the radial distance from the center of the simulation box. The specific weighting function is an exponential with a scale radius of 1 kpc ($\rm w=e^{-R/1 \ kpc}$, where $\mathrm{R}=\sqrt{x^2+y^2}$). This weighting function and scale radius are chosen to match the distribution of star-forming clusters in the simulation, as well as the typical distribution of stars within high-redshift star forming galaxies.

\begin{figure*}[!t]
    \centering
    \subfigure[]{\includegraphics[scale=0.335]{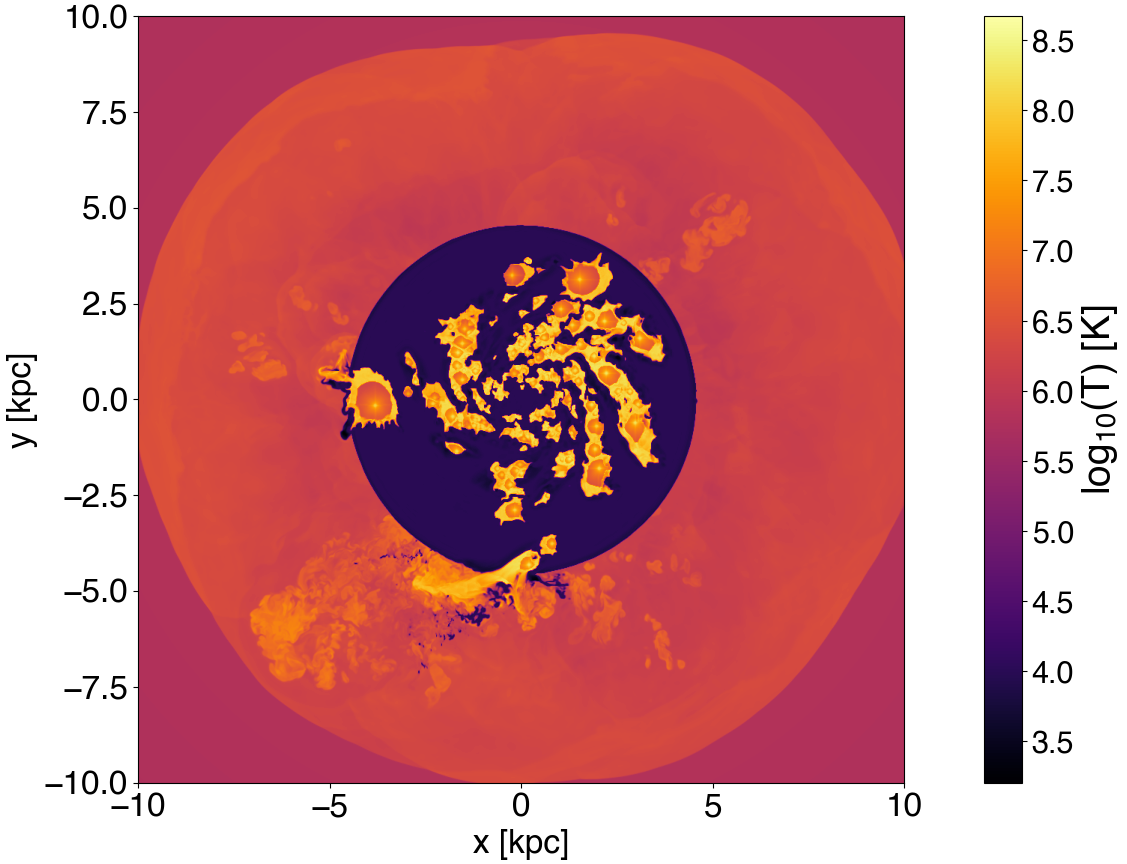}}
    \subfigure[]{\includegraphics[scale=0.331]{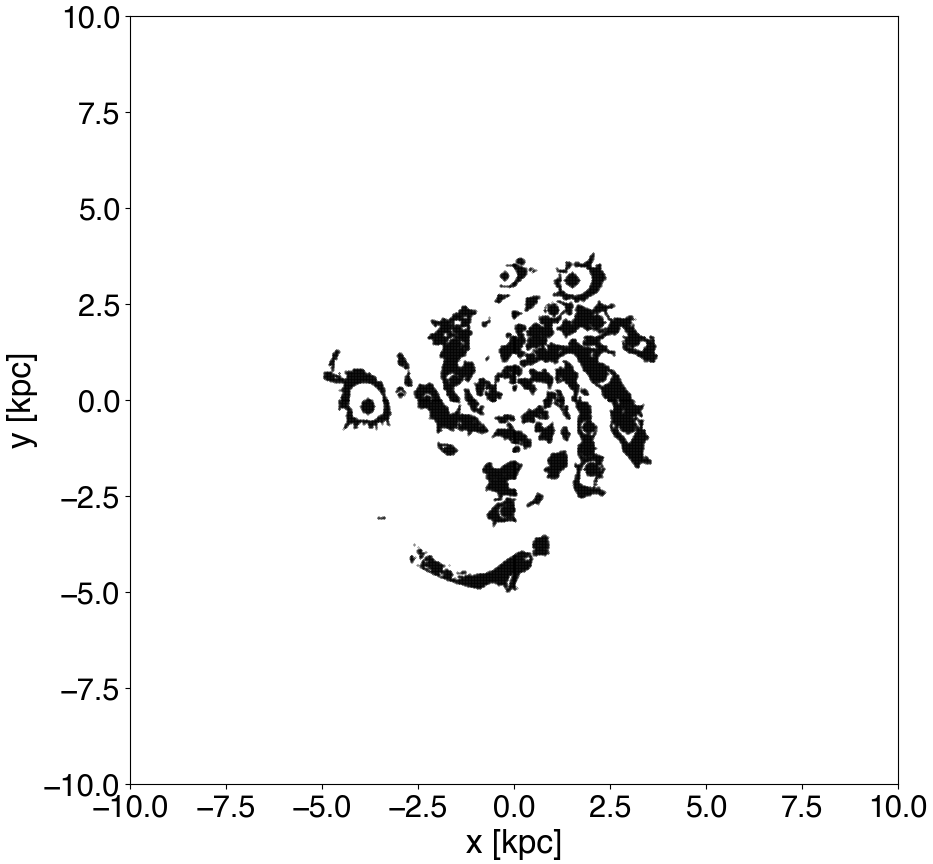}}
    \caption{(a) Gas-temperature slice through the midplane (z=0 kpc). (b) $\rm T>10^7 \ K$ gas-temperature selection mask from the $\Delta x=20 \ \rm pc$ resolution simulation.}
    \label{fig:star regions}
\end{figure*}

One additional weighting scheme we investigate is selecting only the cells in the galaxy aperture that are back-lit by star-cluster regions in the simulation. This selection mask is motivated by the fact that the UV light in a real star-forming galaxy is dominated by the output of young massive stars. By masking for only star-forming regions in our simulation data, we can test to what extent this effect manifests in the synthetic spectra, as it is not entirely clear whether a simple exponential light profile or a light background dominated by young star clusters is the better approximation for simulated data. We make this selection mask via a temperature cut, which selects for cells with an average temperature above $10^7 \ \rm K$ (as shown in Figure \ref{fig:star regions}).

\subsubsection{Ionizing Background}

As explained in the previous section, we use \textit{Trident} to generate the ionic abundances along each ray object for each species we study here. \textit{Trident} does this by referencing precomputed multidimensional tables of ionic abundances, calculated for a range of number densities and temperatures with the radiative transfer code \textit{Cloudy} \citep{2013RMxAA..49..137F}. The fiducial ionizing background spectrum for \textit{Trident} is the metagalactic UV background from \cite{haardt2012radiative}, which we will refer to as ``HM12." Though HM12 is a reasonable choice for mock data focused on galaxy halos, it is not as well suited for our particular use-case. Close to the galaxy, the ionizing spectrum should be dominated by the light from the star clusters, and the characteristic spectral energy distribution (SED) of a young starburst has substantially harder UV radiation than is present in HM12. Although the spectral shape is straightforward to modify, assigning an appropriate normalization of the starburst SED is more challenging, as it should depend both on the overall mass in young clusters and on the distance from the light sources. For this work, we adopt Trident's fiducial HM12 ionizing background spectrum, but we investigate the limitations of this choice in \S\ref{subsec: level pop}.

\subsection{Spectrum Fitting} \label{subsec: spec fit}

In order to compare across galaxies with a range of properties, many observational studies of down-the-barrel spectra measure velocity and EW statistics, with the expectation that these easily-measurable features are reflective of the underlying physical properties of outflows. These ``lower-level" statistics are model-independent probes of the multiphase structure of the outflows they are measured from, and can then be used (in tandem with a line profile model) to derive more complex quantities such as optical depth and column density, and ultimately to estimate outflow rates. Measuring these quantities is key to understanding the structure of observed outflows, but most models are limited by large uncertainties in the extrapolated properties. One promising avenue to disentangle these uncertainties and better constrain the underlying physics are meaningful comparisons between observed and synthetic spectra. Because they represent a ``ground truth" for the mock spectra, high-resolution simulated data allow us to explore the physics behind these models \citep[e.g.][]{hafen2024halo21}.

In order to make a first comparison between our synthetic data and real observations, the features in the mock spectra are fit using similar techniques to those employed in galaxy surveys. For the velocity statistics that we present in the following section, two gaussians are fit to each absorption line, one fixed at $v=0 \ \rm km \ s^{-1}$ to fit the static ISM component (to both the red and blue portions of the line component), and a second that fits for the blue-shifted outflow component with $v<0 \ \rm km \ s^{-1}$. Informed by the prescription outlined in \cite{weiner2009ubiquitous}, we divide out the static fit from the spectrum before fitting for the outflow component in order to isolate the outflowing component of each line. We also fit each line with a single gaussian with no constraints on the velocity. We then calculate the residual from each fit ($\rm \frac{\Sigma |y-f|}{N}$, where y is the spectrum, f is the fit to the spectrum, and N is the number of data points) and adopt the fit with the lowest value. All fits are constrained to regions without obvious contamination from other lines, which were set by visually inspecting the lines and attempting to isolate them from neighboring features. The resulting mean velocity from the adopted fit (the outflow component for the double gaussian fit) provides us with a ``central velocity", $v_{\rm cen}$. We also find the velocity where the fit reaches 90\% of the continuum (i.e. where the fit equals 0.9) for each line. This fitting routine has been adopted to match that of \cite{xu2022classy} in order to best facilitate our comparison between the statistics derived from our mock observables and those derived from actual observations. A key difference between our fitting routine and theirs, however, is that our single gaussian fit is still intended to capture the structure of the outflow, while that of \cite{xu2022classy} is to fit absorption lines that are thought to not host significant outflowing material.

\section{Results} \label{sec:results}

\begin{figure*}[!ht]
    \centering
    \includegraphics[scale=0.575]{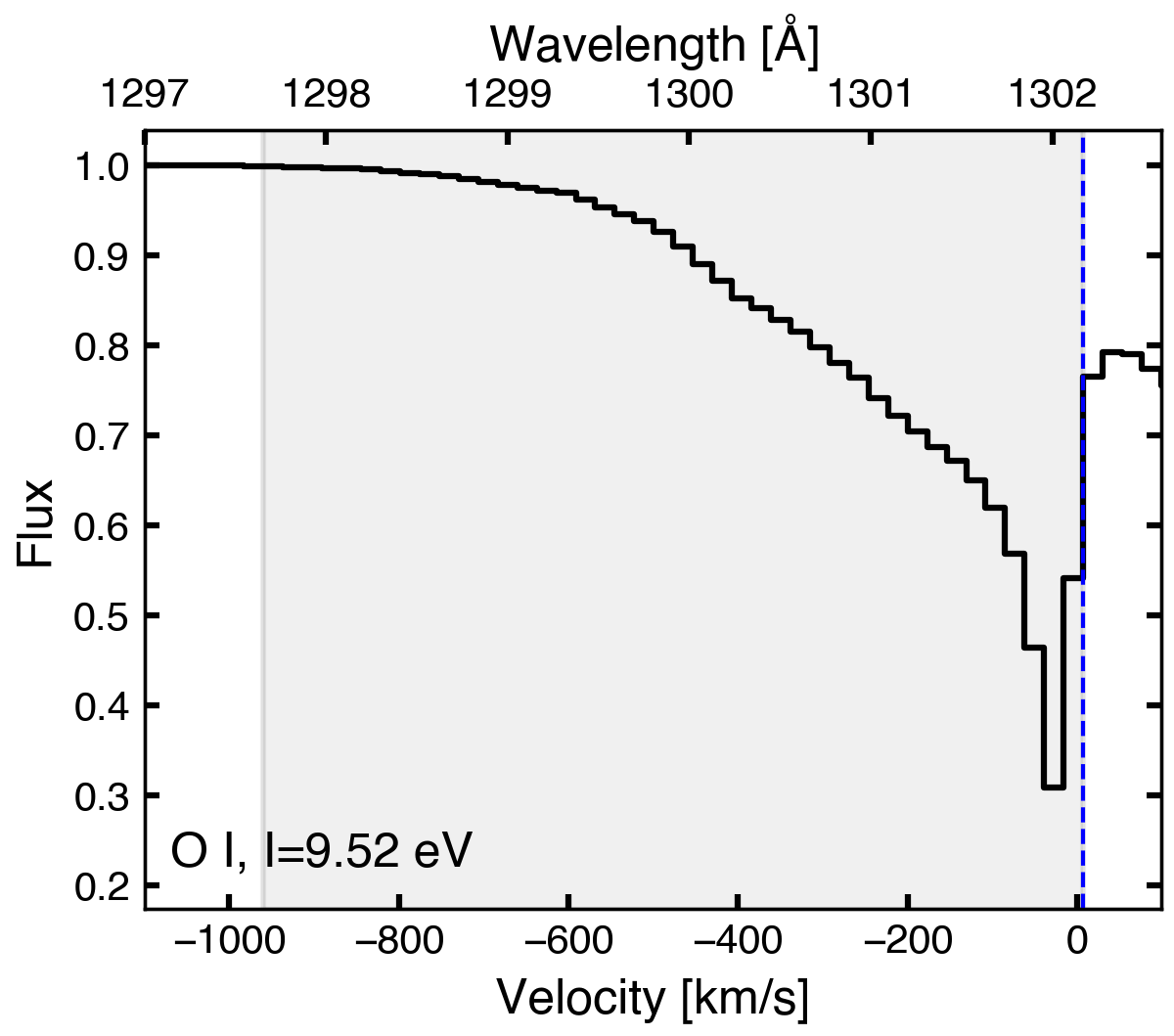}
    \includegraphics[scale=0.575]{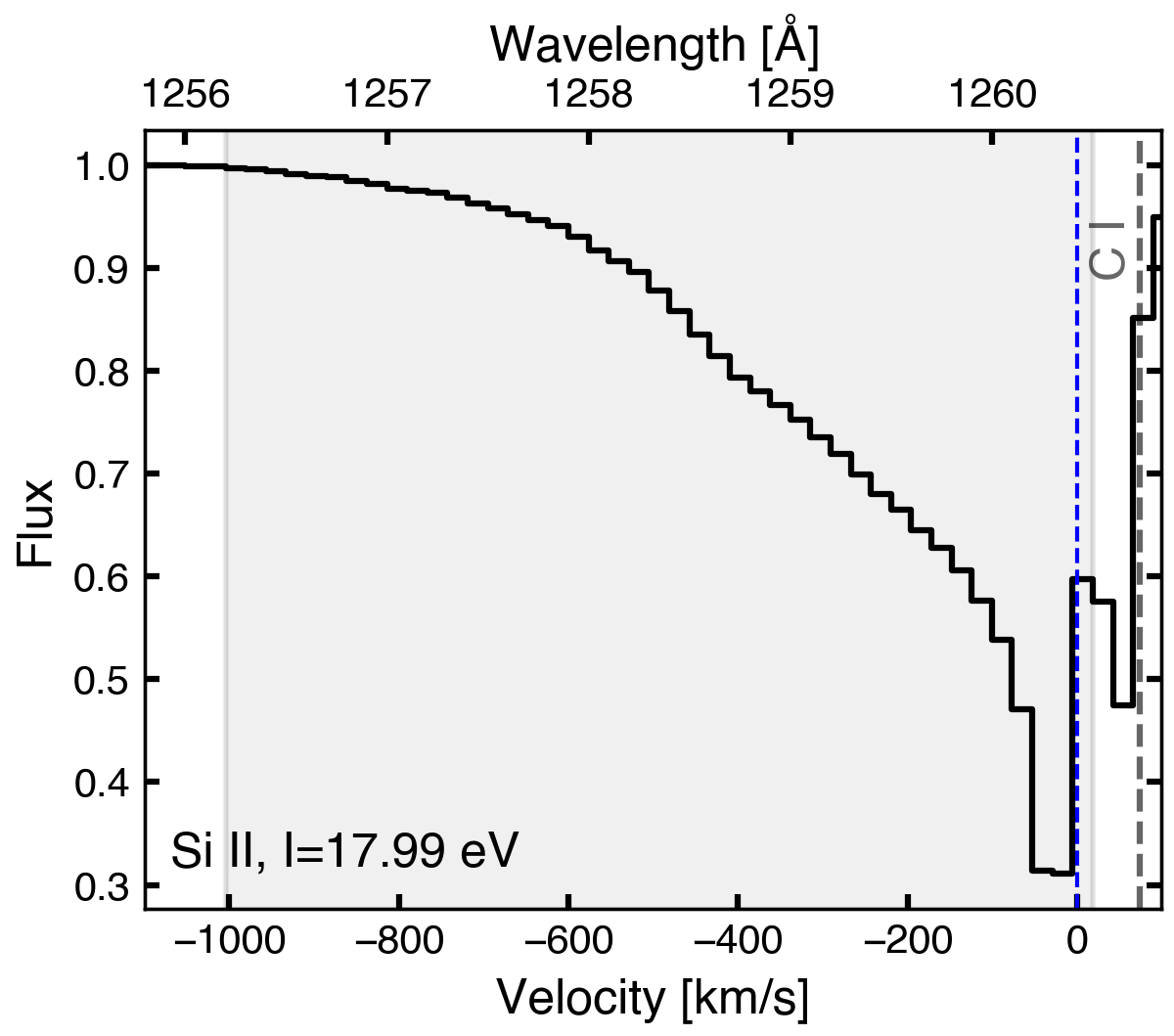}
    \includegraphics[scale=0.575]{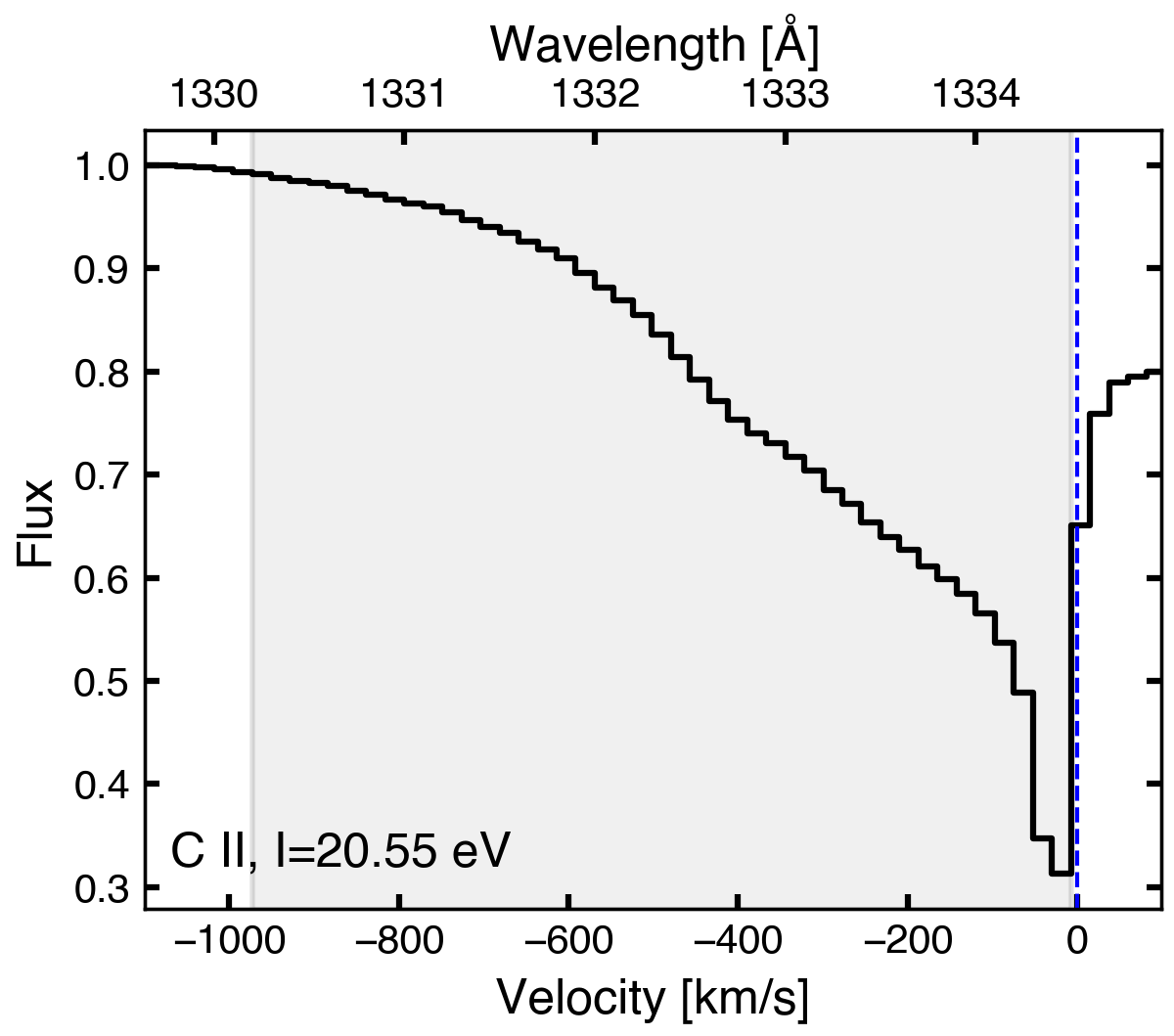}
    \includegraphics[scale=0.575]{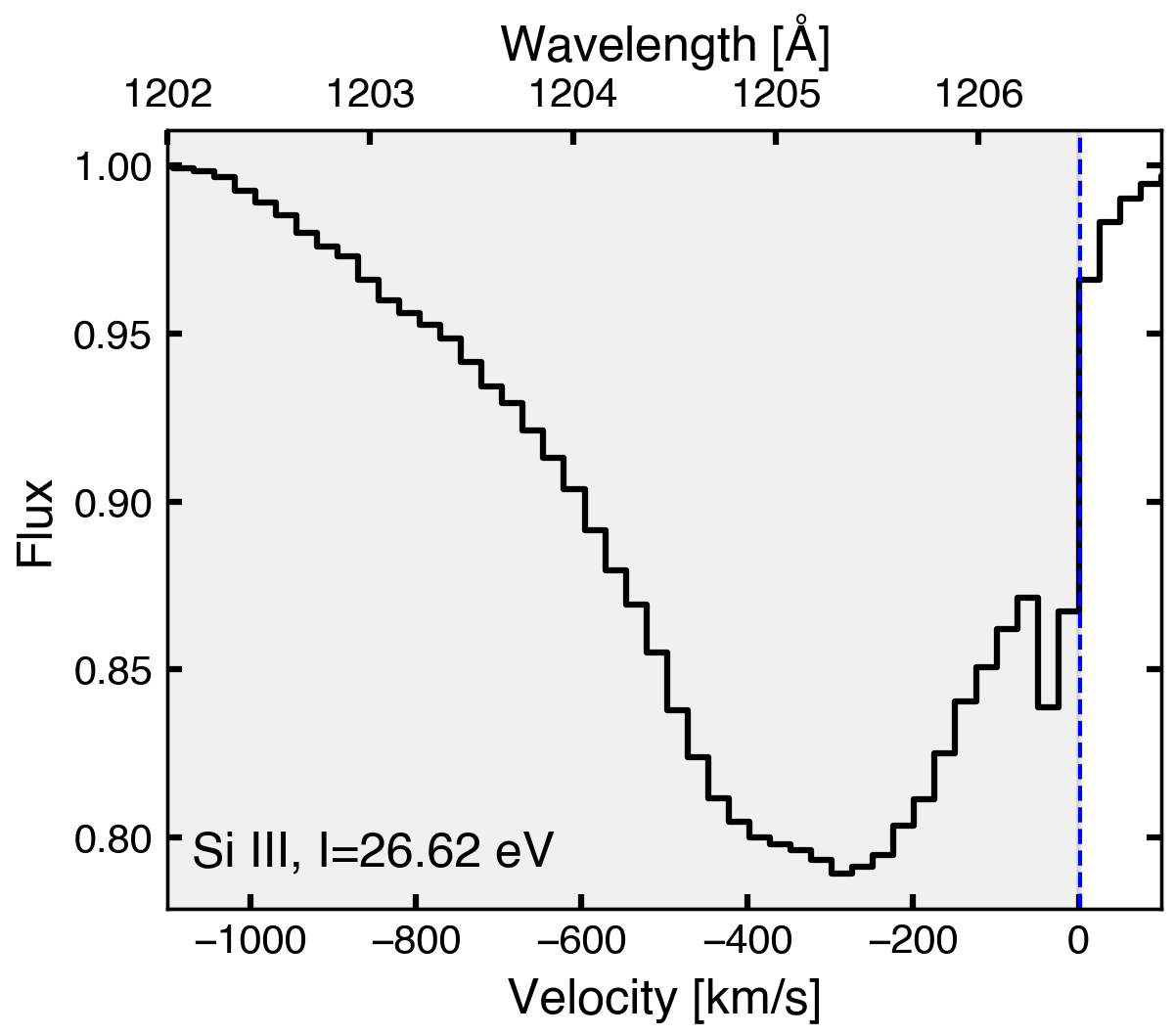}
    \includegraphics[scale=0.575]{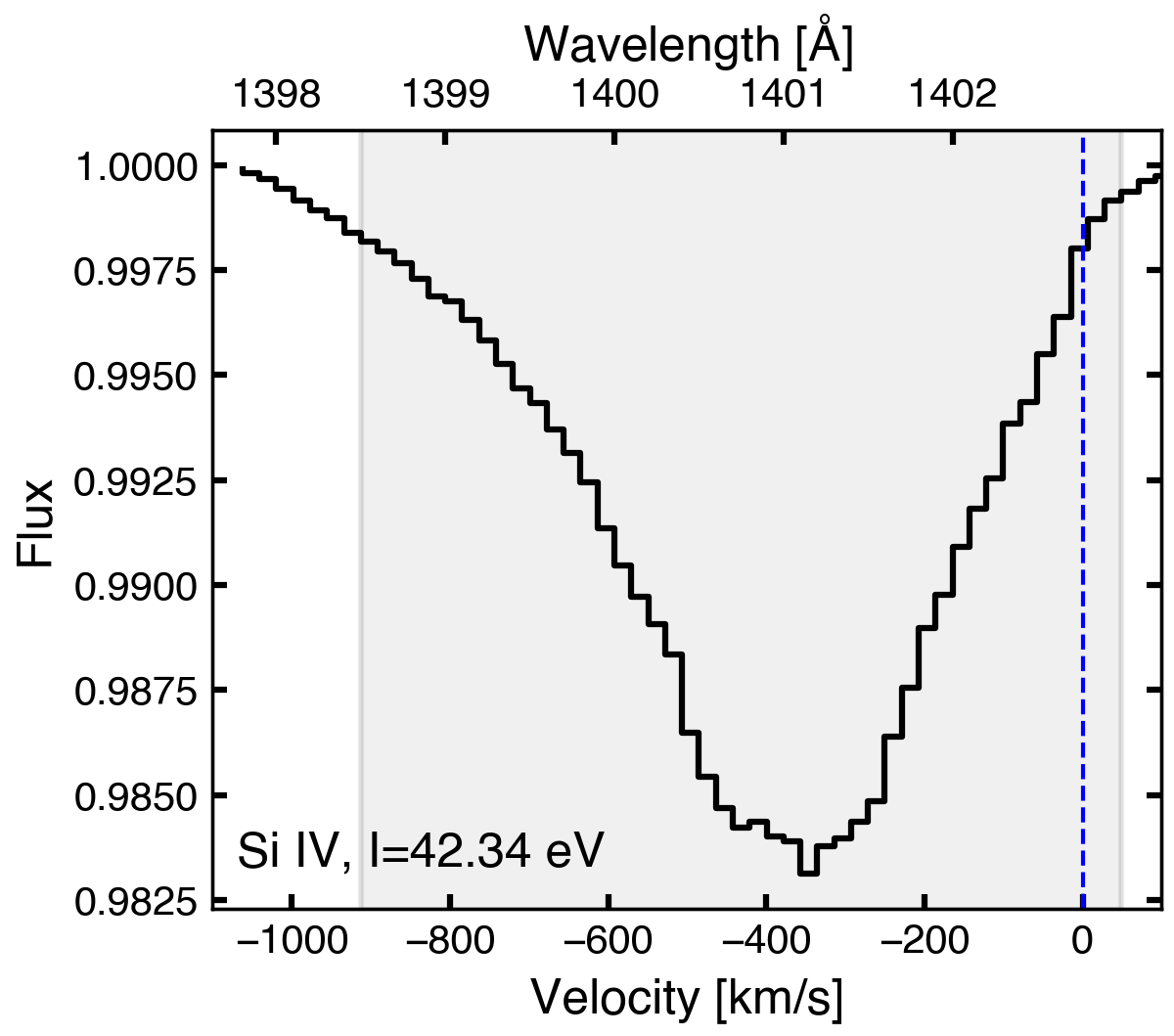}
    \includegraphics[scale=0.575]{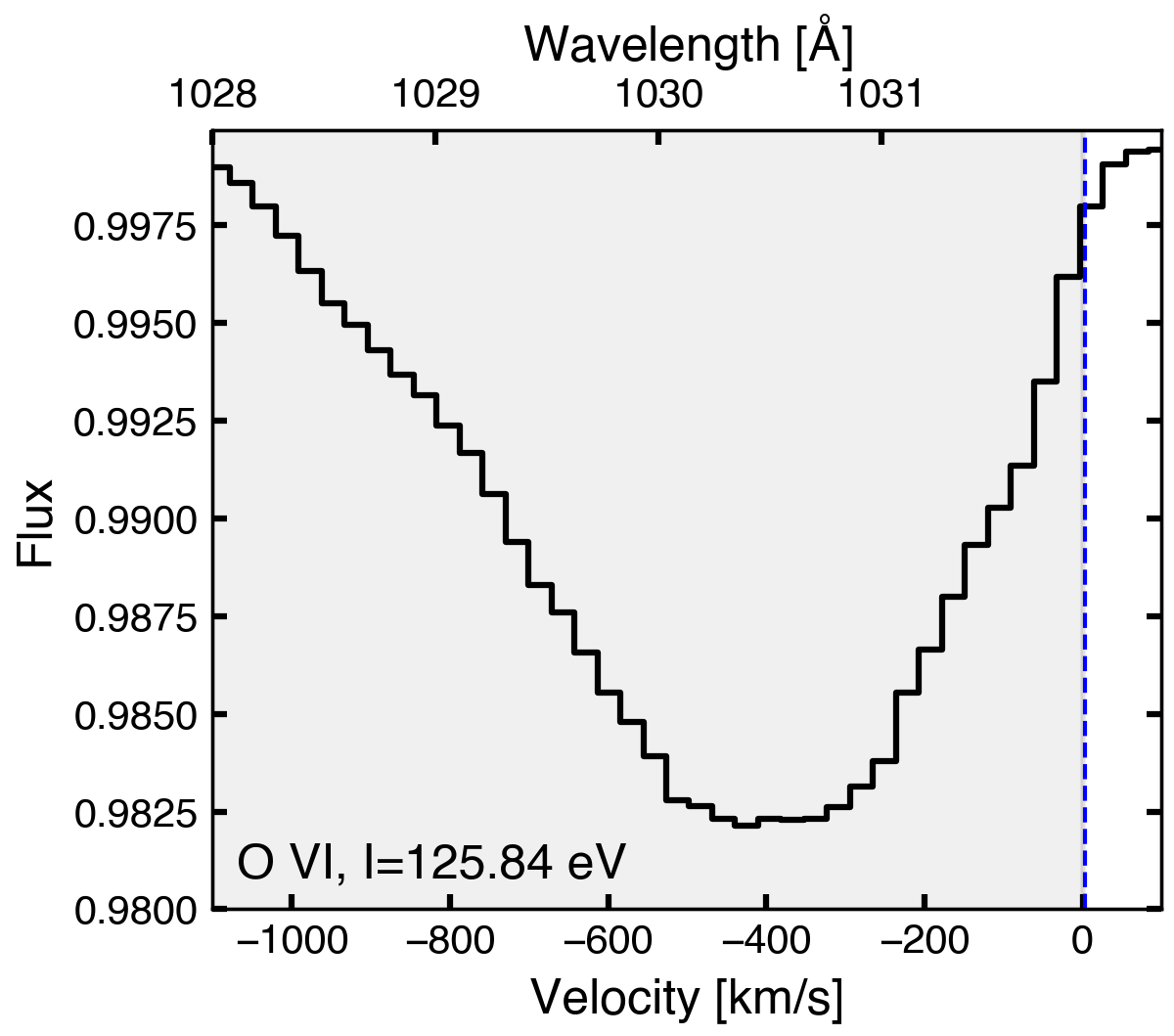}
    \caption{All absorption lines focused on in this work, generated from the high resolution ($\Delta x\simeq20 \ \rm pc$) simulation data using an exponential weighting function with a characteristic radius $\rm R=1\ kpc$. Spectra are arranged in ascending order of excitation potential. Also shown are the regions used to measure the relevant line statistics (in gray), as well as the rest velocity (dashed, in blue) from the Trident line list database. Lines that could be potentially contaminating the absorption profile are labeled and marked with a dashed black line. The spectra are generated in normalized flux units, with a constant wavelength resolution of $\Delta \lambda=0.1 \ \text{\AA}$.}
    \label{fig:all_spec}
\end{figure*}

In order to make comparisons between synthetic observations of simulated data and real down-the-barrel observations of galactic outflows, in this first study we choose to model a representative set of lines based off of those studied in the CLASSY survey (see \citealt{xu2022classy}). CLASSY is an HST treasury program which has produced a catalog of high-resolution, far-UV spectra of 46 star-forming galaxies at z$\sim0$, which were selected to span a wide range of masses, SFRs, metallicities, and ISM structures (e.g. densities and ionization parameters; \citealt{berg2019cos}). Table 1 lists the particular lines we model, including rest wavelengths ($\lambda_{\rm rest}$) and ionization potentials ($I [\rm eV]$). We present these absorption line spectra from our fiducial simulation snapshot (SFR $20 \ \rm M_{\odot}\,\mathrm{yr}^{-1}$ at $\rm t=30 \ \rm Myr$) in Figure \ref{fig:all_spec}, generated from the $\Delta x=20 \ \rm pc$ simulation in the CGOLS SFR suite. We see that, across all line species, there is evidence of absorption at outflow velocities ranging from $\sim0-800 \ \rm km \ s^{-1}$, with maximum velocities reaching $\sim 1000\rm \ km\,s^{-1}$ (consistent with the gas velocities measured directly from the simulation). The first three lines shown in Figure \ref{fig:all_spec}, which represent the lower-energy end of our ionization potential distribution (presumably tracing photoionized gas at $\rm T\simeq10^{4-4.4}$), host a significant absorption component centered near $\rm v\simeq0 \ km~s^{-1}$, which indicates significant ISM absorption. The remaining higher-ionization potential lines (presumably tracing collisionally ionized gas at $\rm T\simeq10^{4.5-5.5}$) are dominated by an outflowing component with higher velocity. Because the lines in our study span over an order of magnitude in ionization potential, this enables the study of trends in outflow properties as the primary mode of ionization transitions from predominately photoionization (e.g. Si II) to collisional ionization (e.g. O VI).

\begin{deluxetable}{cccc}[h]
\tabletypesize{\footnotesize}
\tablecaption{Table of the spectral features focused on in this work, with the rest-frame wavelength $\lambda_{\mathrm{rest}}$ (in $\AA$), ionization potential $\mathrm{I}$ (in $\mathrm{eV}$), and the oscillator strength \textit{f}-value \citep{NIST_ASD}. \label{tab:lines}}
\tablehead{\colhead{Line} & \colhead{$\lambda_{\mathrm{rest}}$ [\AA]} & \colhead{$I$ [eV]} & \colhead{\textit{f}-value}}
\startdata
O I    & 1302.1680 & 9.52   & 0.052 \\
Si II  & 1260.422  & 17.99  & 1.18  \\
C II   & 1334.532  & 20.55  & 0.129 \\
Si III & 1206.500  & 26.62  & 1.63  \\
Si IV  & 1402.770  & 42.34  & 0.255 \\
O VI   & 1031.912  & 125.84 & 0.133 \\
\enddata
\end{deluxetable}
\vspace{-12pt}

\begin{figure}
    \centering
    \includegraphics[scale=0.85]{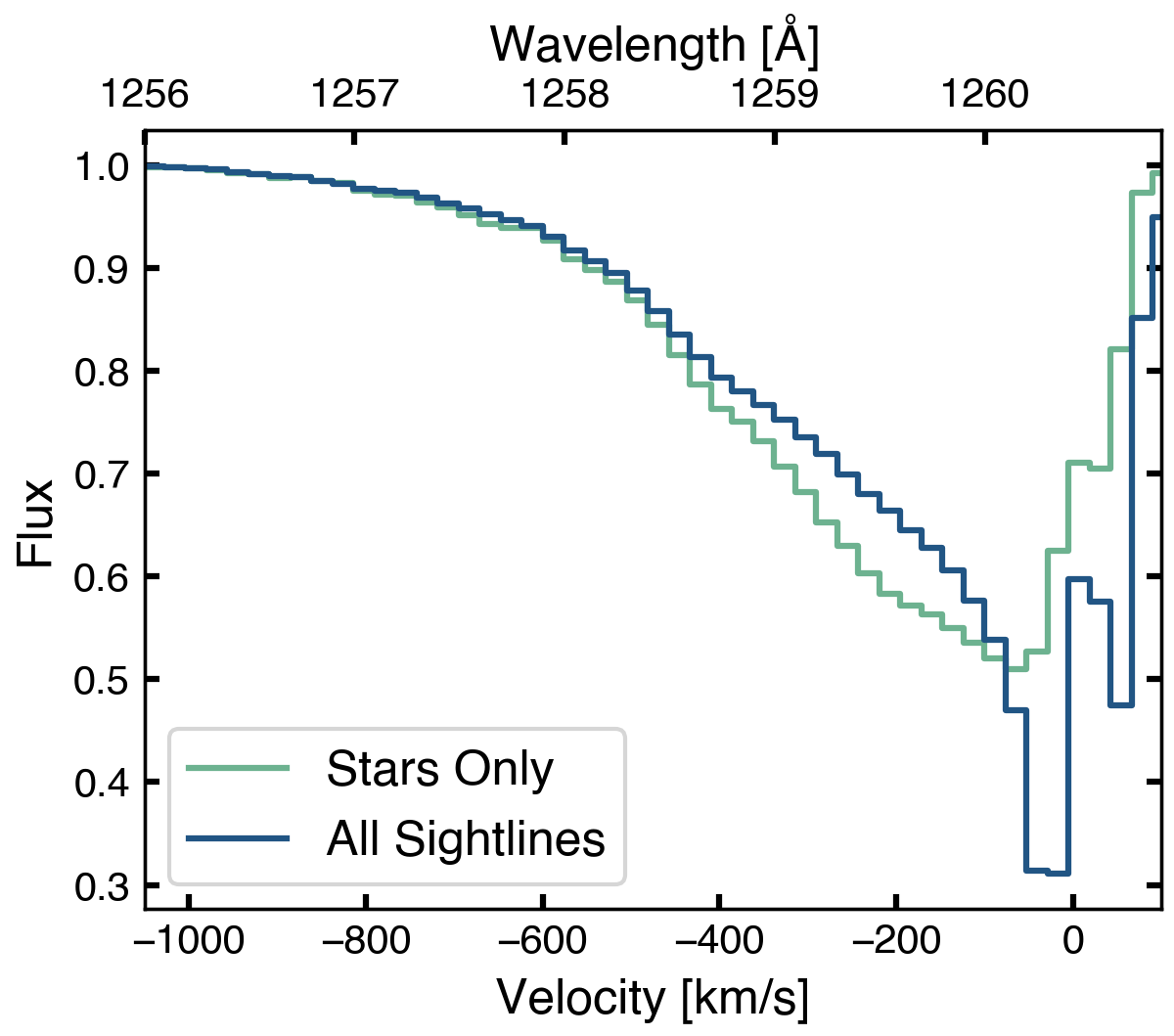}
    \caption{Si II $\sim1260\text{\AA}$ absorption line spectra, generated from the $\Delta x=20 \ \rm pc$ resolution simulation, with the full aperture (in blue), and the ``Stars Only" mask (in green).}
    \label{fig:stars spec}
\end{figure}

Figure \ref{fig:stars spec} shows a pair of Si II spectra comparing the results of the temperature selection mask (referred to hereafter as ``Stars Only") to the fiducial exponential weighting function (hereafter ``All sightlines"). The Stars Only spectrum has a significantly reduced static ($v\sim0 \rm \ km \ s^{-1}$) gas component compared to that of the ``All sightlines" spectrum, and enhanced relative absorption from high velocity outflowing gas. When comparing our spectra to observed spectra which typically have a large static ISM component, it is clear that the ``Stars Only" mask does not do as good a job producing this feature as the exponential light profile adopted in the ``All sightlines" spectra. In real galaxy spectra, the large static ISM component may be a result of embedded star formation. This phase is not represented well in the CGOLS SFR simulations, which only model supernova feedback and do not capture other important forms of early stellar feedback, such as ionizing radiation. As a result, most areas in the ``Stars Only" mask represent star clusters that have already cleared out the intervening ISM along the line-of-sight. For this reason, we proceed with the exponential weighting for the remainder of this work.

\subsection{Velocity Statistics} \label{subsec: velocity}

\begin{figure*}[!t]
    \centering
    \includegraphics[scale=0.8]{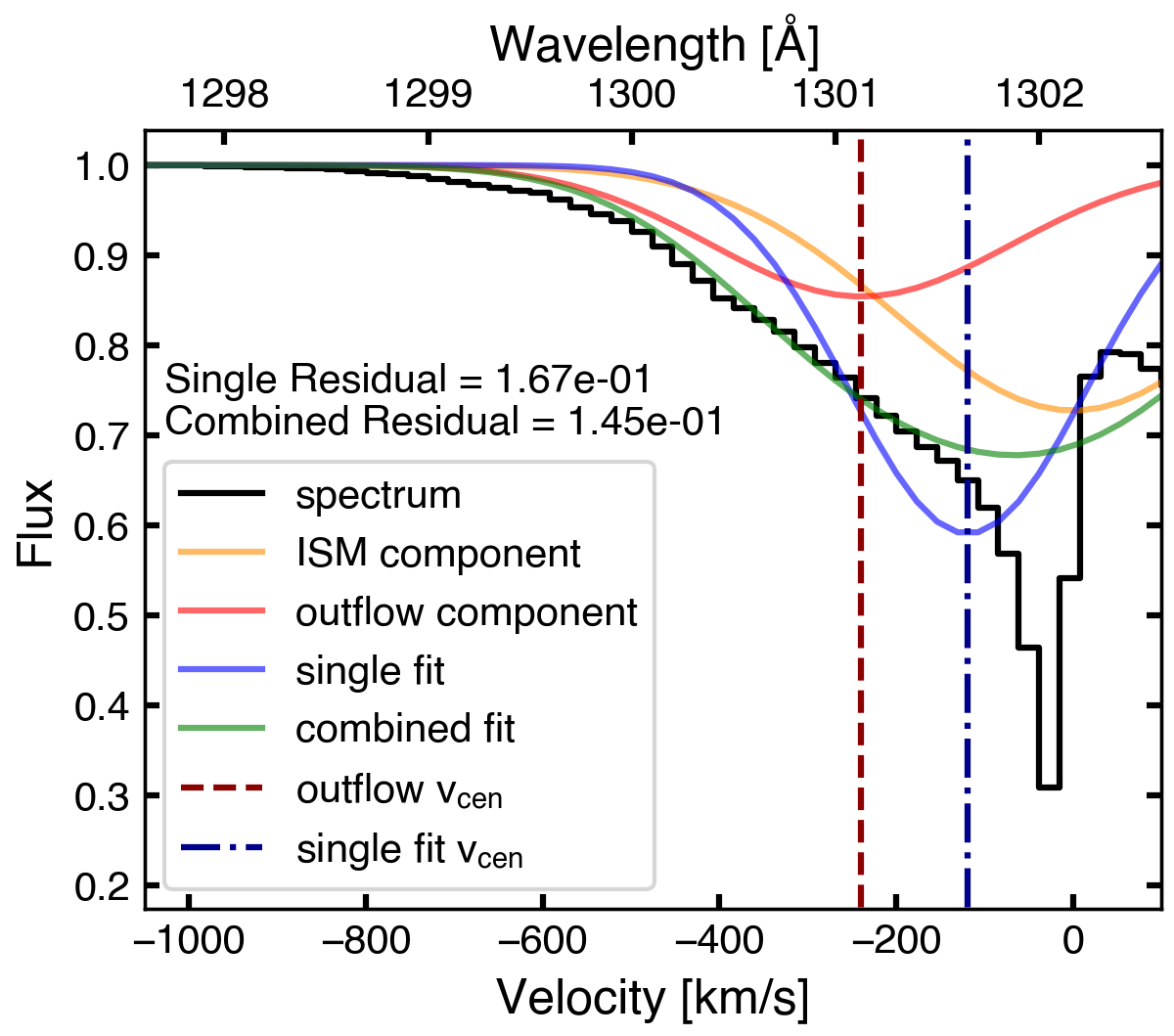}
    \includegraphics[scale=0.8]{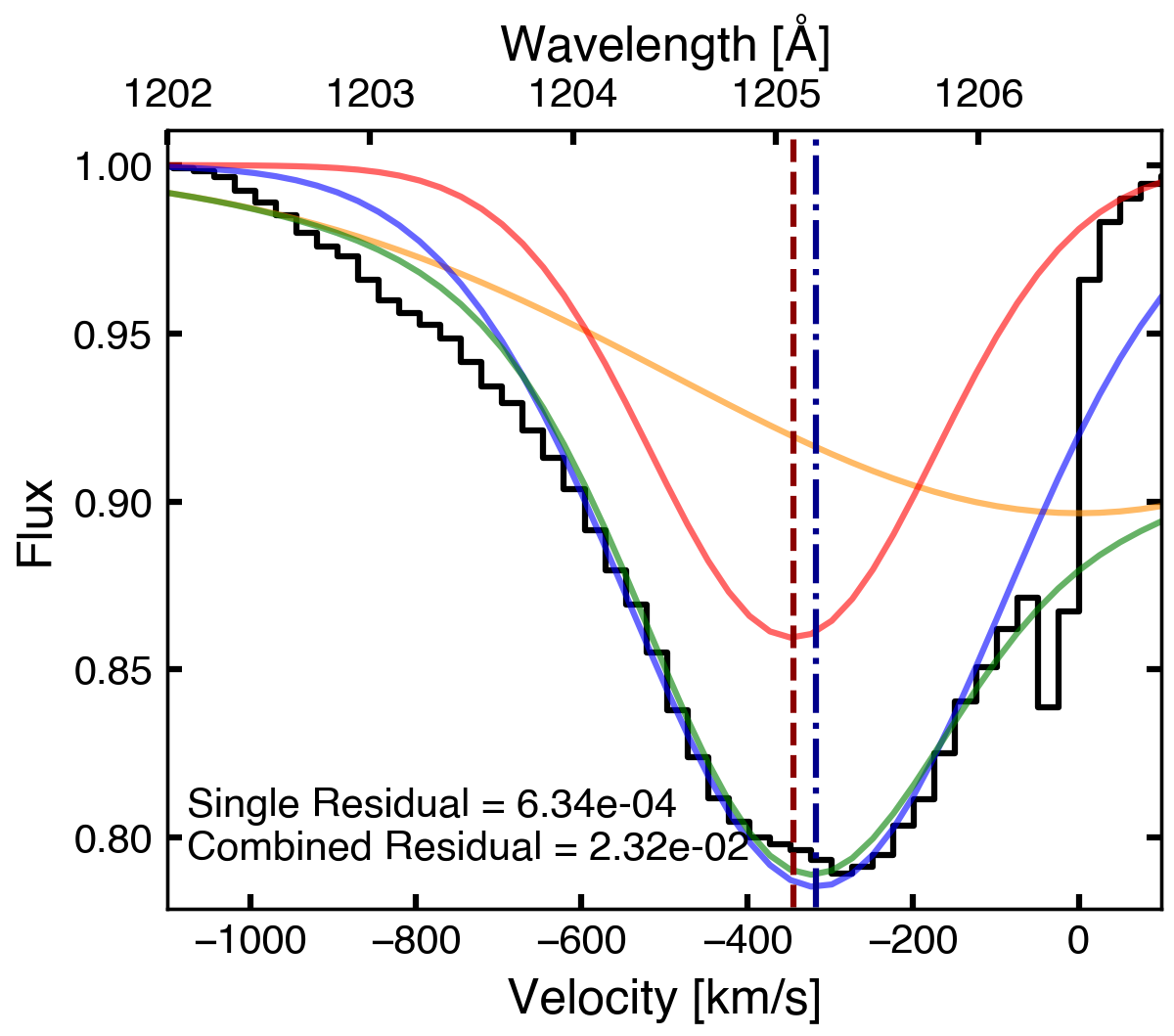}
    \caption{Examples of both single (solid blue) and double (solid green) gaussian fits to OI $~\sim1302 \ \AA$ (left) and Si III  $~\sim1206 \ \AA$ (right), to the mock spectrum (black). We also plot the Gaussian that is fit to the outflow component only (in solid red), which is one of the two components of the ``combined" double gaussian fit. The $v_{\rm cen}$ for the single fit is plotted in the blue dashed lines, and the $v_{\rm cen}$ for the outflow component in the red dashed lines. The left panel is an example of a fit where the residual is minimized for the combined model, while the right is a fit where residual is minimized for the single Gaussian fit.} 
    \label{fig:spec fit}
\end{figure*}

In observational studies of outflows, it is common practice to quantify velocity structure using single-number statistics in order to facilitate comparisons between galaxies. Two commonly-used statistics are the centroid velocity ($v_{\rm cen}$) and the 90th percentile velocity ($v_{90}$), as measured by fitting a Gaussian distribution to the outflowing absorption component, after fitting and removing the ISM component from the spectrum (see \S\ref{subsec: spec fit}). To measure these values from our suite of absorption lines, we fit a Gaussian to the line profile within the $\sim\Delta4\text{\AA}$ regions shown in Figure \ref{fig:all_spec} using a \textit{scipy} curve-fit routine. The mean of this fit is our $v_{\rm cen}$, and we set $v_{90}$ as the velocity where the value of the fit reaches $90\%$ of the continuum (the fitting routine is fully described in \S\ref{subsec: spec fit}). We show examples of this fit for two characteristic absorption lines in Figure \ref{fig:spec fit}.

\begin{deluxetable}{ccccc}[h]
\tablecaption{$v_{\rm cen} \mathrm{[km \ s^{-1}]}$ for each absorption line focused on in this work, measured from all simulation resolutions at $t=30 \ \rm Myr$. $v_{\rm cen}$ is measured as the mean from a Gaussian fit to each absorption line. \label{tab:v_cen}}
\tablehead{
\colhead{Line} & \colhead{$20 \ \mathrm{pc}$} & \colhead{$40 \ \mathrm{pc}$} & \colhead{$80 \ \mathrm{pc}$} & \colhead{$160 \ \mathrm{pc}$}
}
\startdata
O I    & -239.8 & -254.3 & -231.4 & -262.9 \\
Si II  & -253.9 & -273.8 & -255.4 & -300.8 \\
C II   & -254.5 & -263.7 & -260.3 & -290.5 \\
Si III & -316.7 & -327.5 & -349.1 & -351.1 \\
Si IV  & -367.1 & -369.3 & -381.0 & -390.0 \\
O VI   & -448.3 & -456.3 & -458.1 & -457.3 \\
\enddata
\end{deluxetable}
\vspace{-12pt}

Table \ref{tab:v_cen} displays the $v_{\rm cen}$ fits for all lines in this study, in order of ionization potential. The left-most column contains the values for the $\Delta x =\rm 20 \ pc$ simulation data (the lines shown in Figure \ref{fig:all_spec}), while subsequent columns show the fits for lower-resolution simulation data. The most prominent trend evident in the $v_{\rm cen}$ statistic is the increase in magnitude of $v_{\rm cen}$ with ionization potential. The higher ionization potential lines in this study are presumably tracing increasingly hotter gas within the outflow, and it is expected that the hotter gas hosting these absorbers is propagating at larger velocities on average than the cooler phases of the outflow \citep{schneider2020physical, schneider2024cgols, abruzzo2024taming}. If this assumption is true, one would indeed expect to see these higher ions moving faster on average than the lower energy ions, which is precisely the trend that is seen in Table \ref{tab:v_cen}.

\begin{figure*}[!t]
    \centering
    \includegraphics[scale=0.8]{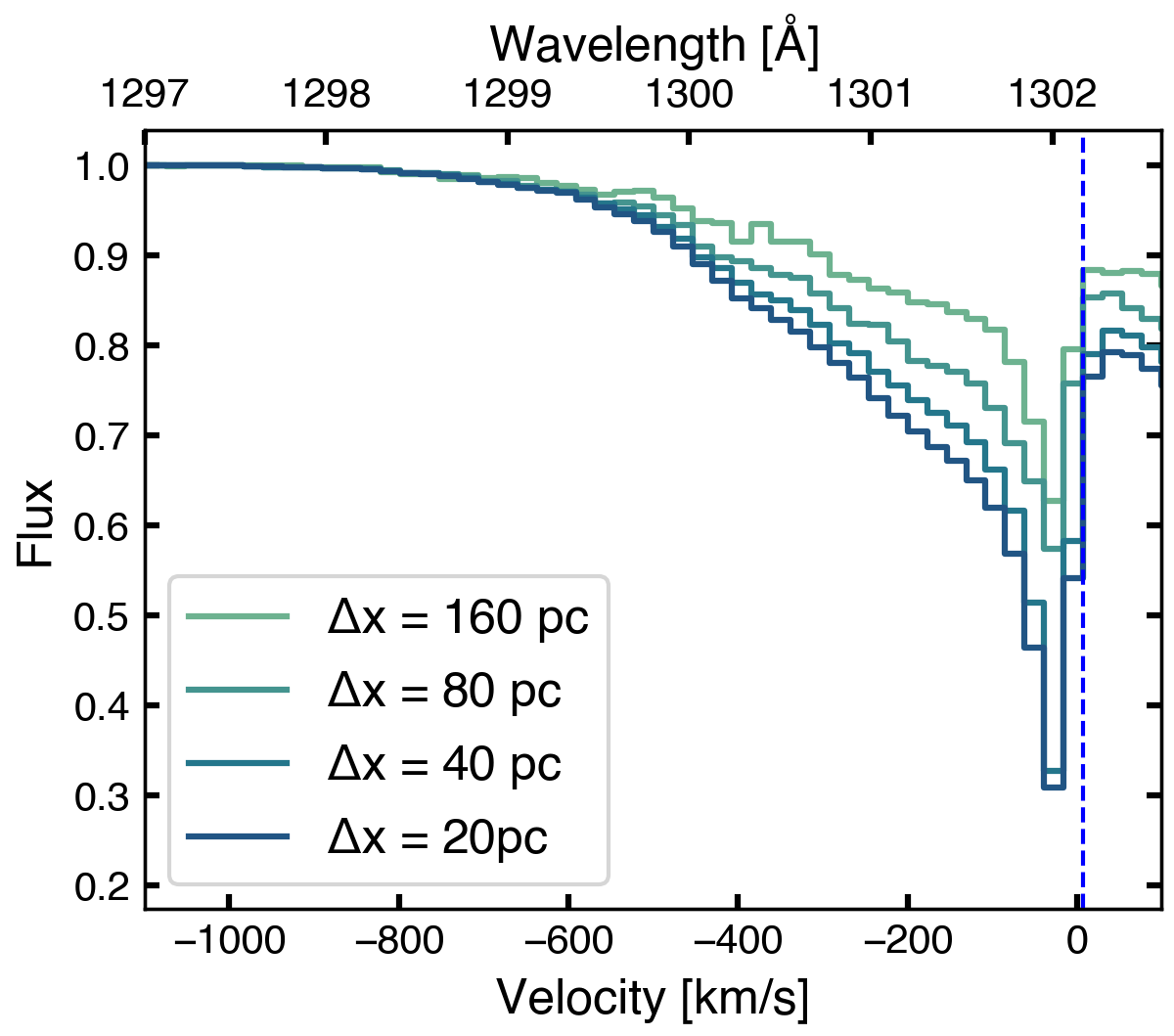}    
    \includegraphics[scale=0.8]{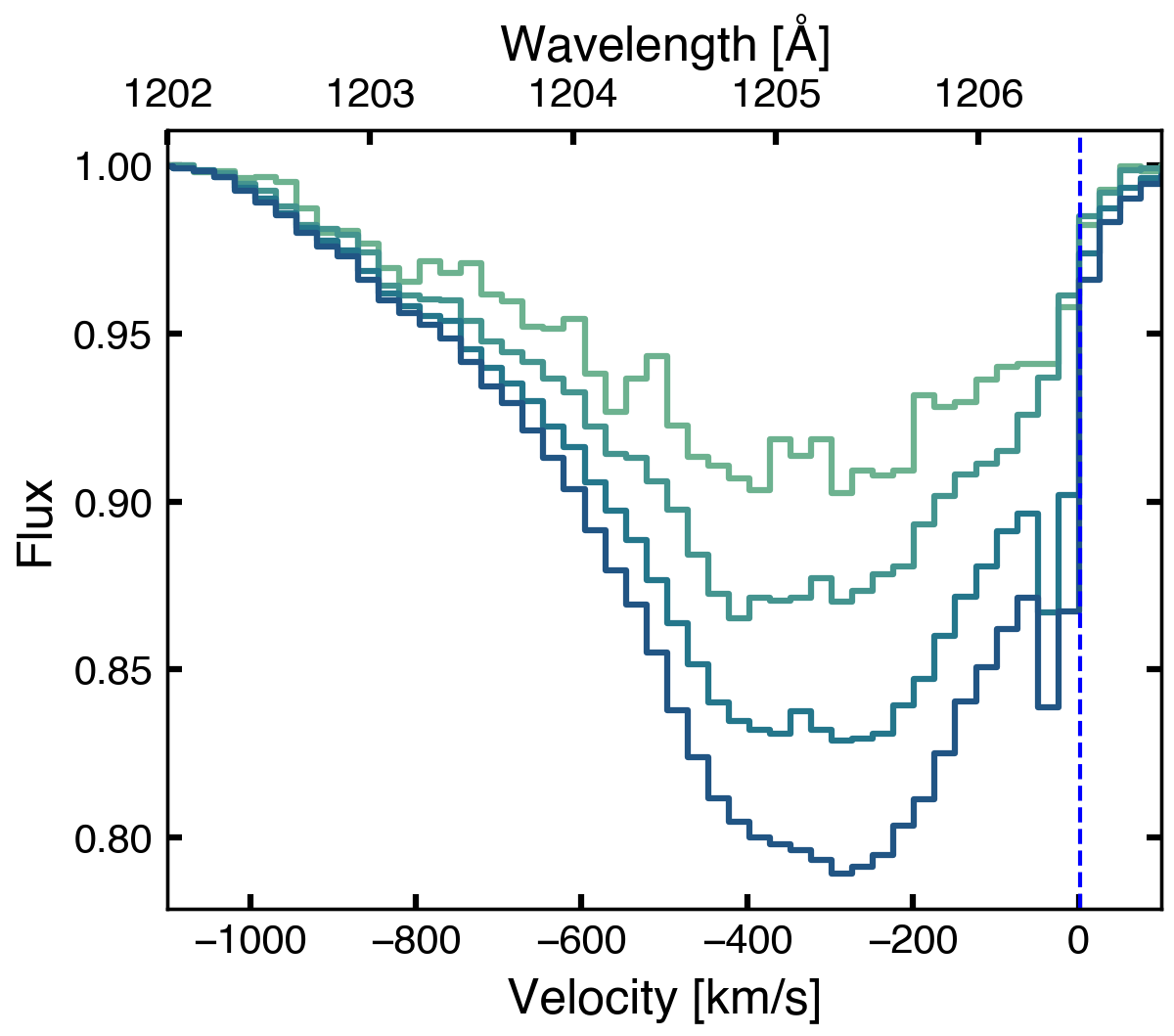}
    \caption{O I $\sim1301\text{\AA}$ (left) and Si III $\sim1206\text{\AA}$ (right) absorption line spectra, both generated from the $\Delta x=160 \ \mathrm{pc}$ (green), $\Delta x=80 \ \mathrm{pc}$ (light blue), $\Delta x=40 \ \mathrm{pc}$ (blue), and $\Delta x=20 \ \mathrm{pc}$ (dark blue) resolution simulations} 
    \label{fig:all res si III}
\end{figure*}

For the higher ionization potential lines in this study, the absolute value of the centroid velocity for the outflowing material decreases slightly with increasing simulation resolution (as shown in Table \ref{tab:v_cen}). In the lower ionization potential lines, the absolute value of the centroid velocity for the outflowing material first increases between the $\Delta x = 20 \ \& \ 40 \ \rm pc$ resolutions, then decreases between the $\Delta x = 40 \ \& \ 80 \ \rm pc$ resolutions, then increasing substantially for the $\Delta x = 160 \ \rm pc$ resolution. For the higher ionization potential lines, $v_{\rm cen}$ increases between the first three resolution steps, then increases slightly between the two coarsest resolutions. Figure \ref{fig:all res si III} shows two examples of this trend, which plots the O I ($\sim 1302 \ \AA$) (left) and Si III (right) lines for four different simulation resolutions, going from the fiducial simulation resolution of $\Delta x=\rm 20\ pc$ to the lowest resolution, $\Delta x=\rm 160\ pc$. In general, the values of $v_{\rm cen}$ for the three finest resolutions agree within $1-10\%$ of one another. Overall, all of the spectra have velocity structures that are largely dominated by an outflowing component ($\rm |v|\sim[100,\ 800] \ km \ s^{-1}$), with a relatively prominent ISM component ($\rm |v|\sim[0,\ 100]\ km \ s^{-1}$). The $\Delta x=160 \ \mathrm{pc}$ spectrum is overall shallower than, but still hosts a qualitatively similar velocity structure to, the higher resolution spectra. Absorption from the outflowing component is typically strongest in the $\Delta x=20 \ \mathrm{pc}$ resolution spectrum. These trends appear to be independent of the ionization potential of the species, and tend to be more pronounced in lines that are relatively isolated from neighboring absorption features. In addition to $v_{\rm cen}$, we also report the values of $v_{90}$ for all lines and all simulation resolutions in Table \ref{tab:v_90}. Similar to $v_{\rm cen}$, we find that the magnitude of $v_{90}$ increases with ionization potential, with the highest value of $v_{90}$ found in the O VI line. There is minimal variation in this statistic between the two finest simulation resolutions (remaining approximately the same for lower ionization potential lines and Si IV), with the values for the remaining coarser resolutions dropping more significantly, except for Si IV, which increases by approximately $5\%$. 

\begin{deluxetable}{ccccc}[h]
\tabletypesize{\footnotesize}
\tablewidth{0pt}
\tablecaption{$v_{90} \mathrm{[km \ s^{-1}]}$ for each absorption line focused on in this work, measured from all simulation resolutions at $t=30 \ \rm Myr$. $v_{90}$ is measured from the same Gaussian fit used to measure $v_{\rm cen}$, and represents where this fit reaches 90\% of the continuum (e.g. 0.9). \label{tab:v_90}}
\tablehead{
\colhead{Line} & \colhead{$20 \ \mathrm{pc}$} & \colhead{$40 \ \mathrm{pc}$} & \colhead{$80 \ \mathrm{pc}$} & \colhead{$160 \ \mathrm{pc}$}
}
\startdata
O I    & -430.1 & -430.1 & -384.0 & -315.0 \\
Si II  & -504.7 & -504.7 & -480.9 & -409.6 \\
C II   & -546.3 & -546.3 & -523.9 & -478.9 \\
Si III & -596.4 & -571.5 & -521.8 & -397.5 \\
Si IV  & -399.7 & -399.7 & -421.0 & -421.0 \\
O VI   & -875.1 & -846.0 & -817.0 & -787.9 \\
\enddata
\end{deluxetable}
\vspace{-12pt}

By studying the effects of simulation resolution on the velocity statistics and structure of the absorption line spectra, we can assess whether $20$ pc resolution is sufficient to generate realistic spectra, or if higher resolution is still necessary to accurately capture the underlying velocity structure of the outflow. Although the velocity statistics have not fully converged at $\Delta x=20 \ \rm pc$ resolution, the differences between measured values for $v_\mathrm{cen}$ are relatively consistent across all simulation resolutions. More specifically, the percentage difference between the fiducial values for $v_\mathrm{cen}$ (i.e. $\Delta x=20$) and the values from the $\Delta x=40, \ 80, \& \ 160 \ \rm pc$ resolution range from $<1-15\%$. High simulation resolution requires significant computational resources (both to run the simulation and analyze the data after the fact), and these statistics suggest that our spectra are nearly converged at the $\Delta x=160 \ \rm pc$ resolution. Though the velocity statistics suggest that our fiducial simulation resolution is sufficiently high for the type of statistical comparisons to real data we intend to do with this work, the velocity statistics alone do not convey the full picture of how resolution affects our absorption lines.

\subsection{Equivalent Width} \label{subsec: EW}

Measurements of equivalent width (EW) are one of the lowest order probes of the amount of gas that resides in different phases of outflows. Though EWs are relatively straightforward to measure, because the quantity depends on the column density, velocity, and covering fraction of the gas, it is difficult to disentangle the exact properties of the absorbing gas that give rise to a specific EW value. In this work, we use EW as a first-order check on the realism of these mock absorption lines, with a future goal to use the underlying distribution of gas in the simulation to untangle these complex trends.

Many of the lines in this study are contaminated by neighboring absorption features, so we define a $\sim\Delta4\text{\AA}$ region around each line to minimize contamination in the measurements (see Figure \ref{fig:all_spec}). To calculate the EW, we tabulate the normalized flux value in each velocity bin that falls within the measurement region, and multiply the sum of those values by the size of the spectral bins in $\AA$. The measured EWs for each line are reported in Table \ref{tab:ews}. For the fiducial simulation resolution ($\Delta x=20 \ \rm pc$), the EWs range between $0-0.9 \ \text{\AA}$, and there is no obvious correlation between the EWs and the ionization potentials or oscillator strengths of the lines. The EWs for Si IV and O VI are over an order of magnitude lower than those of all of the remaining lines in our study, and we discuss these anomalously shallow ions further in \S\ref{subsec: level pop}.

\begin{deluxetable}{ccccc}[h]
\tabletypesize{\footnotesize}
\tablewidth{0pt}
\tablecaption{Equivalent Width from each absorption line focused on in this work, for each simulation resolution at $\rm t=30 \ \mathrm{Myr}$. All values listed in units of [$\text{\AA}$]. \label{tab:ews}}
\tablehead{
\colhead{Line} & \colhead{$20 \ \mathrm{pc}$} & \colhead{$40 \ \mathrm{pc}$} & \colhead{$80 \ \mathrm{pc}$} & \colhead{$160 \ \mathrm{pc}$}
}
\startdata
O I    & 0.64 & 0.58 & 0.45 & 0.33 \\
Si II  & 0.81 & 0.73 & 0.57 & 0.43 \\
C II   & 0.90 & 0.82 & 0.65 & 0.49 \\
Si III & 0.49 & 0.41 & 0.31 & 0.23 \\
Si IV  & 0.04 & 0.04 & 0.03 & 0.03 \\
O VI   & 0.04 & 0.04 & 0.03 & 0.03 \\
\enddata
\end{deluxetable}
\vspace{-12pt}

In each line species, the $\Delta x=20 \ \mathrm{pc}$ resolution has the largest EW value, with the value decreasing in the $\Delta x=40$ and $\Delta x=80 \ \mathrm{pc}$ resolution lines, and the lowest EW values for the $\Delta x=160 \ \mathrm{pc}$ resolution spectrum. When examining the EW trend more closely between the $\Delta x=80$ 40, and 20 pc resolution simulations, it appears that there is a trend towards convergence at higher simulation resolutions, though the values are not yet fully converged. Across the full suite of absorption lines, there is an approximate increase in EW of $\sim15-25\%$ when increasing from the $\Delta x=80 \ \rm pc$ to the $\Delta x=40 \ \rm pc$ resolution simulation, while the percentage increase between the $\Delta x=40 \ \rm pc$ and $\Delta x=20 \ \rm pc$ resolution simulations ranges between $\sim5-15\%$. Neither the line depth nor the relative change in line depth between resolutions show a strong dependence on ionization potential. The only significant outliers in the EW distribution are the relatively underpopulated Si IV and O VI lines, which we will discuss further in \S\ref{subsec: level pop}.

There is an important caveat for all of the analysis in this section and the sections following with regard to our fit regions. In many of the lines we study here, we set the fit region (shown in grey in Figure \ref{fig:all_spec}) to reduce the contamination of neighboring absorption features. Though this is necessary to make more accurate measurements of these spectral features, it does reduce the strength of the measurements because these lines do not reach the continuum flux (see \cite{chisholm2016robust} for measurements of EWs with similar fit regions). Though our spectrum generation pipeline enables us to generate spectra that contain only one atomic species, which would reduce the overall influence of contamination across the full suite of lines, we choose to not leverage this capability to better match the line contamination seen in observations. Additionally, isolating a particular atomic species in spectrum generation would not significantly affect the contamination in the triplet (O I $\sim$ 1302, 1304, \& 1306 $\AA$) lines.

\subsubsection{Covering Fraction} \label{subsubsec: covering fraction}

\begin{figure}[ht!]
    \centering
    \includegraphics[scale=0.8]{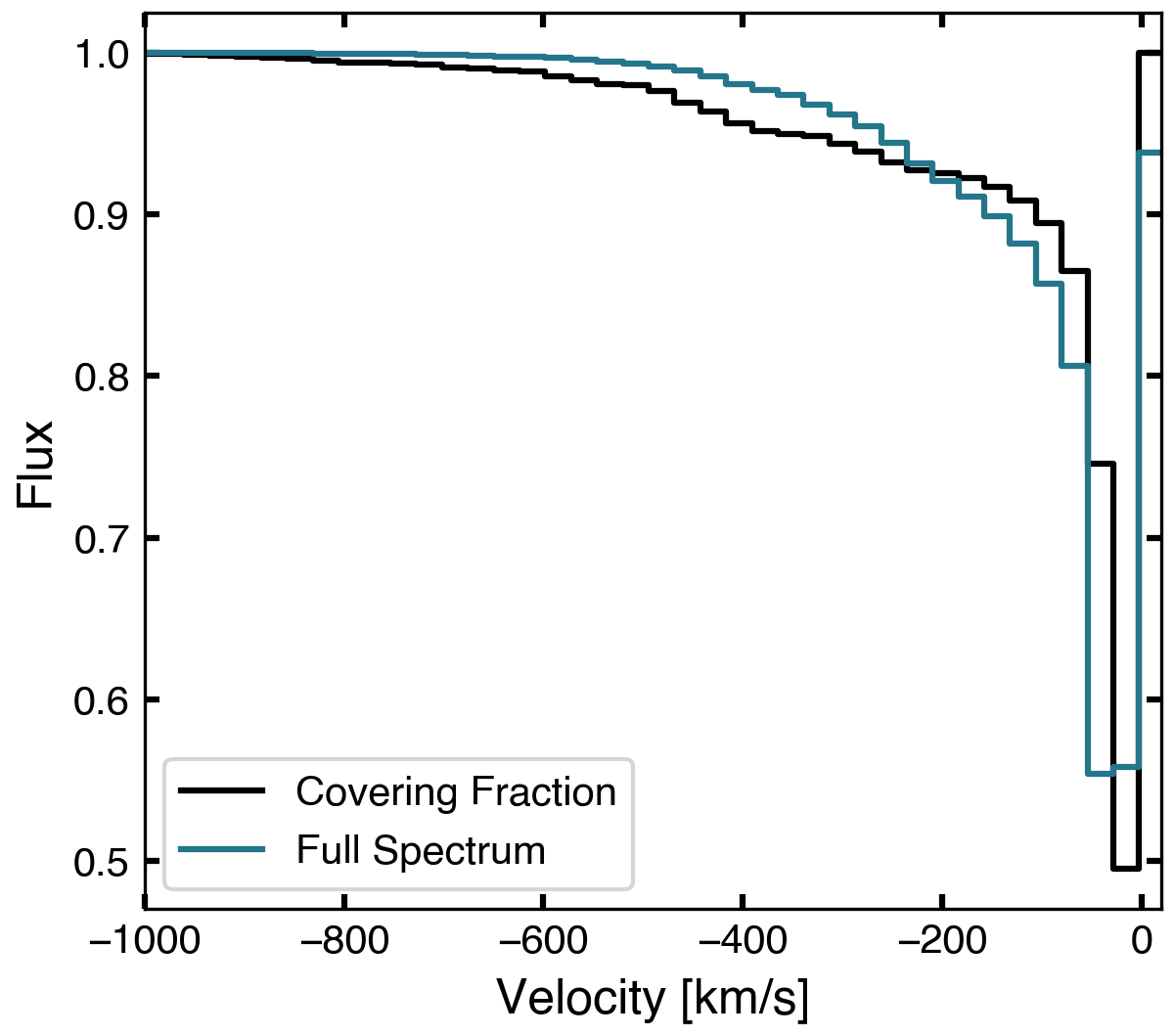}
    \caption{The full average weighted spectrum from the $\Delta x=20 \ \rm pc$ resolution simulation (blue) and the weighted, normalized covering fraction from those same sightlines (black) for C I $\sim1158 \ \AA$.}
    \label{fig:CF plot}
\end{figure}

One major advantage of simulated data is that it enables an investigation into the role of covering fraction in the cumulative EW trends. In particular, we are able to study how absorption changes across individual lines of sight, and how those different lines of sight come together to form the aggregate EWs we report here. We show an example of how the covering fraction compares to the average absorption in Figure \ref{fig:CF plot}. Here we plot the covering fraction and accompanying spectrum of C I $\sim1158 \ \AA$ (left) \footnote{We choose C I for this plot because this line is among the most isolated from neighboring absorption lines included in the \textit{Trident} line list, allowing us to definitively identify absorption features in the individual lines-of-sight.}.

The covering fraction is measured by recording the temperature and velocity of all cells from just above the mid-plane of the simulation box to the edge of the box, and xy-coordinates that lie within the central 5 kpc (i.e. the full aperture used to generate the spectrum). A temperature cut of $\rm T \leq 10^{4.1} \ K$, which is approximately the temperature of gas that we expect to be responsible for the C I absorption, is then applied. The cells that are not masked by the temperature cut are sorted into velocity bins that match those in the average spectrum, and each cell that falls within a given velocity bin contributes a fully saturated absorption line in that bin (times the spatial weight of the cell, see \S\ref{subsubsec:weighting}). Once all absorbing cells have been accounted for, the covering fraction is normalized by dividing by the total number of sightlines in the aperture.

In general, the individual lines of sight exhibit both fully and partially saturated absorption features in discrete velocity bins, and are not representative of the aggregate, fully averaged spectrum. The lines of sight that are fully saturated are less abundant than the optically thin lines of sight, and thus their strength is diluted by the averaging process -- i.e. the covering fraction is the primary cause of partial absorption in the average spectrum, while the intrinsic optical depth is responsible for partial absorption in an individual line of sight. The investigation into how EW evolves with simulation resolution in the previous section revealed that EW increases with simulation resolution, suggesting that the influence of the saturated lines of sight is also increasing with simulation resolution. This implies that, as a saturated line of sight in the $\Delta x=160 \ \rm pc$ resolution simulation gets progressively subdivided as resolution increases, the lines of sight that remain saturated in the higher resolution simulations are able to contribute more to the aggregate absorption in their respective spectra. We would expect, then, that the covering fraction from these saturated lines of sight would trace the resulting absorption profiles well. Indeed, as is shown in the left panel of Figure \ref{fig:CF plot}, the covering fraction traces the absorption profile well (though deviating slightly from the profile at higher-magnitude velocities). As described above, the calculation of the covering fraction is agnostic to the intrinsic optical depth of the absorbing lines of sight, thus further cementing our assertion that the partial absorption seen in our spectra is primarily a reflection of the geometry of the covering fraction.

\subsection{Temporal Evolution} \label{subsec:time evolution}

Thus far, our analysis has focused solely on the $\rm t=30 \ Myr$ snapshot, which we chose to capture the outflow after it has had time to develop a multiphase structure, while still being actively fueled by ongoing star-formation. In order to determine whether the observed trends in velocity statistics and EWs are sensitive to this choice of snapshot, and thereby evaluate their significance, we now repeat our analysis for the fiducial simulation at two additional times: $\rm t=20$ and 40 Myr.

\begin{figure*}[!ht]
    \centering
    \includegraphics[scale=0.8]{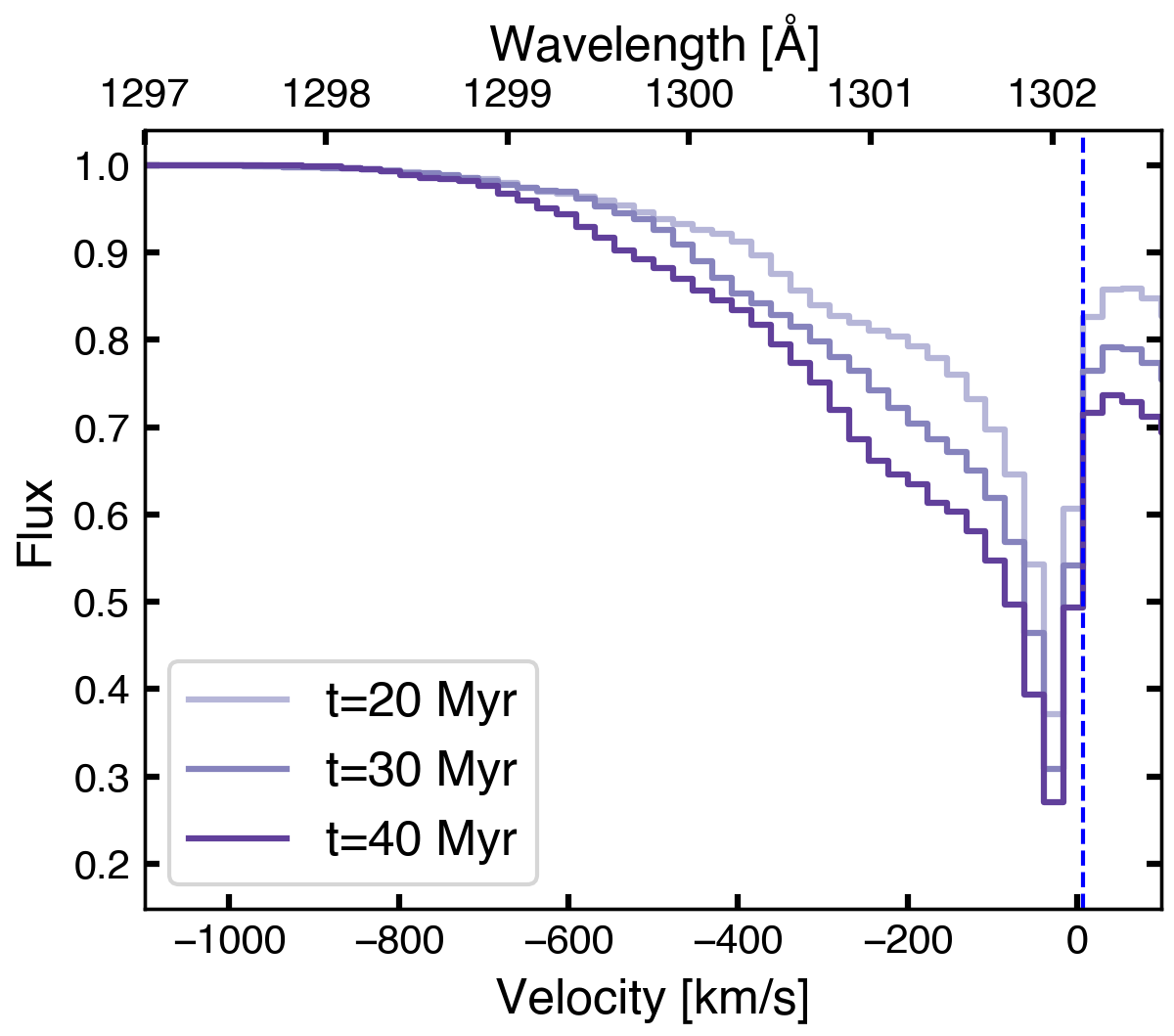}
    \includegraphics[scale=0.8]{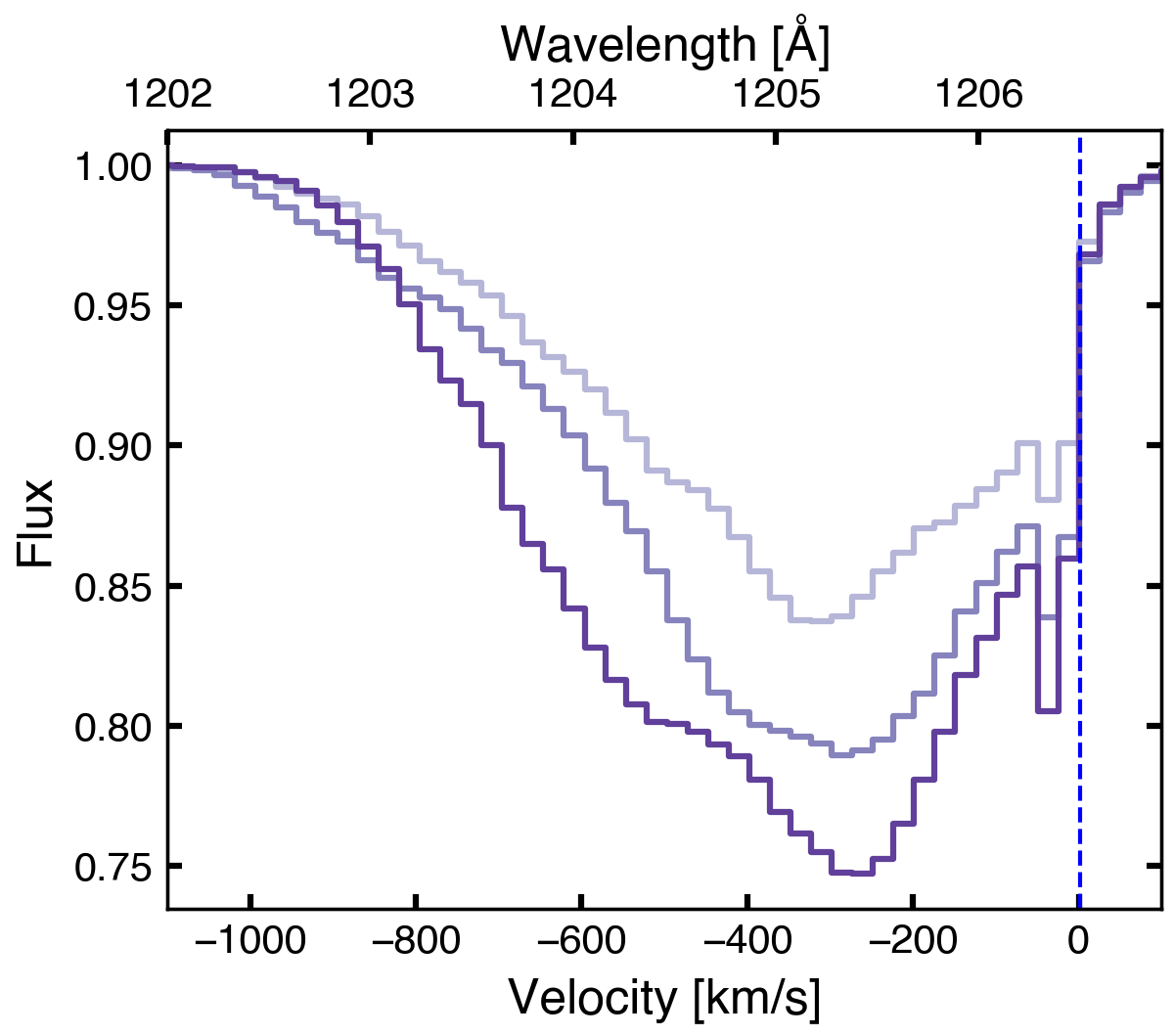}
    \caption{O I $\sim1302 \ \AA$ (left) and Si III $\sim1206 \ \AA$ (right) absorption line spectra generated from the $\Delta x=20 \ \rm pc$ resolution simulation at time snapshots of $\rm t=20, \ 30, \ \& \ 40 \ \mathrm{Myr}$, shown in light purple, purple, and dark purple, respectively. The line center is also plotted in dashed blue.}
    \label{fig:time evo}
\end{figure*}

Figure \ref{fig:time evo} shows absorption line spectra from all three snapshots for O I $\sim1302 \ \AA$ (left) and Si III $\sim1206 \ \AA$ (right). The most prominent qualitative trend in these spectra is the increase in the depth of the features at nearly all velocities as time progresses. This trend reflects not only that more material is being swept up into the outflow as a function of time, but also that the cool material preferentially traced by these lines is being accelerated as it moves outward, causing the lines to grow in depth across the full range of outflow velocities. The EWs for each time snapshot in Table \ref{tab:time_evo_EWs} to better quantify this trend. Table \ref{tab:time_evo_EWs} illustrates that there is indeed a clear and strong trend between the EW values and time across all lines in this study. More specifically, EW increases between each 10 Myr snapshot by $\sim20-35\%$, and this trend is independent of ionization potential. 

\begin{deluxetable}{cccc}[h]
\tabletypesize{\footnotesize}
\tablewidth{0pt}
\tablecaption{EW from each absorption line measured from the $\Delta x=20 \ \rm pc$ resolution simulation at time snapshots of $\rm t=20, \ 30, \ \& \ 40 \ \mathrm{Myr}$. All values listed in units of [$\text{\AA}$]. \label{tab:time_evo_EWs}}
\tablehead{
\colhead{Line} & \colhead{$\rm t = 20 \ Myr$} & \colhead{$\rm t = 30 \ Myr$} & \colhead{$\rm t = 40 \ Myr$}
}
\startdata
O I    & 0.51 & 0.64 & 0.79 \\
Si II  & 0.64 & 0.81 & 0.98 \\
C II   & 0.71 & 0.90 & 1.10 \\
Si III & 0.36 & 0.49 & 0.60 \\
Si IV  & 0.03 & 0.04 & 0.05 \\
O VI   & 0.03 & 0.04 & 0.05 \\
\enddata
\end{deluxetable}
\vspace{-12pt}

\begin{deluxetable}{cccc}[h]
\tabletypesize{\footnotesize}
\tablewidth{0pt}
\tablecaption{$v_{\rm cen}$ from each absorption line measured from the $\Delta x=20 \ \rm pc$ resolution simulation at time snapshots of $\rm t=20, \ 30, \ \& \ 40 \ \mathrm{Myr}$. All values listed in units of [$\rm km \ s^{-1}$]. \label{tab:time_evo_vs}}
\tablehead{
\colhead{Line} & \colhead{$\rm t = 20 \ Myr$} & \colhead{$\rm t = 30 \ Myr$} & \colhead{$\rm t = 40 \ Myr$}
}
\startdata
O I    & -272.6 & -239.8 & -225.4 \\
Si II  & -285.6 & -253.9 & -247.8 \\
C II   & -294.2 & -254.5 & -262.5 \\
Si III & -312.5 & -316.7 & -342.5 \\
Si IV  & -385.3 & -367.1 & -394.9 \\
O VI   & -474.1 & -448.3 & -496.3 \\
\enddata
\end{deluxetable}
\vspace{-12pt}

Though there is clearly more material being entrained/mixed into the outflow as a function of time, it is not immediately obvious whether or not the central velocity of these lines grows with time. The values of $v_{\rm cen}$ for the fiducial simulation resolution at time snapshots of $\rm t=20$, 30, and 40 Myr are reported in Table \ref{tab:time_evo_vs}. Overall, it appears that values of $v_{\rm cen}$ for the lower ionization potential lines (O I, Si II, and C II) tend to decrease with time, while the values for the higher ionization potential lines (Si III, Si IV, and O VI) tend to increase with time. However, it is important to note that the value of $v_{\rm cen}$ for each of these lines differs by less than $\sim11\%$ for all lines (approximately the same increase between simulation resolution steps in \S\ref{subsec: velocity}), and much of this variation can be attributed to the inherent short-comings of our fitting routine. More specifically, the significant asymmetry of absorption profiles caused by outflows is not fully captured by our double-gaussian fitting routine, so these trends in $v_{\rm cen}$ may not solely be reflecting the changes in the velocity structure of the outflows in the simulation snapshots. Certainly, the structures of the outflows are changing significantly with time, but that change is much more clearly reflected in the EWs of the absorption lines.

\subsection{Inclination Effects} \label{subsec: inclination}

\begin{figure*}[!ht]
    \centering
    \includegraphics[scale=0.8]{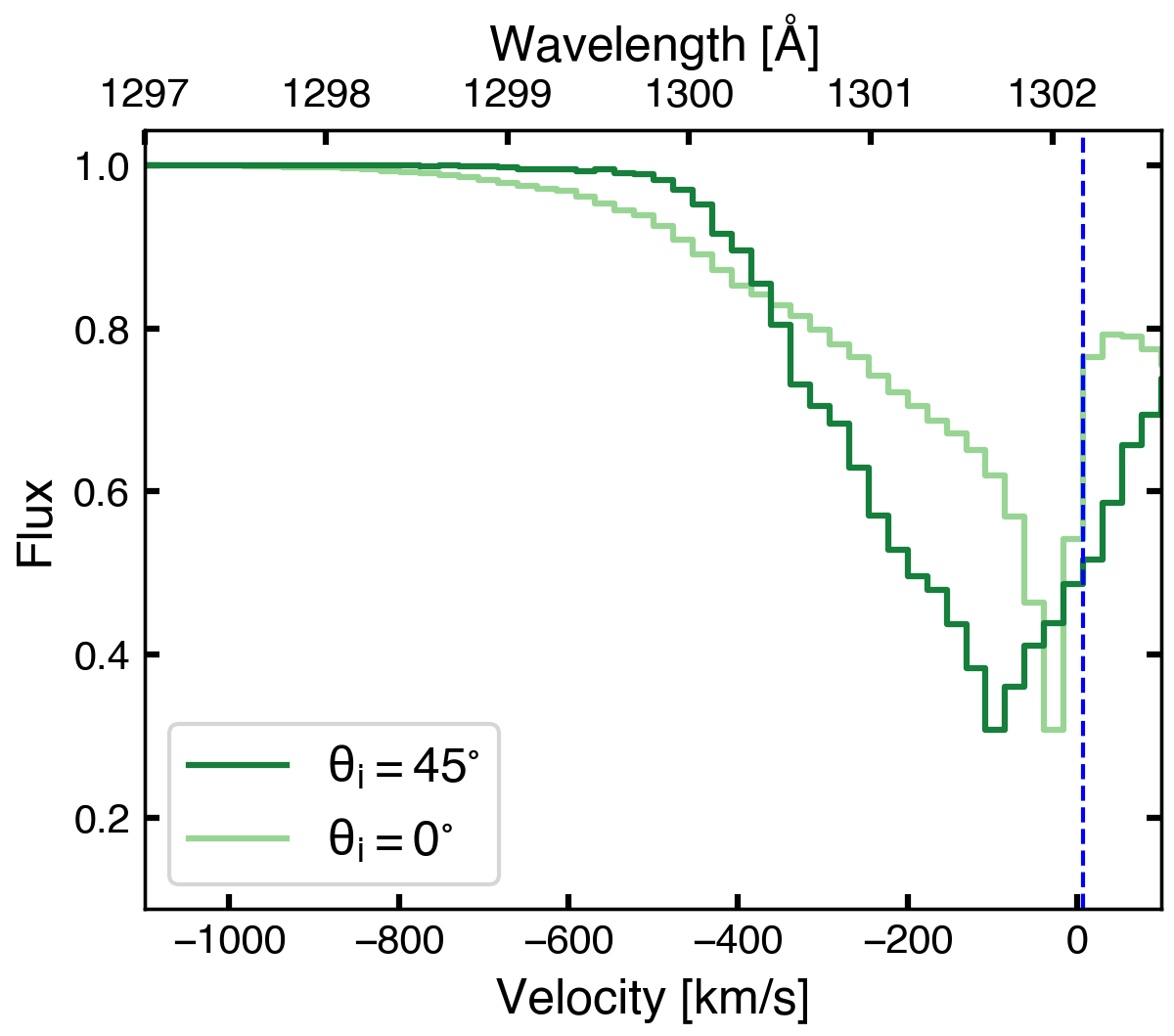}
    \includegraphics[scale=0.8]{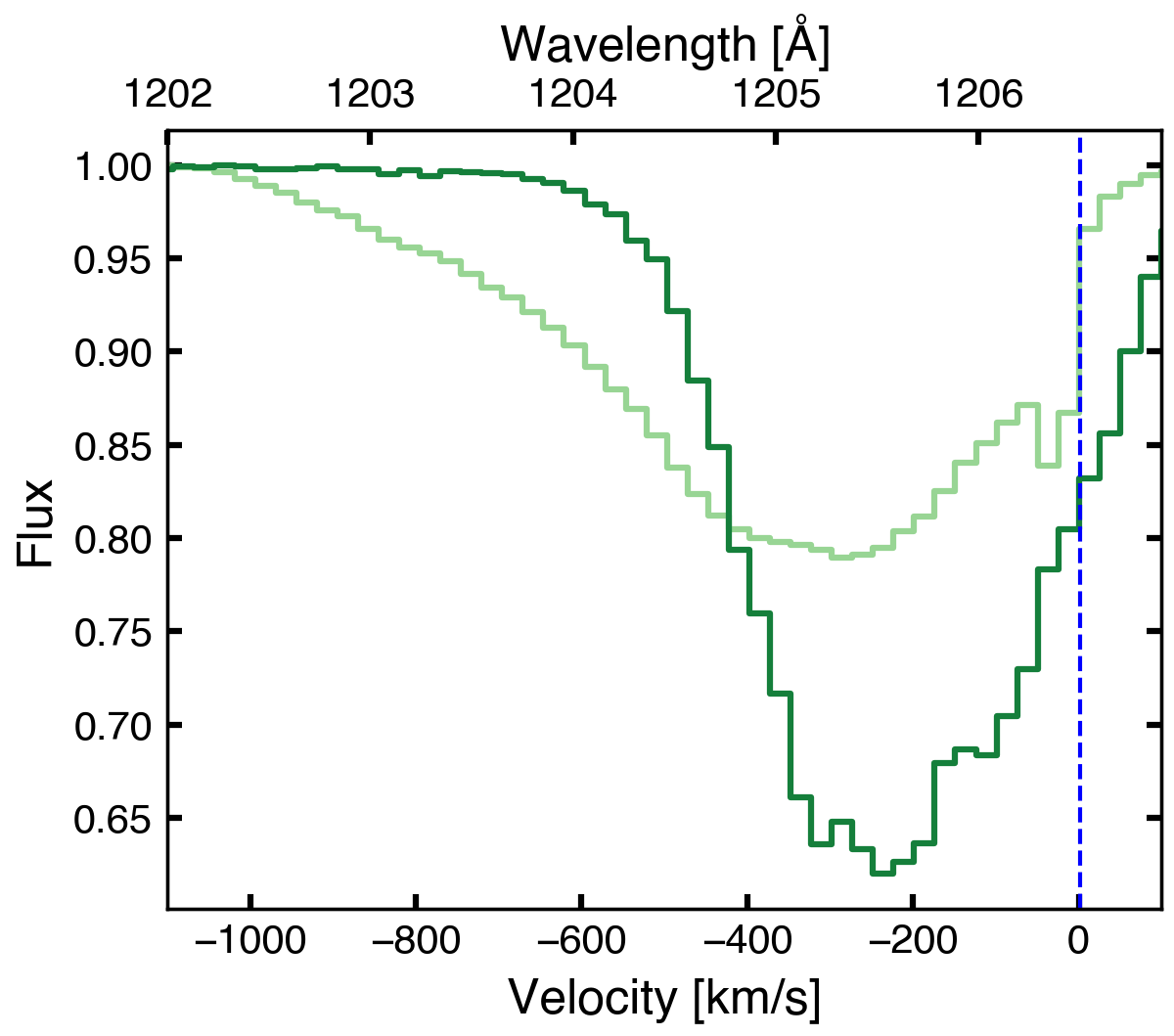}
    \caption{O I $\sim1302 \ \AA$ (left) and Si III $\sim1206 \ \AA$ (right) absorption line spectra generated from the $\Delta x=20 \ \rm pc$ resolution simulation, with a ``face-on" inclination (i.e. the fiducial inclination angle) in light green, and $\theta_i=45^{\circ}$ in dark green. Both are generated at the fiducial time snapshot ($\rm t=30\ \mathrm{Myr}$). The line center is also plotted in dashed blue.}
    \label{fig:inclination}
\end{figure*}

As mentioned in \S\ref{subsec:spec-gen}, down-the-barrel observations often exhibit a range of inclination angles. Thus far this work has focused on mock observations with ``face-on" inclination angles; in  this section, we investigate how inclination angle affects the absorption line features in the fiducial spectrum. To do so, we generate a spectrum representing a galaxy with an inclination angle $\theta_i=45^\circ$ (``$\theta_{45}$" hereafter), by propagating the LightRay objects at that angle from the mid-plane of the simulation domain. Figure \ref{fig:inclination} shows the resulting O I $\sim1302 \ \AA$ and Si III $\sim1206 \ \AA$ absorption features, as well as the corresponding features from the fiducial spectrum (referred to as ``$\theta_0$" for the remainder of this section).

The spectral features shown in Figure \ref{fig:inclination} illustrate the aggregate effects of inclination, which are predominantly increases in line depth, particularly at velocities near zero. Because the LightRay objects used to generate the $\theta_{45}$ spectrum propagate through the simulation domain at an angle, they intersect a larger portion of the ISM (now rotating from this angle) than their fiducial counterparts. At the same time, the inclined LightRay objects pass through a smaller portion of the high-velocity outflowing material, leading to a decrease in the central velocities of the spectral features in the resulting spectrum. The effects of inclination are quantified in Table \ref{tab:inclination}.

\begin{deluxetable}{ccccc}[h]
\tabletypesize{\footnotesize}
\tablewidth{0pt}
\tablecaption{EW and $v_{\rm cen}$ from each absorption line measured from the fiducial resolution simulation with inclination angles of $\theta = 0^{\circ}$ (or ``face on") and $\theta = 45^{\circ}$. We call these spectra $\theta_0$ and $\theta_{45}$, respectively. All values listed in units of [$\text{\AA}$] and $km \ s^{-1}$. \label{tab:inclination}}
\tablehead{
\colhead{Line} & \colhead{$\theta_{45} \ \rm EW$} & \colhead{$\theta_{0} \ \rm EW$} & \colhead{$\theta_{45} \ v_{\rm cen}$} & \colhead{$\theta_{0} \ v_{\rm cen}$}
}
\startdata
O I    & 0.80 & 0.64 & -153.6 & -239.8 \\
Si II  & 1.06 & 0.81 & -162.0 & -253.9 \\
C II   & 1.05 & 0.90 & -180.7 & -254.5 \\
Si III & 0.58 & 0.49 & -219.0 & -316.7 \\
Si IV  & 0.04 & 0.04 & -248.6 & -367.1 \\
O VI   & 0.04 & 0.04 & -279.5 & -448.3 \\
\enddata
\end{deluxetable}
\vspace{-12pt}

The EWs from the $\theta_{45}$ spectrum are $\sim0-31\%$ larger than those from the fiducial spectrum, while the $v_{\rm cen}$ values are smaller than the fiducial values by $\sim29-38\%$. The most dramatic increase in line depth occurs in the lower ionization potential lines, and conversely the higher ionization potential lines exhibit the largest decrease in the values of $v_{\rm cen}$. These results indicate that in the fiducial ``face-on" inclination, it is likely that the EWs will be smaller than those predicted by empirical scaling relations, and the velocity statistics will be larger than those predictions. We address this point further in \S\ref{subsec: obs comp}.

\subsection{Star-Formation Rate Dependence} \label{subsec: SFR dep}

\begin{figure*}[!ht]
    \centering
    \includegraphics[scale=0.575]{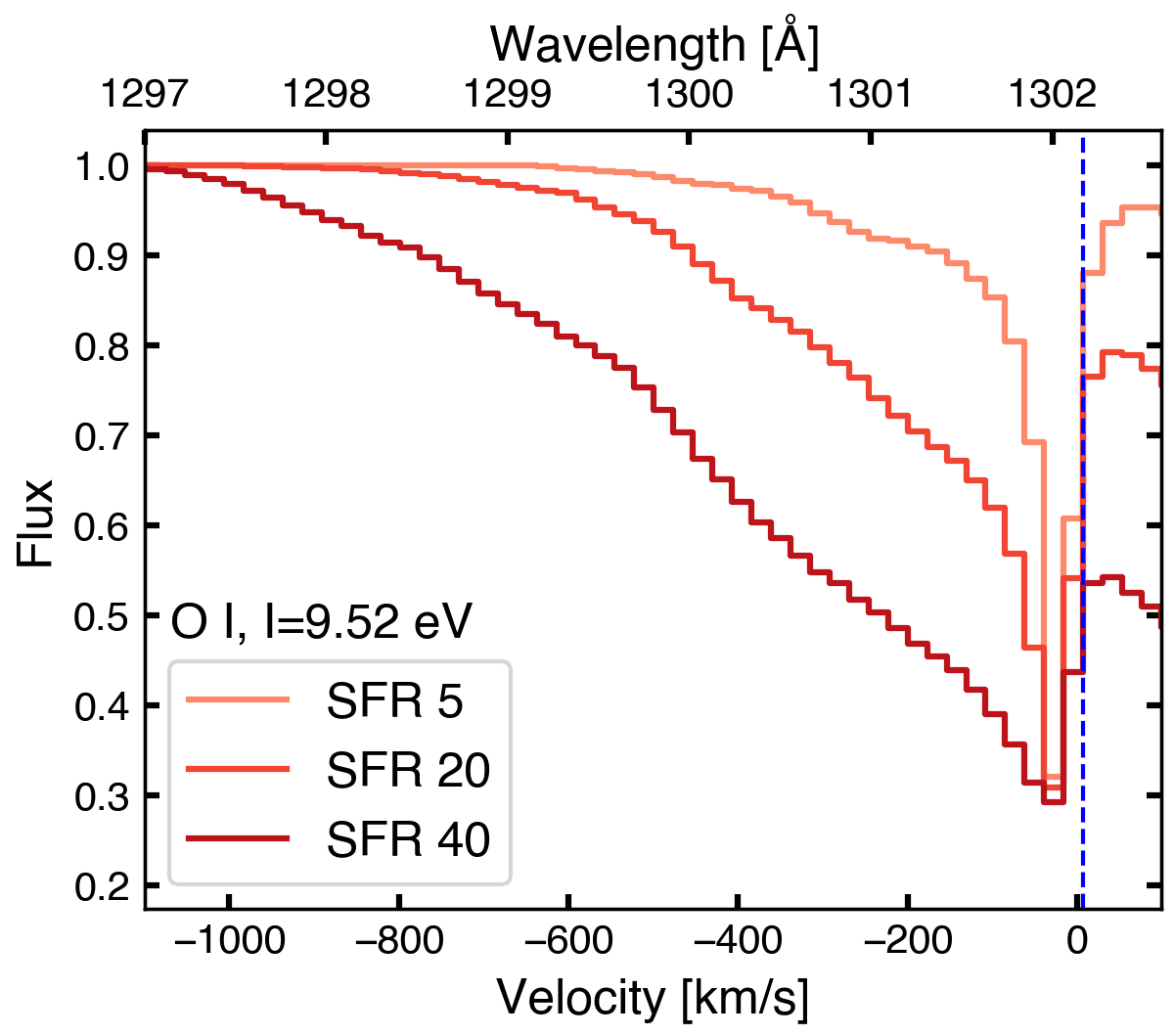}
    \includegraphics[scale=0.575]{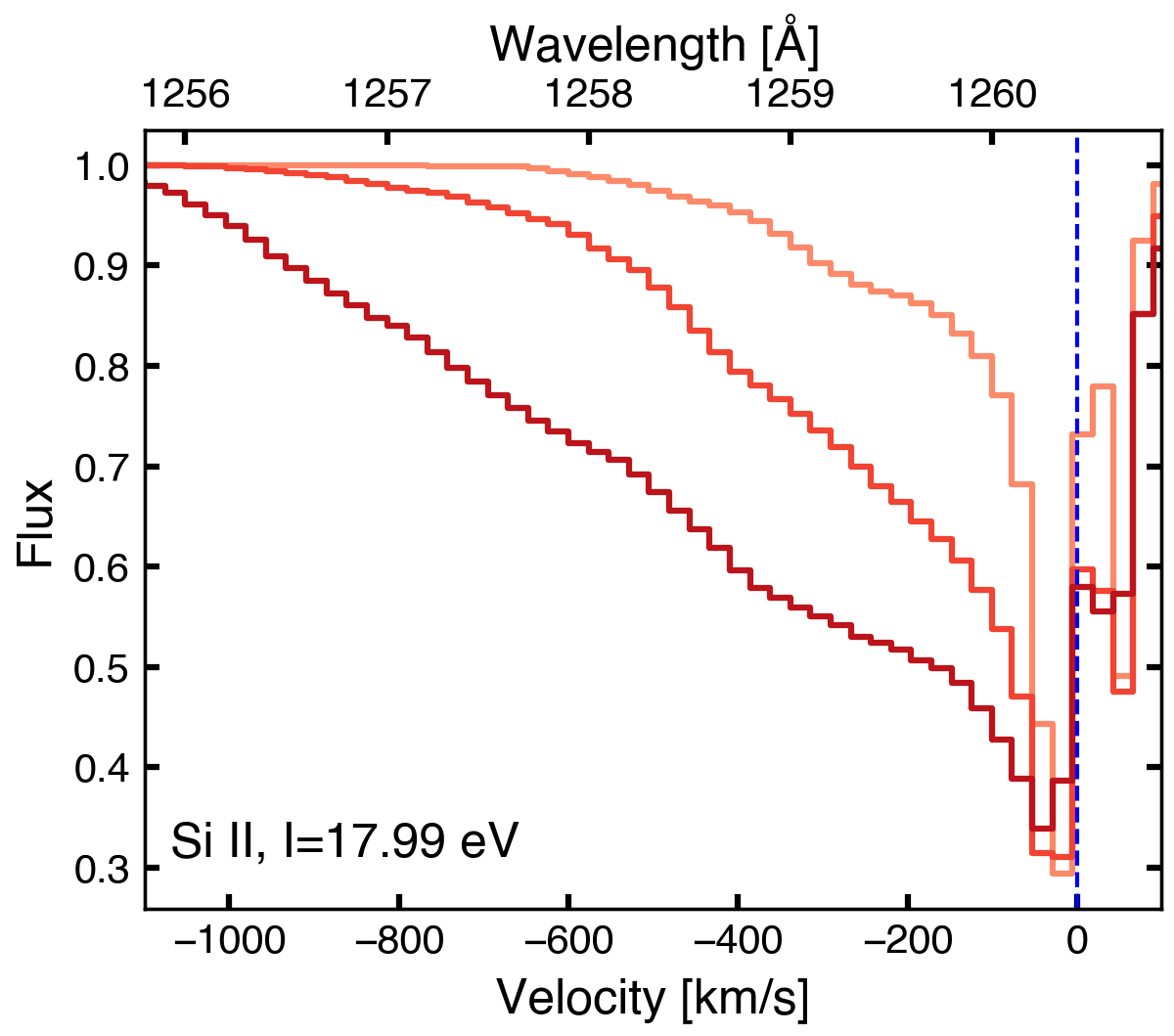}
    \includegraphics[scale=0.575]{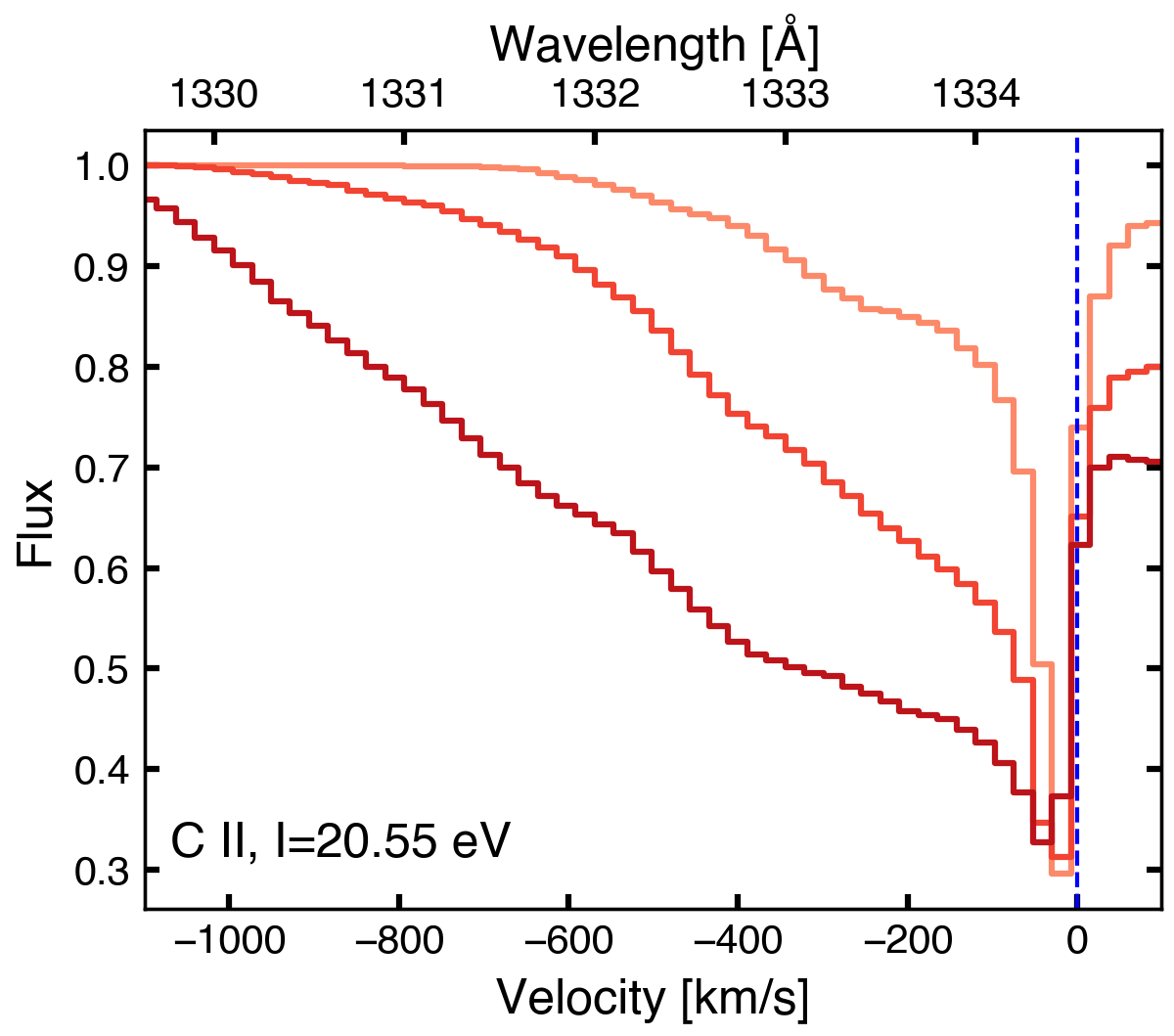}
    \includegraphics[scale=0.575]{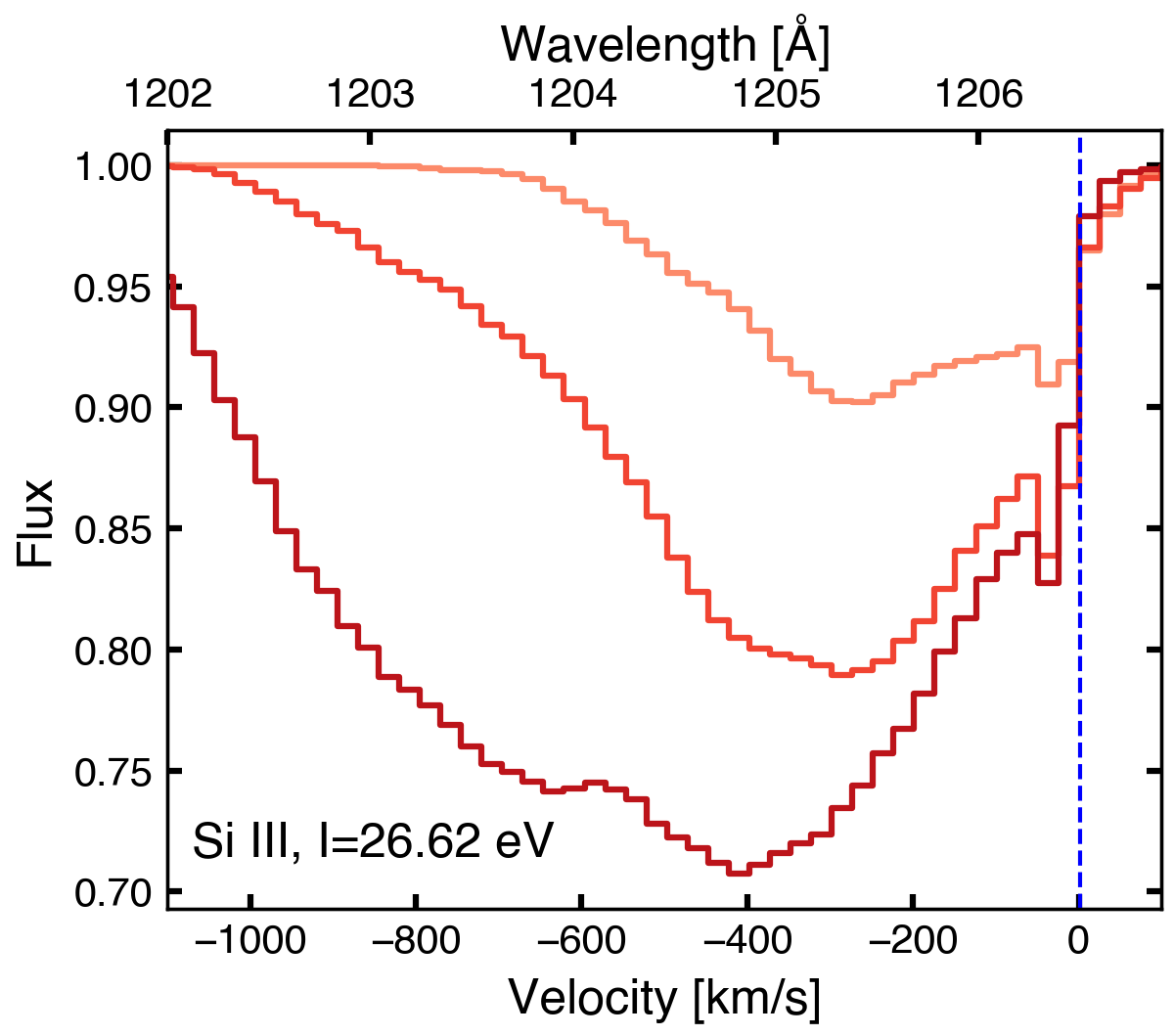}
    \includegraphics[scale=0.575]{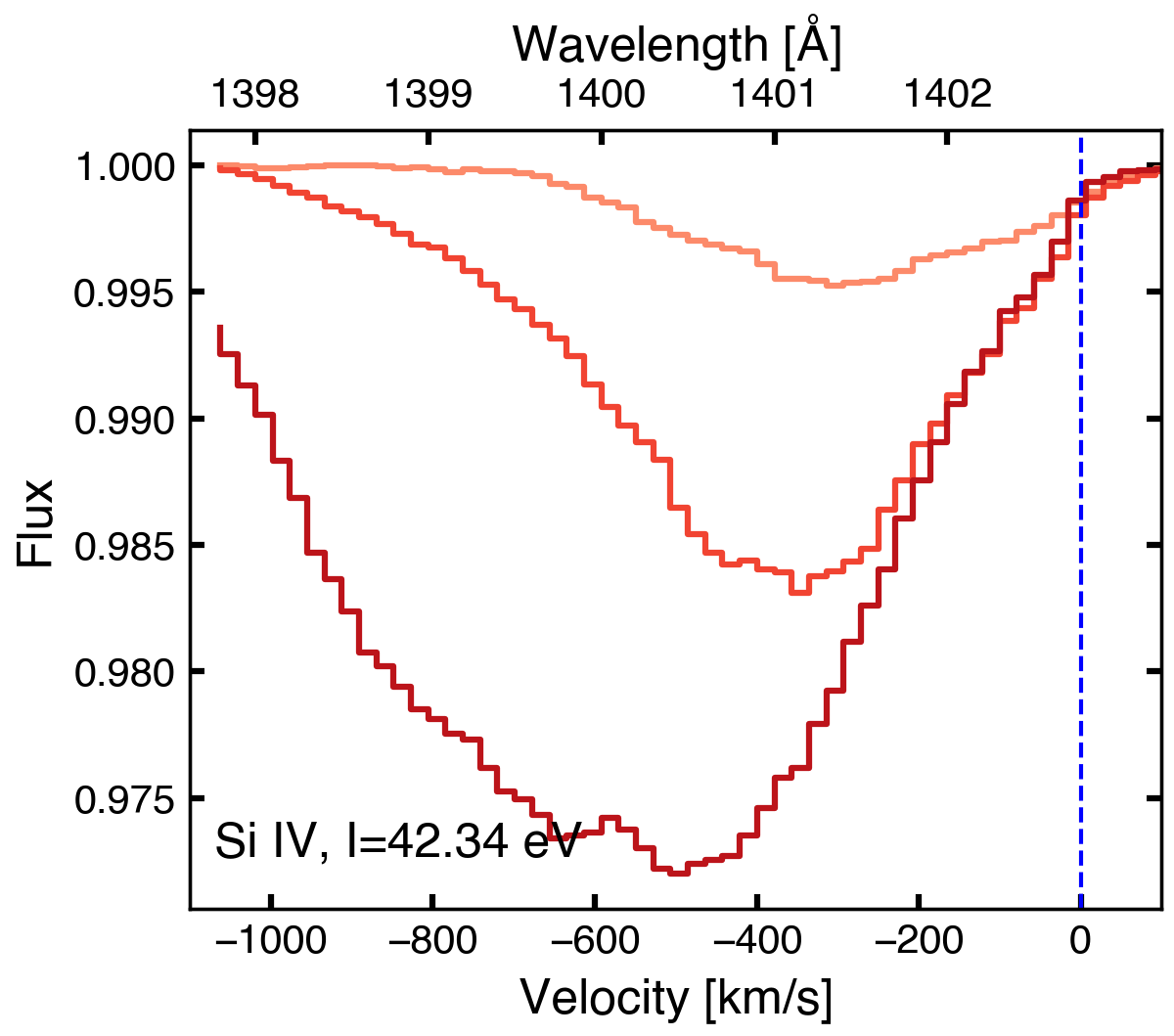}
    \includegraphics[scale=0.575]{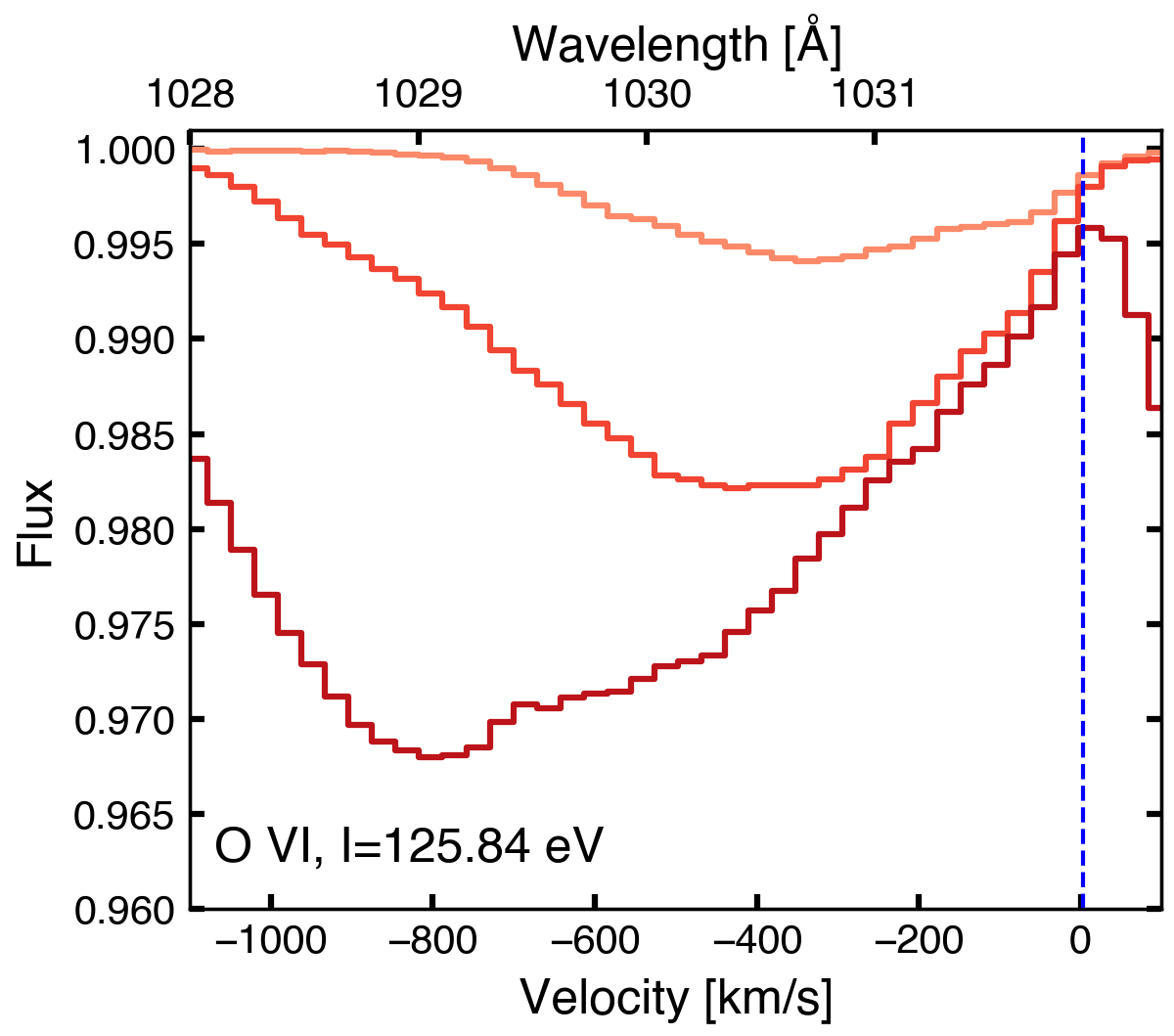}
    \caption{Mock absorption spectra generated from three simulations with SFRs of $5, \ 20, \ \& \ 40 \ \rm M_{\odot}yr^{-1}$, shown in orange, red, and dark red, respectively. All simulations have a spatial resolution of $\Delta x=20 \ \rm pc$. Spectra are arranged in ascending order of excitation potential. The spectra are generated in normalized flux units, with a constant wavelength resolution of $\Delta \lambda=0.1 \ \text{\AA}$. The rest wavelength of each line is plotted in dashed blue.}
    \label{fig:all_spec SFR}
\end{figure*}

The CGOLS SFR suite covers a range of star formation rates, which enables us to investigate into how SFR impacts the development of multiphase outflows in a controlled setting. In this section, we study how larger and smaller SFR (40 and $5 \rm \ M_{\odot} \ yr^{-1}$, hereafter ``SFR 40" and ``SFR 5," respectively) using the same simulation set up affects the resulting absorption profiles. The SFR 40 and SFR 5 spectra, along with the fiducial SFR 20 spectrum, are shown in Figure \ref{fig:all_spec SFR}. The features in the SFR 40 spectrum indicate that there is significantly more mass in this outflow than in the fiducial counterpart, particularly at high velocities. Conversely, the features in the SFR 5 spectrum are much shallower than those in the fiducial spectrum, and exhibit much less high-velocity absorption. These trends reflect the expected result that the more (less) energy is driving the outflows, the more (less) accelerated that material will be in the same amount of time. We quantify the differences between the velocity structure and depths of these spectra in Tables \ref{tab:SFR_vs} and \ref{tab:SFR_EWs}, respectively.

\begin{deluxetable}{cccc}[h]
\tabletypesize{\footnotesize}
\tablewidth{0pt}
\tablecaption{$v_{\rm cen}$ from each absorption line measured from the $\Delta x=20 \ \rm pc$ resolution simulations with SFRs of $5, \ 20, \ \& \ 40 \ \mathrm{M_{\odot}yr^{-1}}$, respectively, all measured at a time snapshot of $\rm t=30 \ Myr$. All values listed in units of [$\rm km \ s^{-1}$]. \label{tab:SFR_vs}}
\tablehead{
\colhead{Line} & \colhead{$\rm 5 \ M_{\odot}yr^{-1}$} & \colhead{$\rm 20 \ M_{\odot}yr^{-1}$} & \colhead{$\rm 40 \ M_{\odot}yr^{-1}$}
}
\startdata
O I    & -152.9 & -239.8 & -249.9 \\
Si II  & -169.8 & -253.9 & -316.0 \\
C II   & -226.0 & -254.5 & -349.4 \\
Si III & -231.4 & -316.7 & -496.5 \\
Si IV  & -282.8 & -367.1 & -535.7 \\
O VI   & -352.2 & -448.3 & -709.6 \\
\enddata
\end{deluxetable}
\vspace{-10pt}

\begin{deluxetable}{cccc}[h]
\tabletypesize{\footnotesize}
\tablewidth{0pt}
\tablecaption{EWs from each absorption line measured from the $\Delta x=20 \ \rm pc$ resolution simulations with SFRs of $5, \ 20, \ \& \ 40 \ \mathrm{M_{\odot}yr^{-1}}$, respectively, all measured at a time snapshot of $\rm t=30 \ Myr$. All values listed in units of [$\text{\AA}$]. \label{tab:SFR_EWs}}
\tablehead{
\colhead{Line} & \colhead{$\rm 5 \ M_{\odot}yr^{-1}$} & \colhead{$\rm 20 \ M_{\odot}yr^{-1}$} & \colhead{$\rm 40 \ M_{\odot}yr^{-1}$}
}
\startdata
O I    & 0.27 & 0.64 & 1.34 \\
Si II  & 0.38 & 0.81 & 1.44 \\
C II   & 0.39 & 0.90 & 1.70 \\
Si III & 0.17 & 0.49 & 0.95 \\
Si IV  & 0.01 & 0.04 & 0.09 \\
O VI   & 0.01 & 0.04 & 0.10 \\
\enddata
\end{deluxetable}
\vspace{-12pt}

Table \ref{tab:SFR_vs} reports the $v_{\rm cen}$ values for all three SFRs, and these statistics further illuminate the direct influence that SFR has on the properties star-formation driven outflows. The values of $v_{\rm cen}$ measured from the SFR 5 spectrum are lower than the fiducial values by $\sim17-42\%$, and the values from the SFR 40 spectrum are higher than the fiducial values by $\sim4-58\%$. These trends are mirrored, and more pronounced, in those of the EWs of these spectra. The EWs reported in Table \ref{tab:SFR_EWs} make it abundantly clear that the amount of cool gas entrained in outflows, and how fast that material is moving, is directly reflected in the depth of spectral absorption features. The EWs measured from the SFR 5 spectrum are lower than the fiducial values by factors of $\sim50-75\%$, while those measured from the SFR 40 spectrum are larger than the fiducial values by factors between $\sim60-150\%$. These trends are insensitive to ionization potential, highlighting that the outflows driven by a higher (lower) SFR are both more (less) efficiently dredging-up ISM material into the flow, while also accelerating that material more (less) dramatically. Taken in concert, the differences in the statistics of these spectra demonstrate that SFR has a very strong and direct influence on the mass entrained in the outflow, as well as its acceleration.

\subsection{Model Dependence} \label{subsec: sim comp}

The results presented thus far have investigated the dependence of the statistics of the spectra with respect to various parameters in one suite of simulations, and have shown compellingly that the velocity statistics are converged. However, it is still unclear whether the spectrum from a simulation in a different suite would show similar absorption features to those from the CGOLS SFR suite. To answer this question, we have repeated our analysis on the previous generation of the CGOLS project: CGOLS V. The initial conditions of CGOLS SFR, and the fiducial SFR ($20 \rm \ Myr$), are identical to that of CGOLS V (as outlined in \S\ref{sec:simulation}), but the volume of the simulation box for CGOLS SFR is 32 times larger than that of CGOLS V, while the fiducial resolution of CGOLS SFR ($\Delta x\sim20 \rm \ pc $) is four times coarser than that of CGOLS V ($\Delta x\sim5 \rm \ pc $). In this section we investigate the effects that higher \textit{native} simulation resolution in a smaller simulation domain have on the absorption profiles in our study.

\begin{figure*}[!ht]
    \centering
    \subfigure[]{\includegraphics[scale=0.575]{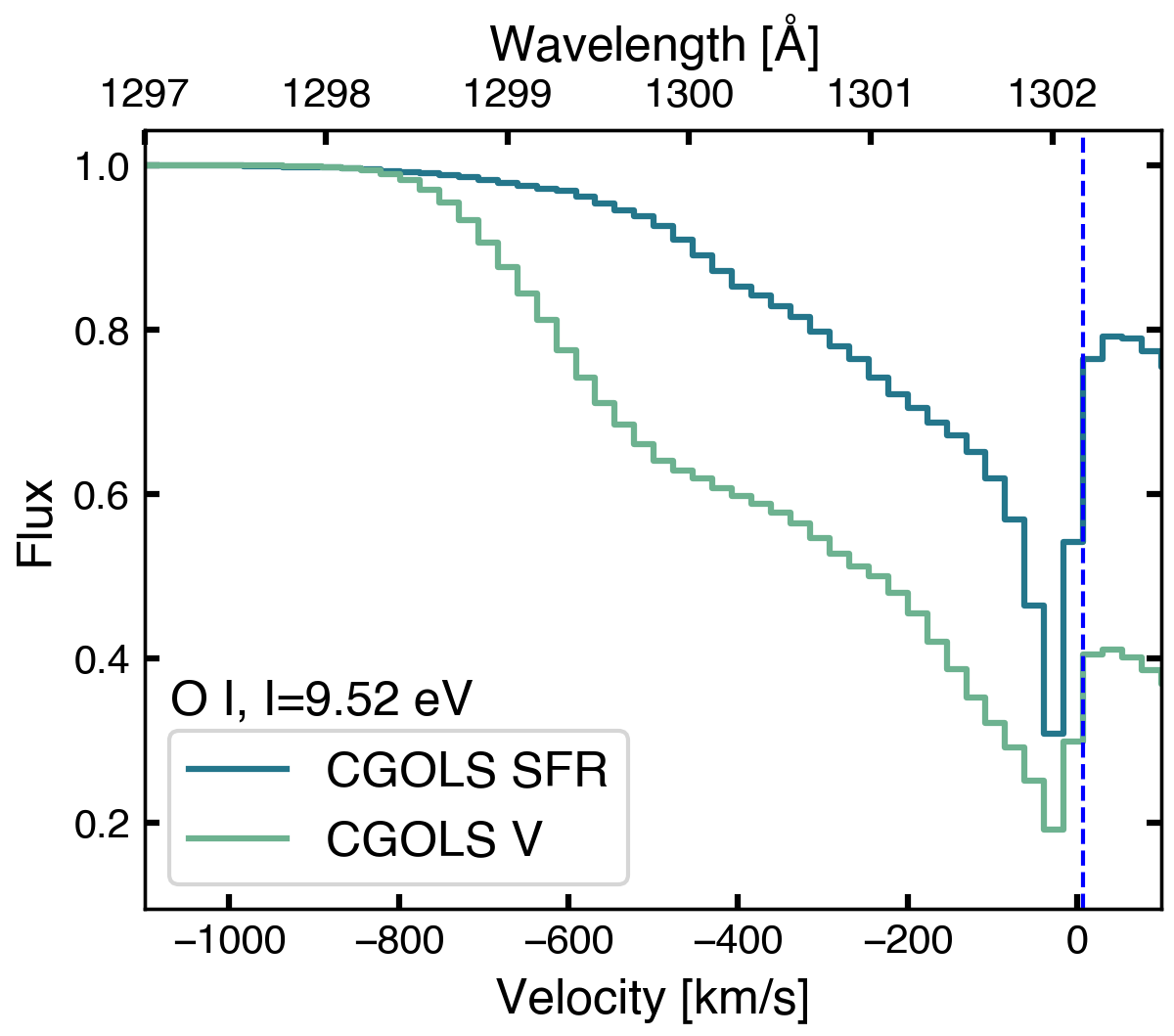}}
    \subfigure[]{\includegraphics[scale=0.575]{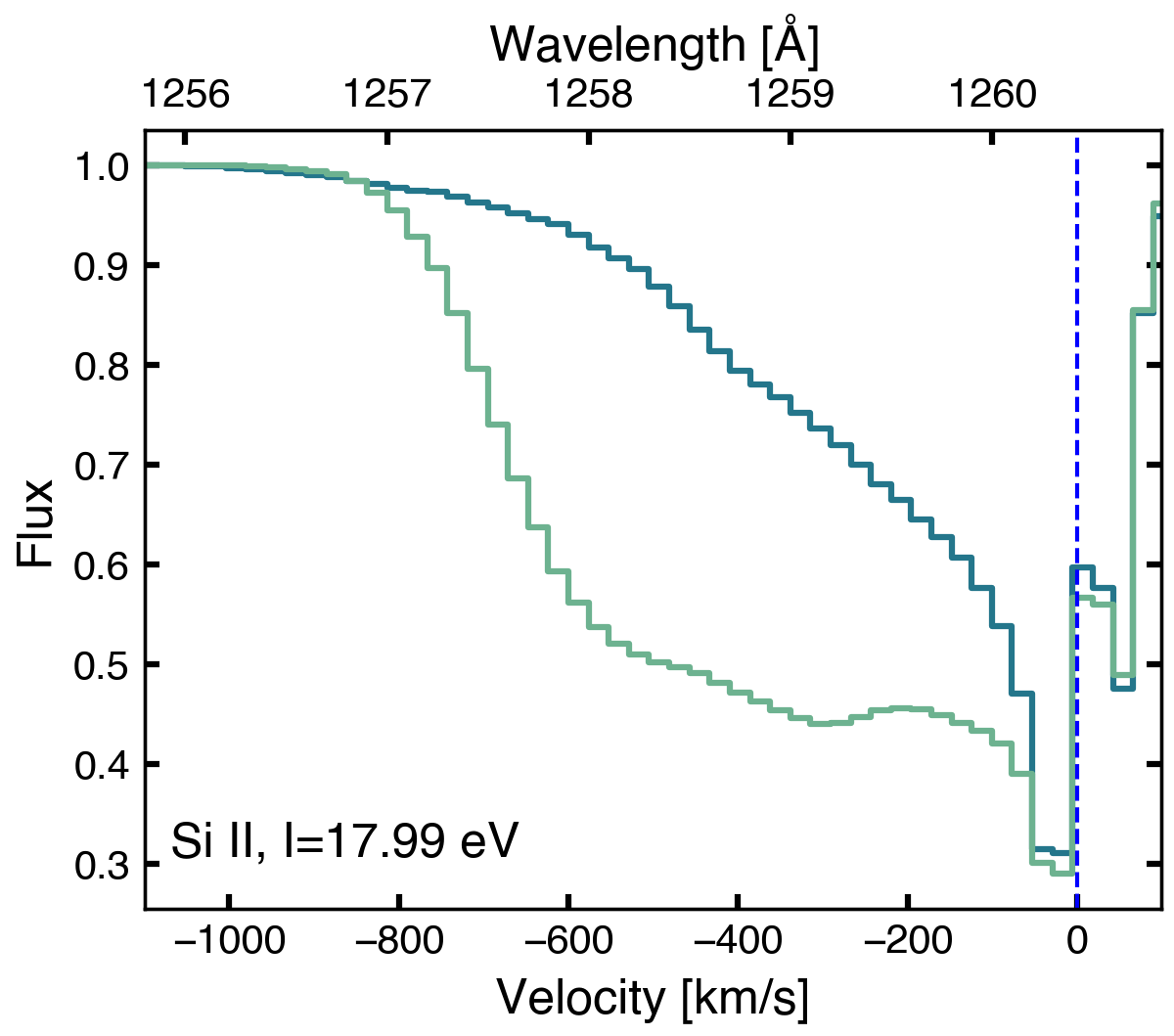}}
    \subfigure[]{\includegraphics[scale=0.575]{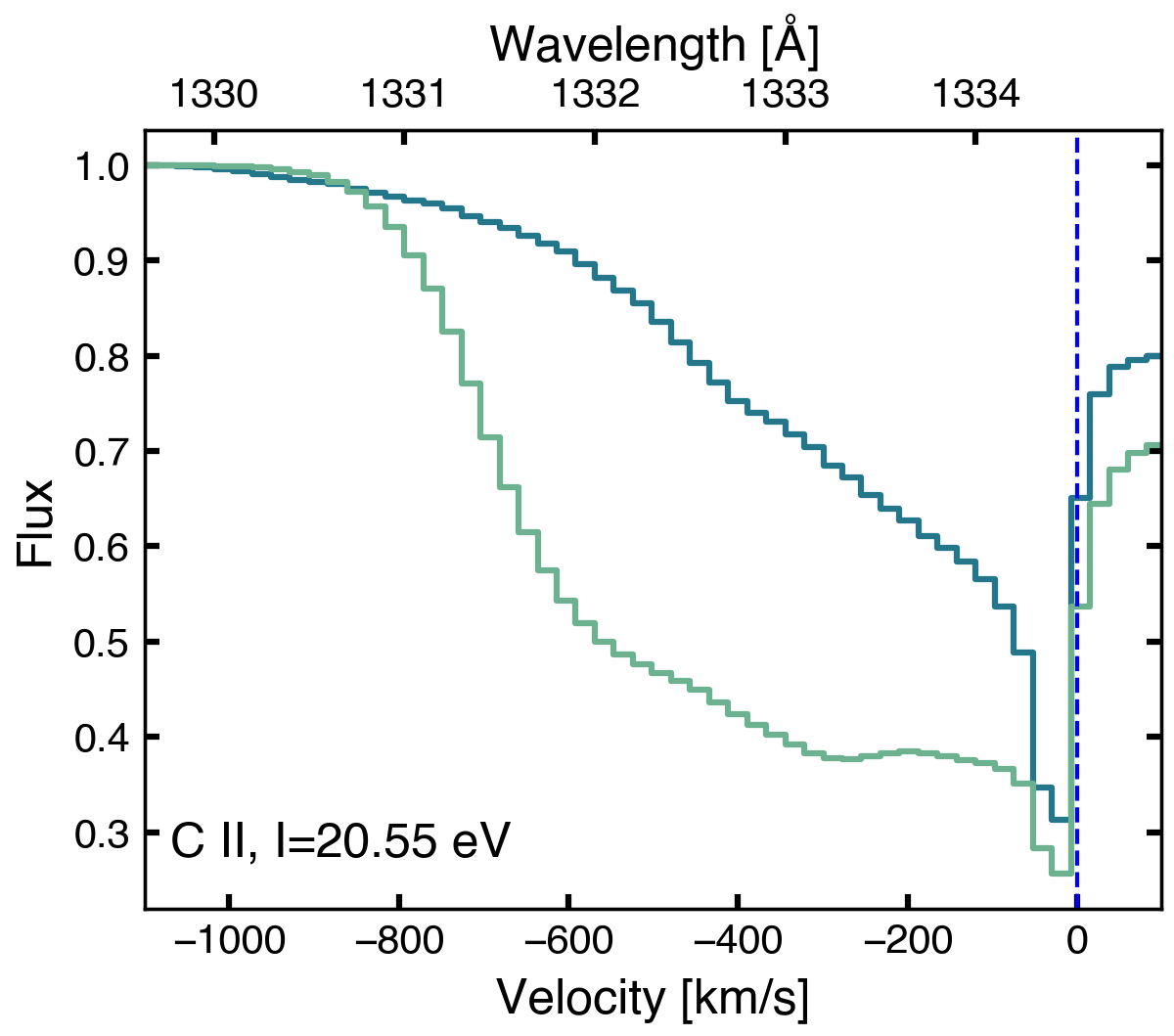}}
    \subfigure[]{\includegraphics[scale=0.575]{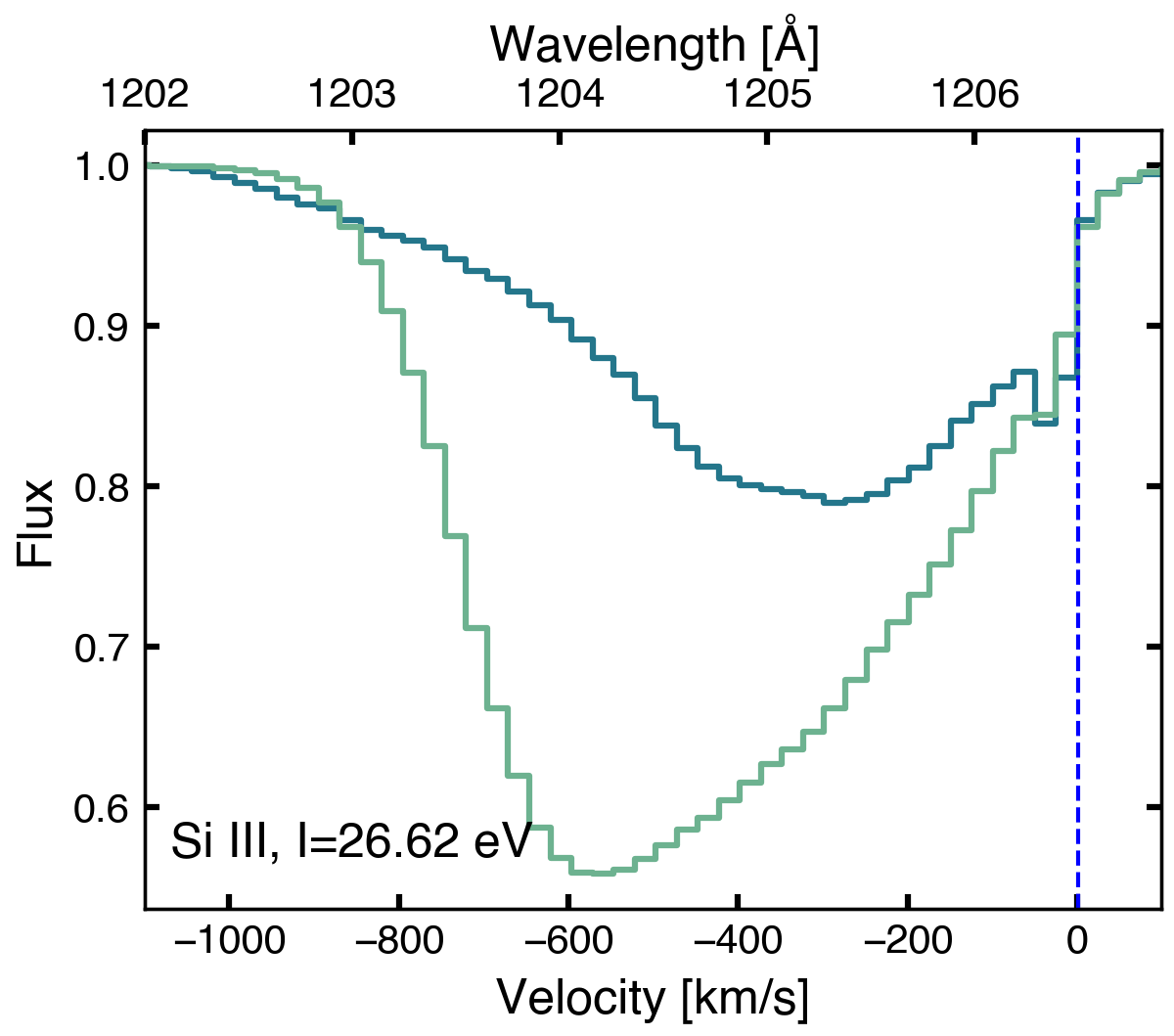}}
    \subfigure[]{\includegraphics[scale=0.575]{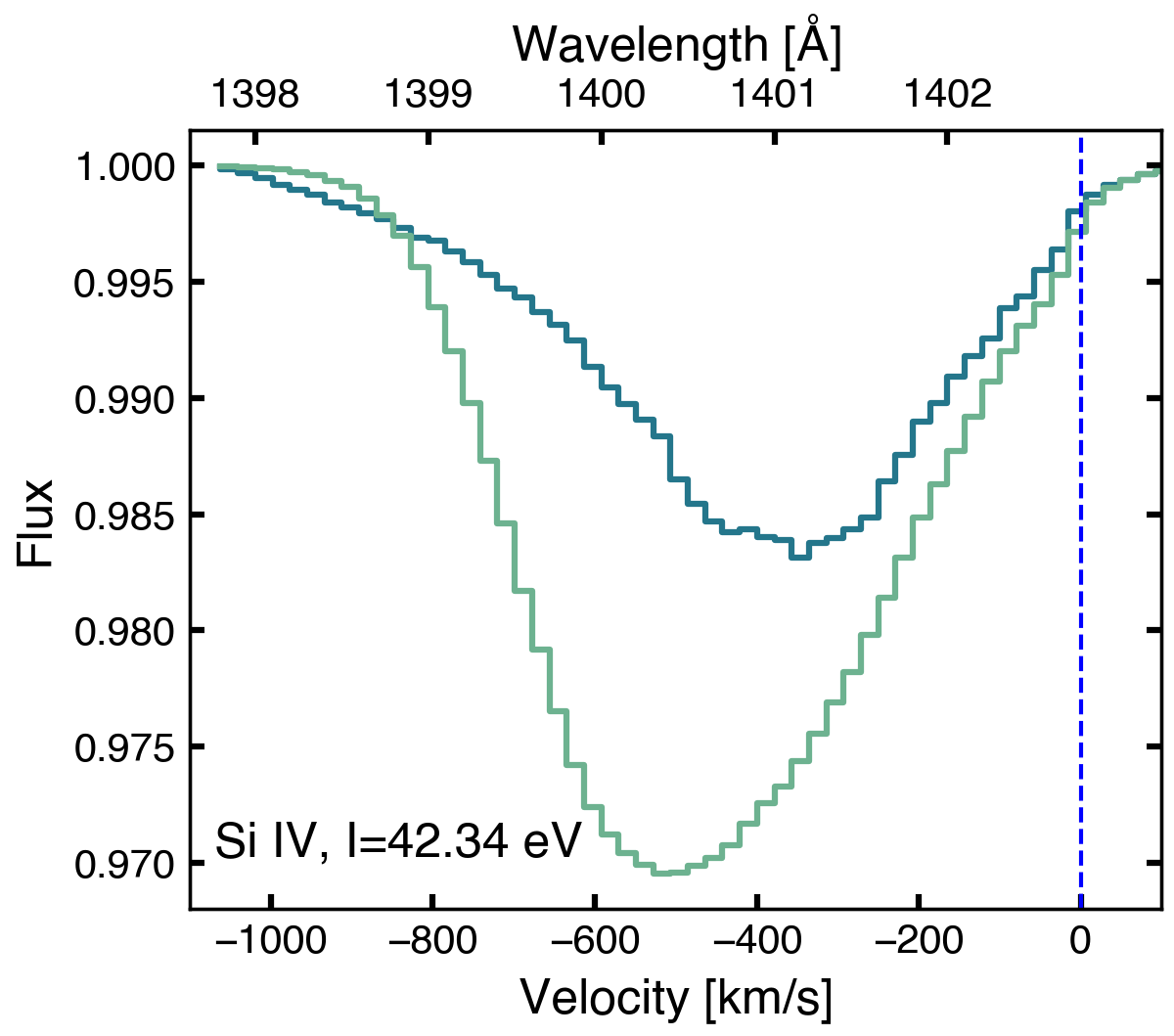}}
    \subfigure[]{\includegraphics[scale=0.575]{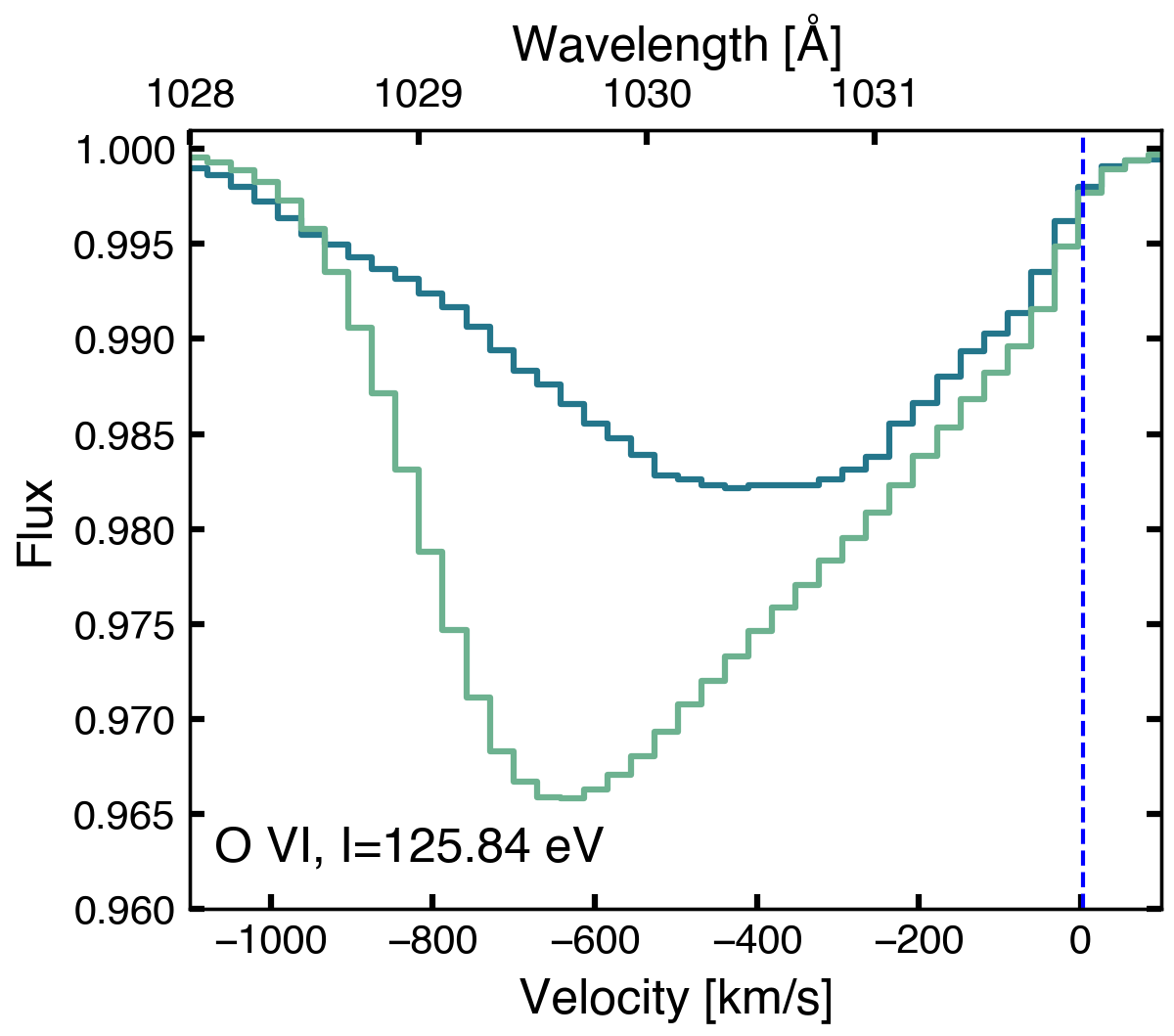}}
    \caption{Mock absorption spectra generated from CGOLS V (green), using an exponential weighting function with a characteristic radius $\rm R=1\ kpc$. Spectra are arranged in ascending order of excitation potential. Also shown is the fiducial ($\Delta x\simeq20 \ \rm pc$) CGOLS SFR spectrum (dark blue). Both simulations have the same SFR ($20 \ \rm M_{\odot} \ yr^{-1}$) and are measured at $\rm t=30 \ Myr$, but CGOLS V has a four times finer resolution than that of CGOLS SFR. The spectra are generated in normalized flux units, with a constant wavelength resolution of $\Delta \lambda=0.1 \ \text{\AA}$. The line center is plotted in dashed blue.}
    \label{fig:all_spec CGOLS V}
\end{figure*}

We present the full suite of absorption lines generated from the CGOLS V dataset, along with the corresponding lines from the fiducial CGOLS SFR 20 spectrum, in Figure \ref{fig:all_spec CGOLS V}. Figure \ref{fig:all_spec CGOLS V} qualitatively illustrates the major differences between the CGOLS V and the fiducial CGOLS SFR spectra. Across the full suite of absorption lines, we find that the CGOLS V absorption features are deeper at almost all velocities, with the majority of the excess absorption at outflowing velocities. This increased presence of outflowing material becomes more pronounced in the higher ionization potential lines (the bottom row of Figure \ref{fig:all_spec CGOLS V}). By contrast, all of the spectra exhibit similar features near $v\sim-100-0 \ \rm km \ s^{-1}$, indicating that this feature primarily traces gas in or near the ISM, which should be very similar for all models. Both the Si IV and OVI lines are still anomalously shallow in the CGOLS V spectrum (and change by a similar amount as the lower ions), which further supports our conclusion that this missing gas is likely not coming mainly from unresolved collisionally-ionized intermediate temperature features.

\begin{deluxetable*}{ccccccc}[!ht]
\tabletypesize{\footnotesize}
\tablewidth{0pt}
\tablecaption{$v_{\rm cen} \mathrm{[km \ s^{-1}]}$, $v_{\rm 90} \mathrm{[km \ s^{-1}]}$, EW, for the CGOLS V spectrum, with all percentage increases between the corresponding values from the fiducial CGOLS SFR simulation for each statistic. \label{tab:V_vs_SFR}}
\tablehead{
\colhead{Line} & \colhead{$v_{\rm cen} \mathrm{[km \ s^{-1}]}$} & \colhead{$\Delta v_{\rm cen}[\%]$} & \colhead{$v_{\rm 90} \mathrm{[km \ s^{-1}]}$} & \colhead{$\Delta v_{\rm 90}[\%]$} & \colhead{EW} & \colhead{$\Delta \rm EW[\%]$}
}
\startdata
O I    & -272.4 & 14 & -683.3  & 59 & 1.38 & 16 \\
Si II  & -354.4 & 40 & -837.7  & 66 & 1.61 & 99 \\
C II   & -311.2 & 22 & -838.4  & 54 & 1.82 & 102 \\
Si III & -440.5 & 39 & -869.7  & 46 & 1.03 & 110 \\
Si IV  & -447.3 & 22 & -506.5  & 27 & 0.07 & 75 \\
O VI   & -573.2 & 28 & -1049.4 & 20 & 0.07 & 75 \\
\enddata
\end{deluxetable*}
\vspace{-12pt}

In Table \ref{tab:V_vs_SFR} we report the velocity statistics and EWs measured from the CGOLS V spectrum, as well as the percentage difference between the corresponding values from the fiducial CGOLS SFR spectrum. We find that the trends in the velocity statistics and EWs behave similary to those seen in the fiducial CGOLS SFR spectrum (see \S\ref{subsec: velocity}, \ref{subsec: EW}) however the values of all statistics are substantially larger in CGOLS V. In particular, the values of $v_{\rm cen}$ for CGOLS V are larger than those of fiducial (i.e. SFR 20) CGOLS SFR simulation by $\sim15-40\%$, and the EWs are larger still by factors between $\sim50-115\%$.

The main reason for the discrepancy between the EWs of CGOLS V and CGOLS SFR 20 (which have the same star formation rate) appears to be a significant difference in the mass contained in each simulation's outflow. The cool gas mass contained in the CGOLS V outflow at $\rm t = 30 \ Myr$ is approximately $1.5\times10^8 \ \rm M_{\odot}$ \citep{schneider2024cgols}, while the cool gas mass contained in the CGOLS SFR 20 outflow at the same time snapshot is only $\sim5.5\times10^7 \ \rm M_{\odot}$. In addition, the larger outflow velocities measured from the CGOLS V spectrum may be a result of the smaller box size. By 30 Myr, the CGOLS V hot outflow has completely cleared the pre-existing CGM from the simulation domain, and the hot outflow velocities align well with the ``free wind" prediction described by models with no ambient CGM. By contrast, the outflowing material in the CGOLS SFR simulation box is still pushing against the pressure of the ambient medium inside the simulation domain, preventing that material from reaching the high asymptotic velocities seen in CGOLS V at all but the smallest radii. Overall, the discrepancies between CGOLS V and CGOLS SFR 20 reinforce that adequately predicting cool outflow properties requires resolutions higher than 20pc to better resolve the physics driving the evolving structures of multiphase outflows, and that this higher resolution is also needed in larger simulation volumes.

\section{Discussion} \label{sec:discussion}

In the following section, we will contextualize our results with relevant observational data, as well as address some of the limitations of the current methods we have employed to generate our mock observations. First, we compare the results from the fiducial simulation data ($\Delta x=20 \ \rm pc$ resolution) with two relevant observational studies of galactic outflows. We further contextualize these observational comparisons with the data from the SFR 5, SFR 40, and CGOLS V spectra, as well considering temporal evolution and inclination effects. We then investigate the possible causes for the under-population of Si IV and O VI in all of our spectra, paying particular attention to the role of the UV ionizing background, and outline our future plans for improving the physical realism of our analysis pipeline.

\subsection{Comparison to Observations} \label{subsec: obs comp}

When comparing the fiducial velocity statistics to those derived from observation, we find that, on average, our values for both the central and 90th-percentile velocities at the highest simulation resolution agree well with those found in the literature. \cite{xu2022classy} report scaling relations between SFR and outflow velocity. These scaling relations are derived from all of the galaxies that host outflows within their sample, and consider velocities measured from both low and high ionization potential lines (not including O VI). Their scaling relation predicts that a galaxy with a SFR of $20 \ \rm M_{\odot}\ yr^{-1}$ would host outflows with an average velocity of $255 \rm \ km \ s^{-1}$. The magnitude of the average of the values of $v_{\rm cen}$ from our study (for the fiducial simulation resolution \cite[excluding O VI and weighted to better replicate the distribution of line species in][]{xu2022classy} is $282 \rm \ km \ s^{-1}$, which is $10.5\%$ larger than the predicted value from \cite{xu2022classy}. \cite{chisholm2016shining} report similar scaling relations from a similar observational study (48 low-redshift, star-forming galaxies hosting outflows observed with COS) for both $v_{\rm cen}$ and $v_{90}$, but rather than reporting a scaling relation for the average velocity derived from all absorption features, their relations are for individual absorption lines (O I and for Si II-IV). The predictions from these scaling relations for $v_{\rm cen}$ and $v_{90}$ are reported in Table \ref{tab:obs_v}, along with the percentage difference between these values and those from the fiducial spectrum (see \S\ref{subsec: velocity} for more details on these values). 

\begin{deluxetable}{ccccc}[!ht]
\tabletypesize{\footnotesize}
\tablewidth{0pt}
\tablecaption{$v_{\rm cen} \mathrm{[km \ s^{-1}]}$, $v_{\rm 90} \mathrm{[km \ s^{-1}]}$ predictions from the \cite{chisholm2016shining} scaling relations for O I and Si II-IV, with all percentage increases between the corresponding values from the fiducial CGOLS SFR simulation for each statistic. \label{tab:obs_v}}
\tablehead{
\colhead{Line} & \colhead{$v_{\rm cen} \mathrm{[km \ s^{-1}]}$} & \colhead{$\Delta v_{\rm cen}[\%]$} & \colhead{$v_{\rm 90} \mathrm{[km \ s^{-1}]}$} & \colhead{$\Delta v_{\rm 90}[\%]$}
}
\startdata
O I    & 126.5 & 90  & 394.6 & 8   \\
Si II  & 171.5 & 48  & 593.1 & -18 \\
Si III & 185.4 & 71  & 774.7 & -42 \\
Si IV  & 144.3 & 154 & 540.9 & 9   \\
\enddata
\end{deluxetable}
\vspace{-12pt}

The contrast between the \cite{chisholm2016shining} predictions and the fiducial CGOLS SFR statistics are much more pronounced, which is unsurprising, as we are only considering SFR when making these comparisons, and are not accounting for the stellar mass (or any other properties) of the galaxy. It is also not entirely surprising that our values of $v_{\rm cen}$ are larger than these predicted values, as we are making our measurements with a perfectly face-on inclination. While the galaxies in both of the aforementioned studies are selected to be close to face-on inclination, any deviation will lead to outflow velocity measurements that are systematically lower than those that would result from perfectly face-on inclination observations (see \S\ref{subsec: inclination} for a more detailed analysis of inclination effects on our spectra). Additionally, if this comparison were with the $\rm t=20\ Myr$ time snap shot (as opposed to the fiducial $\rm t=30\ Myr$ snap shot), the differences between the predictions and the values from the mock spectrum would not be as pronounced, though the $\sim11\%$ decrease in the magnitude of the values of $v_{\rm cen}$ between these snapshots would not be sufficient to fully alleviate this tension.

\S\ref{subsec: sim comp} presented the statistics of the CGOLS V spectrum, which has the same SFR as the fiducial CGOLS SFR spectrum in this work ($20 \rm \ M_\odot\ yr^{-1}$), and the velocity statistics for this spectrum are systemically larger than those from the fiducial CGOLS SFR spectrum. In this case, increasing the resolution of the simulation does not alleviate the tension between the predictions and our measured statistics. There are similar discrepancies to those found in the fiducial (i.e. SFR 20) velocity statistics when comparing the velocity statistics of the SFR 5 and 40 simulations (see \S\ref{subsec: SFR dep}) to the predictions from \cite{xu2022classy} and \cite{chisholm2015scaling} scaling relations. The values of $v_{\rm cen}$ predicted by the scaling relation from \cite{xu2022classy} for SFRs of 5 and $40 \rm \ M_\odot \ yr^{-1}$, respectively, are $188  \ \rm km \ s^{-1}$ and $297 \ \rm km \ s^{-1}$, and the values of $v_{\rm cen}$ predicted from the \cite{chisholm2015scaling} scaling relations are $133 \rm \ km \ s^{-1}$ and $210 \rm \ km \ s^{-1}$. Similar to the discrepancies seen between the fiducial CGOLS SFR dataset and the observational predictions, the velocity statistics from the SFR 40 and SFR 5 datasets are also larger than the predicted values. Though, as mentioned previously, inclination effects and the choice in time snap shot (which is $\rm t=30\ Myr$ for these comparisons) can reduce the magnitude of these discrepancies, they will not bring these statistics for the individual lines into agreement with \cite{chisholm2015scaling}.

In addition to comparing the velocity statistics to predictions to empirical scaling relations, we now make similar comparisons for the EWs. We find that, when comparing our EWs for O I, Si II, III, and IV to those predicted from scaling relations reported in \cite{chisholm2016shining}, we find that our values are smaller by $52, \ 50, \ 80, \ \& \ 97\%$, respectively. While including more of the static ISM and red-shifted components of our line profiles would increase our current reported EW values, it would not increase them enough to be in perfect agreement with those reported in \cite{chisholm2016robust}, nor those predicted by \cite{chisholm2016shining}. It is important to note that, because our fiducial spectrum (and the SFR 5 and 40 spectra) are generated with ``face-on" inclination, we expect that our EWs will be systematically lower than empirical predictions. However, the differences between our fiducial EWs and those from the $\theta_{45}$ spectrum reported in \S\ref{subsec: inclination} range between $0-30\%$, which are significantly smaller than the differences we report here. Similarly, if one were to these comparisons with a later time snap shot ($\rm t=40 \ Myr$), the EWs would be closer to agreeing with the predicted values, but again would not be a significant enough increase for the values to be meaningfully closer to agreement. However, as is shown in Table \ref{tab:V_vs_SFR}, the EWs from the CGOLS V spectrum agree quite well with these predictions, suggesting that the choice in simulation model can significantly reduce the discrepancies between certain mock observables and empirical predictions. Unfortunately, while the effects from increasing the inclination angle would bring both the EWs and velocity statistics closer to the predictions, the time snap shots must move in the opposite direction (earlier for the velocities, and later for the EWs) to address the respective discrepancies. Similarly, while the EWs from the CGOLS V spectrum agree quite well with the predictions, the velocity statistics are in significantly worse agreement with the predicted values than those from the fiducial CGOLS SFR spectrum. 

Overall, it appears that the CGOLS SFR simulations might not contain as much outflowing material as is observed in nearby star-forming galaxies, or that our analysis pipeline is not accurately populating the ionic species in our spectrum (or perhaps a combination of both effects). Despite the apparent lack of mass in the outflows, this analysis demonstrates that the velocities measure from the outflows hosted in our simulations agree quite well with the \cite{xu2022classy} scaling relations. While the discrepancies in the EWs can be addressed by tuning the various parameters we have investigated in this work, it is not clear that we can adequately address the tension by tuning these parameters alone, and there is likely some missing physics in our analysis pipeline (and in the simulations) that must be accounted for in order for the mock observables to be in agreement with empirical predictions.

\subsection{Level Population and Ionizing Background} \label{subsec: level pop}

\begin{figure*}[!ht]
    \centering
    \includegraphics[scale=0.8]{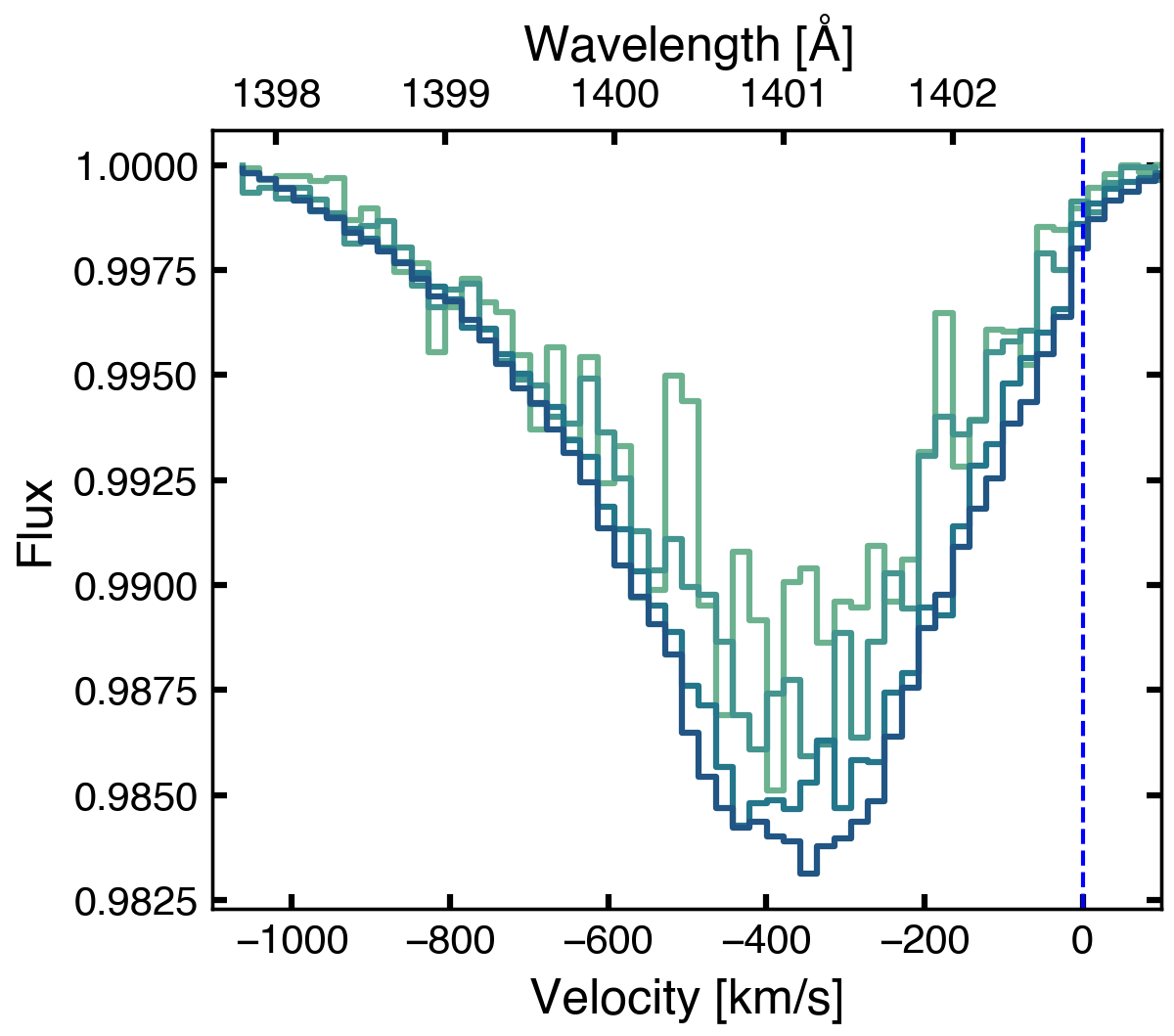}
    \includegraphics[scale=0.8]{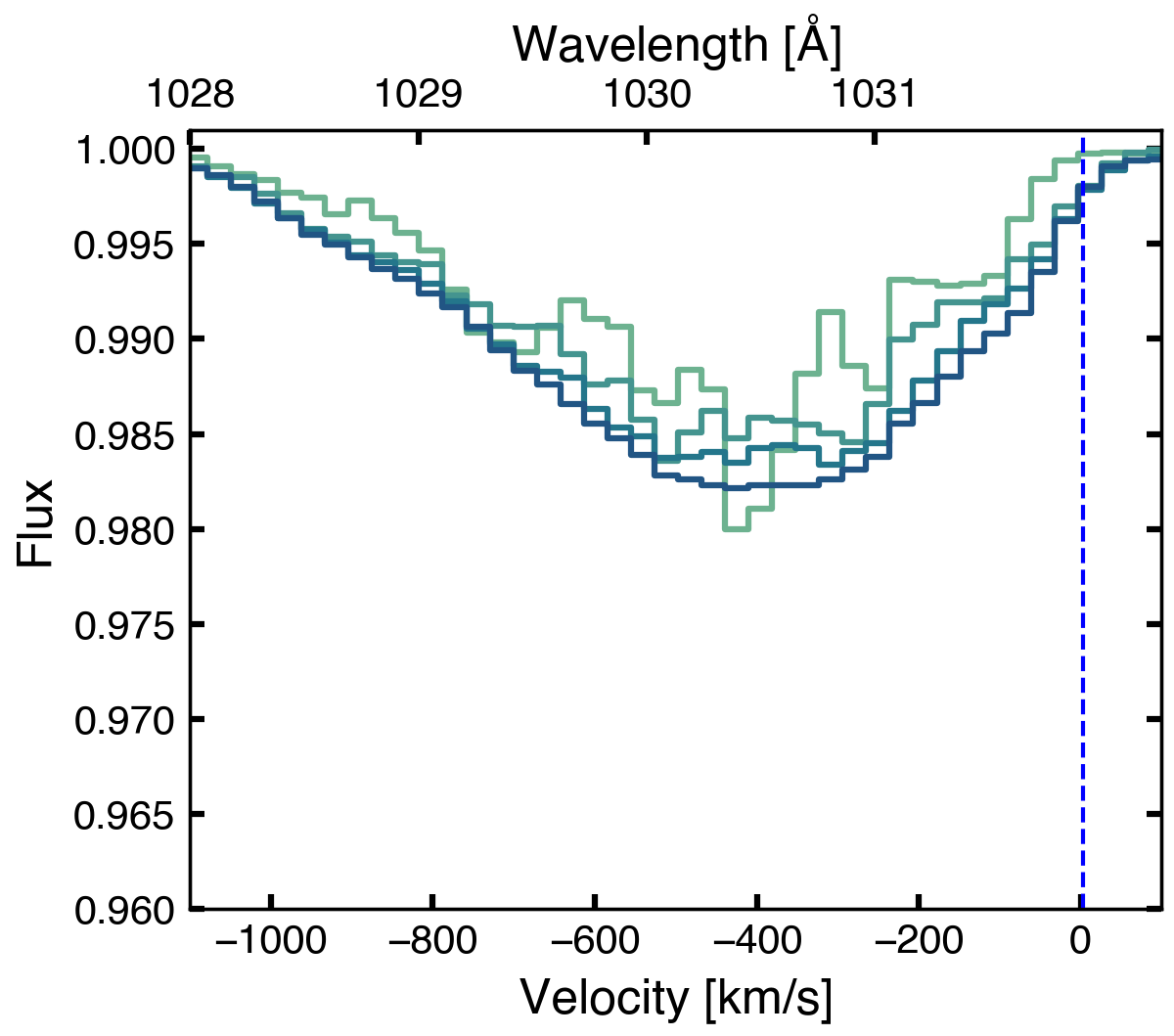}
    \caption{(left) Si IV $\sim1401\text{\AA}$ and (right) O VI $\sim1032\text{\AA}$ absorption line spectra, generated from the $\Delta x=160 \ \mathrm{pc}$ (green), $\Delta x=80 \ \mathrm{pc}$ (light blue), $\Delta x=40 \ \mathrm{pc}$ (blue), and $\Delta x=20 \ \mathrm{pc}$ (dark blue). The line center is also plotted here (in dashed-blue).}
    \label{fig:si IV}
\end{figure*}

As previously mentioned, the EW values for Si IV and O VI (reported in \S\ref{subsec: EW}) are unexpectedly low, in comparison not only to the rest of the absorption features in our study, but, more importantly, to values reported in observational studies. In fact, when comparing the values of Si IV EW from the fiducial CGOLS SFR 20, SFR 40, and CGOLS V spectra to that reported in \cite{chisholm2016shining}, they differ by factors of 15 to 30. Although the lines are clearly underpopulated, the source of the discrepancy is not immediately obvious. One possibility is resolution: if the turbulent mixing layers between the cool and hot phases are insufficiently resolved, there could be missing intermediate temperature gas that would populate these lines via collisional ionization. However, if there is missing mass in these ions from collisionally ionized gas, then there would likely be a strong dependence on simulation resolution in the EWs. In Figure \ref{fig:si IV}, we show the spectra for Si IV and O VI for all simulation resolutions ($\Delta x=20,\ 40,\ 80, \ \& \ 160 \ \rm pc$) from the fiducial simulation snapshot, and do not find any strong relationship between simulation resolution and line-depth, nor do we find such dependence in Table \ref{tab:ews}. In addition to this lack of relation between resolution and EW, we do not expect either ion to be the most predominant species for either element from gas in collisional ionization equilibrium \cite[see Figure 2.1 in][]{gnat2007time}. Therefore, we conclude that the majority of the missing absorption in these ions is likely due to a lack of photoionized gas, which can be explained by the shape and normalization of the background ionizing spectrum.

\begin{figure}[!t]
    \centering
    \includegraphics[scale=0.8]{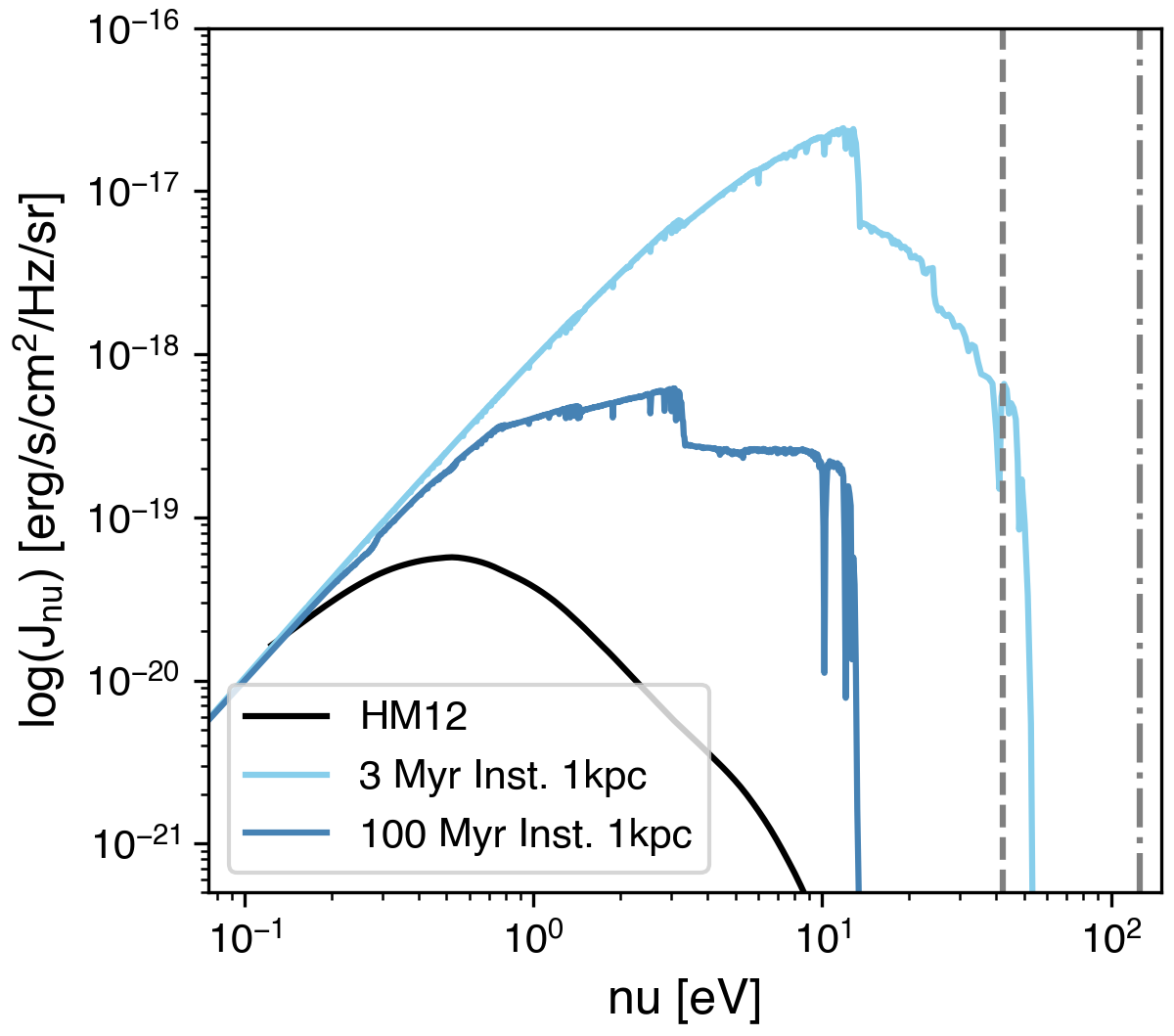}
    \caption{Spectral energy distributions for the ``HM12" Haardt and Madau 2012 metagalactic UV background (black), a $10^6 \ \rm M_{\odot}$ 3 Myr instantaneous starburst model from Starburst99 (light blue), and a $10^6 \ \rm M_{\odot}$ 100 Myr instantaneous starburst model from Starburst99 (blue). Both Starburst99 models are normalized by a distance of 1 kpc from the gas cloud being ionized, and the output spectra are generated from CLOUDY. The ionization energies of Si IV and O VI are marked by the dashed and dot-dashed gray vertical lines, respectively.  Spectra generated via the py4radiation pipeline (see \textit{Software} a link to their GitHub page)} 
    \label{fig:SEDs}
\end{figure}

Figure \ref{fig:SEDs} shows the SEDs for HM12, the fiducial background SED used in \textit{Trident} spectrum generation, as well as two Starburst99 SED models, one from a 3 Myr instantaneous burst, and one from a 100 Myr instantaneous burst. Both Starburst99 models use a $10^6 \ \rm M_{\odot}$ star cluster normalized at 1kpc. In the younger burst models, the bulk of the UV radiation is coming from massive, short-lived young O and B stars, while the older (100 Myr) model represents a more evolved star cluster that is missing the most massive stars, but still has a population of A stars that dominate the UV spectrum. As this comparison shows, there is significantly harder UV radiation present in both of the starburst models than in the HM12 SED, particularly in the 3 Myr model. The 3 Myr instantaneous model is the only one which produces significant radiation at the ionization energy of Si IV (dashed line), and no model produces significant ionizing radiation at that of O VI (dot-dashed line). Though the normalization (i.e. the distance from the clusters) in both of these models is arbitrary, they highlight that, when a sightline is near a (young) star cluster, we expect there to be a much higher flux of UV photons than in the HM12 background.

\begin{figure}[!th]
    \centering
    \includegraphics[scale=0.8]{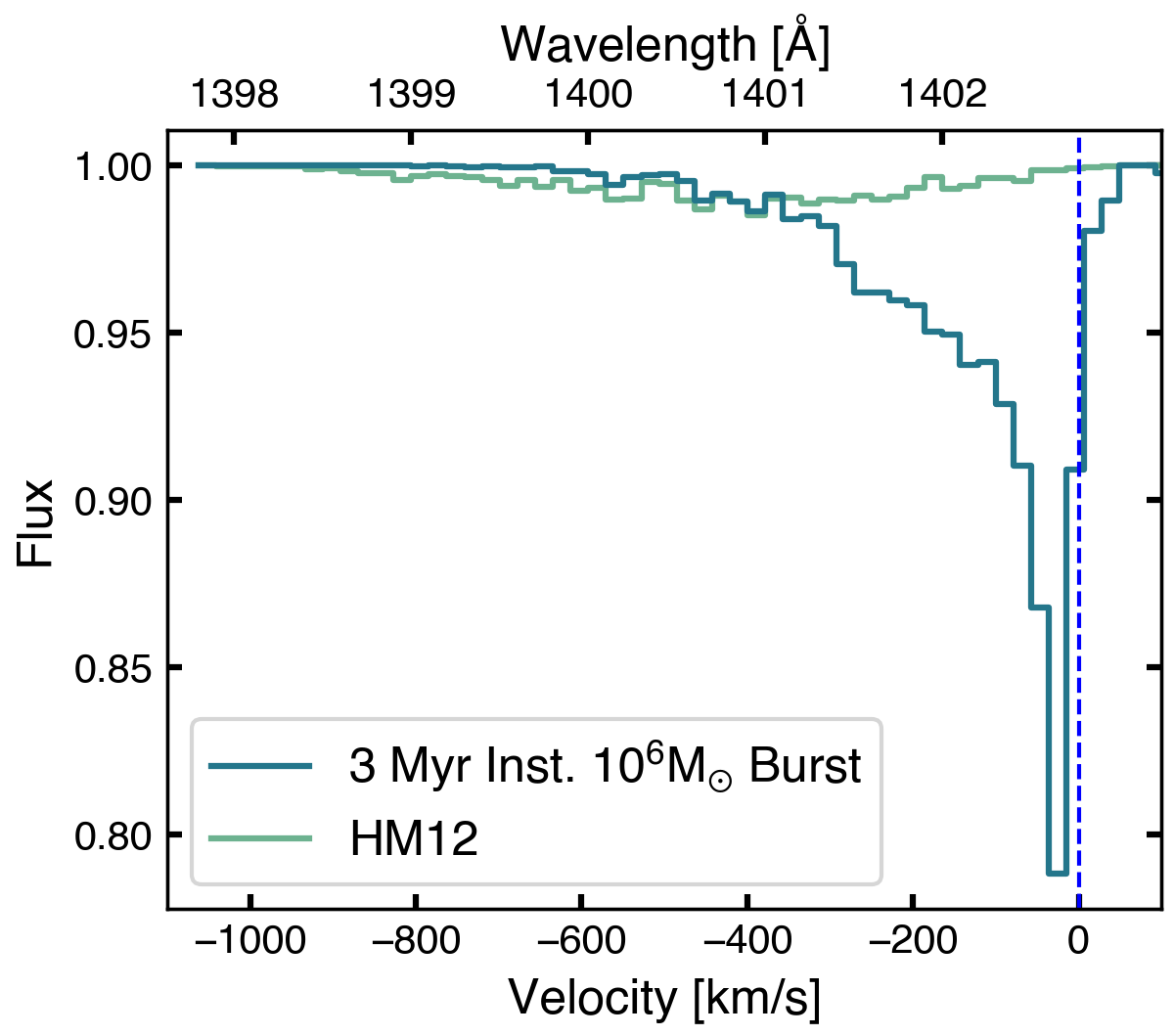}
    \caption{Si IV $\sim1401\text{\AA}$ absorption line spectra, generated from the $\Delta x=160 \ \mathrm{pc}$, with ionization balance set by the fiducial HM12 ionizing background spectrum (green) and a Starburst99 spectrum of a 3 Myr instantaneous $10^6 \rm \ M_{\odot}$ starburst (blue). The line center is also plotted in dashed blue.}
    \label{fig:si IV new BG}
\end{figure}

As a proof-of-concept, in Figure \ref{fig:si IV new BG} we present Si IV absorption lines from two spectra generated from the $\Delta x=160 \ \mathrm{pc}$ resolution simulation, one with ionization states set by the fiducial HM12 ionizing background, and the other with the spectrum of a 3 Myr instantaneous $10^6 \rm \ M_{\odot}$ starburst (see Figure \ref{fig:SEDs}) (ionization table generated via the py4radiation pipeline). It is important to note that this spectrum is a toy model, which assumes that all of the gas in the simulation is 1 kpc away from the ionizing source. This is not intended to be realistic, but rather to demonstrate the role of the ionizing background on the absorption features in our spectra. In this figure, it is clear that the spectrum generated with the star cluster ionizing background produces a significantly stronger (i.e. larger EW) Si IV feature, with a noticeably different velocity structure than our fiducial spectrum. Clearly, the HM12 background is not a sufficient representation of the primary source of ionizing radiation for our dataset, and leads to a significant underestimation of absorption from outflowing, photoionized Si IV. This reflects both the lack of ionizing radiation in our toy model (as shown in Figure \ref{fig:SEDs}), as well as the need for a more sophisticated implementation of our ionizing background spectrum.

\begin{figure*}[!t]
  \centering
  \subfigure[]{\includegraphics[scale=0.8]{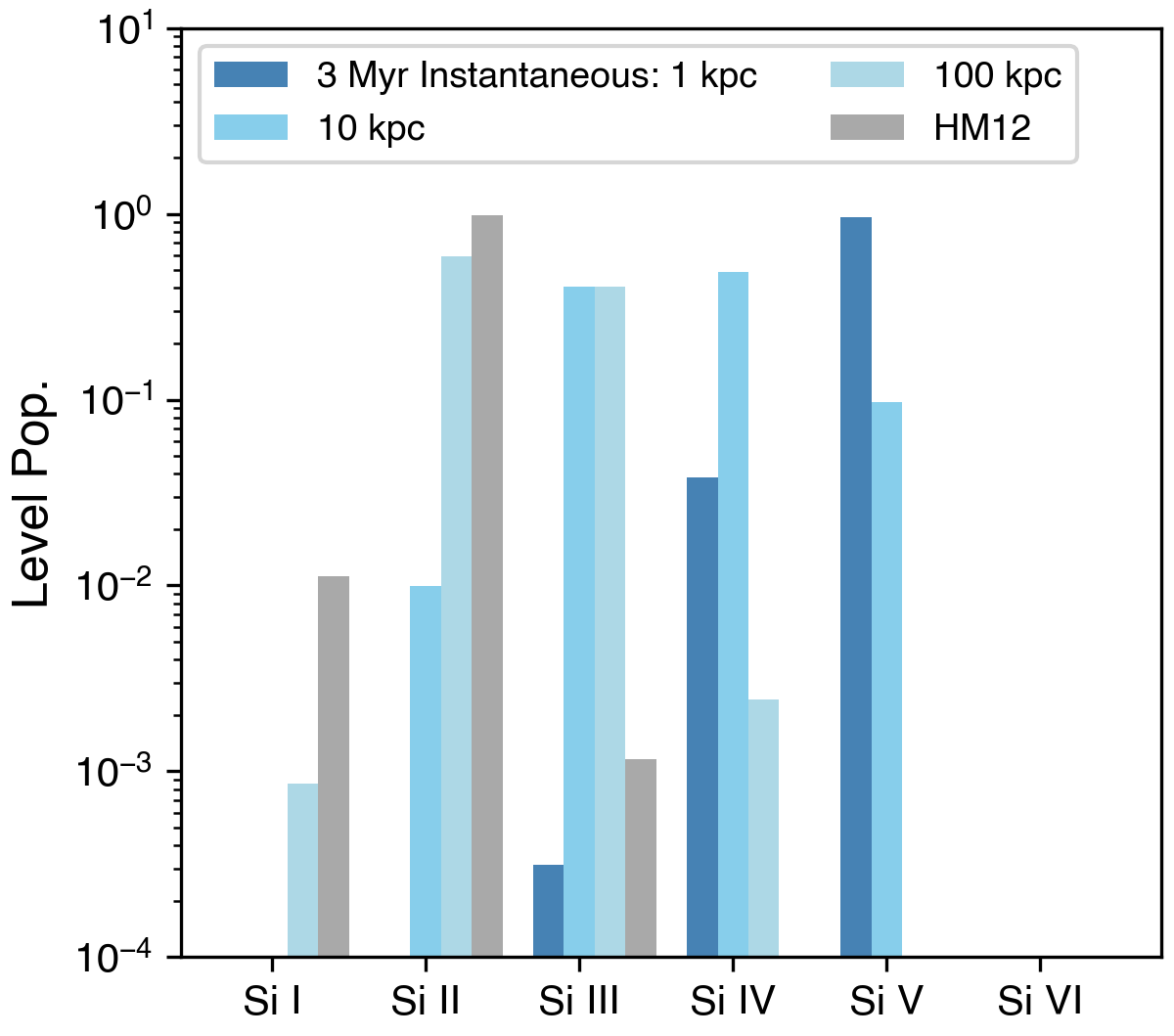}}
  \subfigure[]{\includegraphics[scale=0.8]{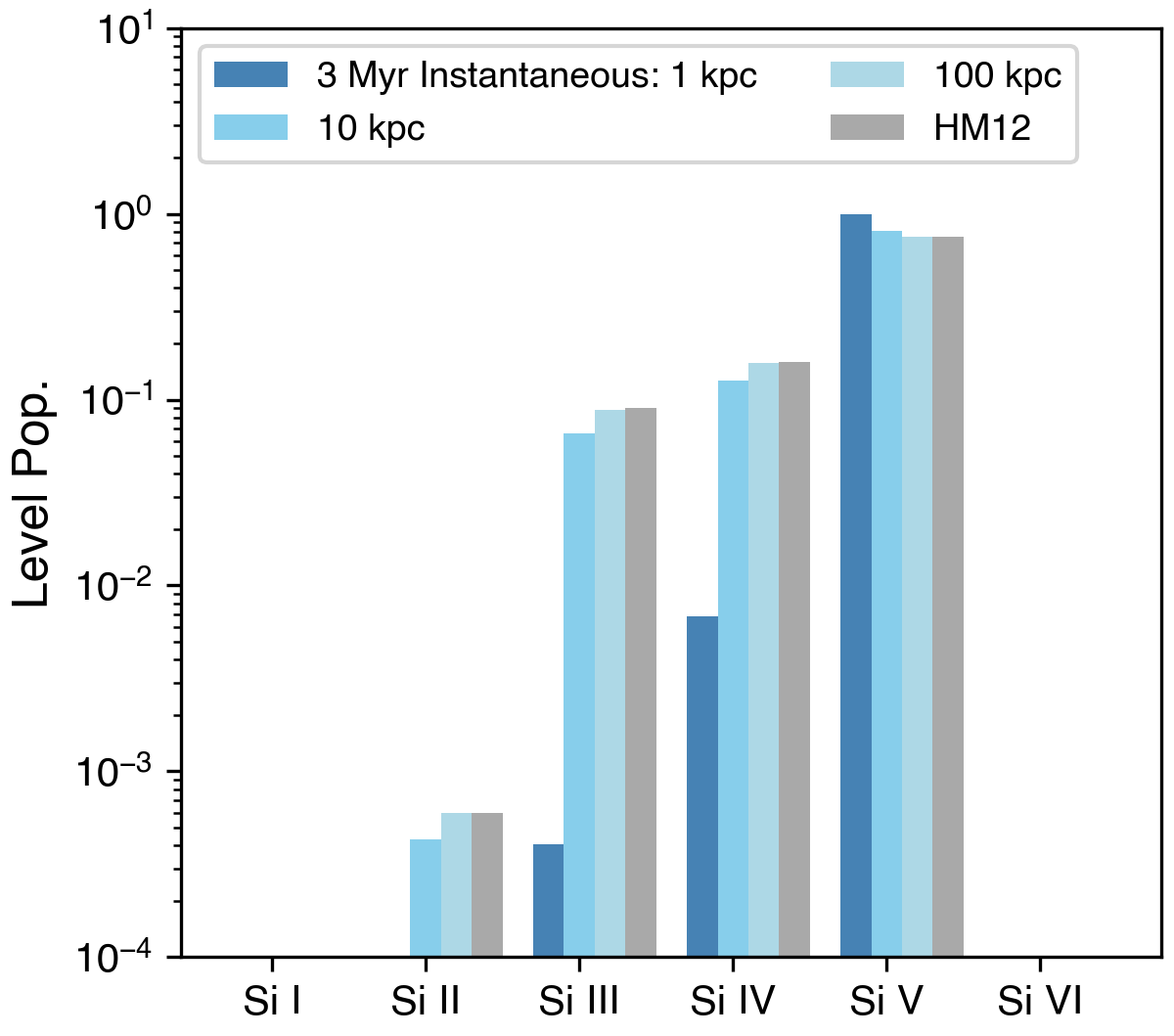}}
  \caption{Fraction of total species population for the first six ionization states of silicon at $\rm T=10^4 \ K$ (left), and at $\rm T=10^5 \ K$ (right). These are generated for four different ionizing background spectra, the first three being the same 3 Myr instantaneous starburst SED model from Starburst99 at a distance of 1, 10, and 100 kpc (steel blue, sky blue, and light blue, respectively), and the Haardt and Madau 2012 metagalactic UV background (gray). Both panels are generated for a $\rm n_H=0.1 \ cm^{-3}$.}
  \label{fig:ions}
\end{figure*}

This effect is further emphasized in Figure \ref{fig:ions}, which shows the level population of silicon generated from CLOUDY models utilizing the HM12 background, as well as the 3 Myr instantaneous starburst SED at distances of 1, 10, and 100 kpc. The HM12 background and the 3 Myr Starburst99 SED produce significantly different level populations of silicon at both $\rm T=10^4$ and $\rm 10^5 \ K$, particularly in Si III and IV. In the left panel of the figure, there is substantial variation in the level populations of the first three ionization states of silicon. Importantly, there is essentially no population of Si IV from the HM12 background, while all of the Starburst99 models produce significant level populations in both Si IV, and the 1 and 10 kpc models also substantially populate Si V. In the right pane of the figure, there is a less drastic contrast in the level population between the 10 and 100 kpc Starburst99 background spectra and HM12, but these models do differ significantly from the 1 kpc Starburst99 model. Though at temperatures near $\rm T=10^5 \ K$ most of the gas should be predominantly collisionally ionized, the more intense background radiation field appears to alter the level population significantly, again highlighting the importance of the ionizing background spectrum for the higher ionization species. The substantial variation in the level population of these elements illustrates that we are indeed missing some physics when generating mock spectra using the HM12 background. While the HM12 background is a reasonable assumption when generating mock pencil-beam spectra at large distances in the CGM, it is not sufficient for modeling the conditions in the ISM of an actively star-forming galaxy. In particular, we need an ionizing background that takes into account the substantial UV flux from young, massive clusters, as well as the optical depth to those clusters along the line of sight. 

In order to address this challenge we must implement a physically motivated ionizing background spectrum that accounts for the distance and optical depth to ionizing sources. This requires generating a suite of \textit{Cloudy} ionization tables \citep{2013RMxAA..49..137F} to be read by \textit{Trident} that can appropriately account for the normalization of the UV backgrounds of the star-forming regions in our simulations, and doing radiative transfer to avoid ``double-counting" photons in the same line of site (i.e. accounting for intervening absorption, escape fraction from star-forming regions, etc.). We plan to address this problem in future work.

\section{Conclusions} \label{sec:conclusions}

In this work we have presented an analysis of the CGOLS SFR simulation suite, including an investigation into how various parameters (SFR, time, and galaxy inclination angle) manifest in the properties of synthetic absorption line spectra. Expanding on the capabilities of the \textit{Trident} package, we have created a resolved aperture from multiple pencil-beam spectra averaged together with a physically-motivated weighting scheme, in order to generate synthetic down-the-barrel spectra of multiphase outflows. We explored the effects of simulation resolution on the resulting mock observables, as well as physical trends in commonly-measured statistics that are tied to the ionization potential of the species in our absorption line suite. Primary findings from this analysis include:

\begin{itemize}
    \item Both $v_{\rm cen}$ and $v_{90}$ increase with ionization potential, and tend to weakly decrease with increasing simulation resolution. Though these statistics can differ from their fiducial values by $\sim10\%$ when the spatial resolution is increased by an order of magnitude, the overall qualitative velocity structure of the absorption features is preserved across all simulation resolutions.
    \item Equivalent width is more sensitive to simulation resolution than the velocity statistics. EWs increase with simulation resolution, primarily as a result of increasing line depth, and the fiducial values of EW decrease by $\sim5-25\%$ with increasing simulation resolution.
    \item As a simulation evolves with time, the EWs of all spectral features increase due to the entrainment of ISM material and mixing between the wind and ambient medium. By contrast, the values of $v_{\rm cen}$ measured from the features grow weakly with time (at most by $\sim10\%$ between time steps), suggesting that, though the velocity structure of the lines does change with time, the overall temporal evolution of these features is better captured by their EWs.
    \item SFR has the strongest effect on the velocity structure and depth of the absorption features in our spectra. The values of $v_{\rm cen}$ and EW increase by factors of $\sim1.6-2$ and $\sim3.8-10$, respectively, when SFR increases from 5 to $40 \ \rm M_{\odot}yr^{-1}$.
    \item When comparing our results to those predicted by empirical scaling relations, we find that the velocity statistics are consistent with the predicted values, especially once inclination effects are taken into account. Though individual lines in the study host velocities ranging between $\sim48-154 \%$ higher than the predicted values from \cite{chisholm2016shining}, the weighted average velocity of all of the lines in our study is $10.5\%$ than the predicted value from \cite{xu2022classy}. 
    \item By contrast, the fiducial EWs are lower than the predictions by $\sim50-97\%$. Some of these discrepancies can be attributed to inclination effects and the choice in time snap shots, but not entirely. For the lower ions, these differences are more likely due to physical differences between the outflows in our simulations, and those observed in real systems. For the higher ions in our study, this is indicative of shortcomings in our current analysis pipeline, motivating our future efforts to improve the physical realism of our pipeline.
    \item The existing UV background ionizing spectrum implemented in \textit{Trident} does not capture the significant UV flux from young stellar clusters. While it is clear that adding a starburst background can increase the amount of photoionized silicon IV, the impact of the assumed ionizing radiation field and its possible contribution to the weak Si IV and O VI lines requires further investigation.
\end{itemize}

In future work, we plan to implement a custom ionizing background spectrum that better reflects the geometry and intensity of the star-forming regions in our simulation, in order to improve the synthetic spectra for intermediate ionization species.

\begin{acknowledgments}

J.P.M. would like to acknowledge Matthew Abruzzo for many helpful conversations, specifically regarding software infrastructure, as well as Helena Richie and Hannah Leary for insightful feedback in the early stages of the writing process for this work. This research was supported in part by the University of Pittsburgh Center for Research Computing and Data, RRID:SCR\_022735, through the resources provided. Specifically, this work used the H2P cluster, which is supported by NSF award number OAC-2117681. E.E.S. acknowledges support from  NASA awards HST-AR-16633 and 80NSSC21K0271, the David and Lucile Packard Foundation, and the Sloan Foundation. S.A.M. acknowledges support from Pitt PACC.

\end{acknowledgments}

\software{\texttt{Cholla} \citep{schneider2015cholla}, \texttt{matplotlib} \citep{Hunter:2007}, \texttt{seaborn} \citep{Waskom2021}, \texttt{numpy} \citep{harris2020array}, \texttt{hdf5} (\href{https://www.hdfgroup.org/}{The HDF Group. Hierarchical Data Format, version 5.}), \texttt{yt} \citep{turk2010yt}, \texttt{Trident} \citep{hummels2017trident}, \texttt{Cloudy} \citep{2013RMxAA..49..137F}, \texttt{astropy} \citep{astropy:2013}, \texttt{py4radiation} (\href{https://github.com/cPhysPlus/py4radiation}{Wladimir Banda Barragan and Daniel Villarruel})}

\vspace{5mm}

\bibliography{sample631}{}
\bibliographystyle{aasjournal}

\end{document}